\documentclass[useAMS]{mn2e}
\usepackage{graphicx}
\usepackage{ulem}
\usepackage{longtable}
\def\eo{SN~2004eo}

\def\a{SN~1992A}

\def\m100{mag/100$^d$}

\def\c57{{$^{57}$Co}\/}

\def\ti44{{$^{44}$Ti}\/}

\begin{document}
 
\title[ESC and KAIT Observations of SN 2004eo] {ESC and KAIT
Observations of the Transitional Type Ia SN 2004eo} 
\author[Pastorello et al. ] {A. Pastorello$^{1,2}$
\thanks{e-mail: pasto@MPA-Garching.MPG.DE}, P. A. Mazzali$^{1,3}$, G. Pignata$^{4}$,
S. Benetti$^{5}$, E. Cappellaro$^{5}$,\and A. V. Filippenko$^{6}$, W. Li$^{6}$, W. P. S. Meikle$^{7}$, 
A. A. Arkharov$^{8}$, G. Blanc$^{5,9}$,\and F. Bufano$^{5,10}$, A. Derekas$^{11}$, M. Dolci$^{12}$,
N. Elias-Rosa$^{1,5,13}$, R. J. Foley$^{6}$,\and 
M. Ganeshalingam$^{6}$, A. Harutyunyan $^{5,10}$, L. L. Kiss$^{11}$,
R. Kotak$^{2,7,14}$,\and  V. M. Larionov$^{8,15}$, J. R. Lucey$^{16}$, N. Napoleone$^{17}$,
H. Navasardyan$^{5}$, F. Patat$^{14}$, \and J. Rich$^{18}$, S. D. Ryder$^{19}$, M. Salvo$^{18}$, B. P. Schmidt$^{18}$,
V. Stanishev$^{20}$, P. Sz\'ekely$^{21}$,\and S. Taubenberger$^{1}$,
S. Temporin$^{22,23}$, M. Turatto$^{5}$, and W. Hillebrandt$^{1}$\\
$^{1}$ Max-Planck-Institut f\"{u}r Astrophysik,
Karl-Schwarzschild-Str. 1, 85741 Garching bei M\"{u}nchen, Germany\\
$^{2}$ Astrophysics Research Centre, School of Mathematics and Physics, Queen's University Belfast, Belfast BT7 1NN, United Kingdom\\
$^{3}$ INAF Osservatorio Astronomico di Trieste, Via Tiepolo 11, 34131 Trieste, Italy\\
$^{4}$ Pontificia Universidad Cat\'olica de Chile, Departamento de
Astronom\'ia y Astrof\'isica, Campus San Joaqu\'in,\\ Vicu\~{n}a Mackenna
4860, Casilla 306, Santiago 22, Chile\\
$^{5}$ INAF Osservatorio Astronomico di Padova, Vicolo dell'Osservatorio 5, 3512
2 Padova, Italy\\
$^{6}$ Department of Astronomy, University of California, Berkeley, CA 94720-3411, USA\\
$^{7}$ Astrophysics Group, Imperial College London, Blackett Laboratory, Prince Consort Road, London SW7 2AZ, United Kingdom\\
$^{8}$ Central Astronomical Observatory of Pulkovo, 196140 St. Petersburg, Russia\\
$^{9}$ APC, UMR 7164 (CNRS, Universit\'e Paris 7, CEA, Observatoire de Paris), 
10, rue Alice Domon et Léonie Duquet, \\ F-75205 Paris Cedex 13, France \\
$^{10}$ Dipartimento di Astronomia, Universit\`a di Padova, Vicolo dell'Osservatorio, 2, I-35122 Padova, Italy\\
$^{11}$ School of Physics A28, University of Sydney, NSW 2006, Australia\\
$^{12}$  INAF Osservatorio Astronomico di Collurania, via M. Maggini, 64100 Teramo, Italy\\
$^{13}$ Universidad de La Laguna, Av. Astrof\'isico Francisco S\'anchez s/n, E-38206 La Laguna, Tenerife, Spain\\
$^{14}$ European Southern Observatory (ESO), Karl-Schwarzschild-Str. 2,
85748, Garching bei M\"{u}nchen, Germany\\
$^{15}$ Astronomical Institute of St. Petersburg University, St. Petersburg, Petrodvorets, Universitetsky pr. 28,\\ 198504 St. Petersburg, Russia\\
$^{16}$ Department of Physics, University of Durham, South Road, Durham DH1 3LE, UK\\
$^{17}$ INAF Osservatorio Astronomico di Roma, Via di Frascati 33, I-00040 Monte Porzio Catone, Italy\\
$^{18}$ Research School of Astronomy and Astrophysics, Australian
National University, Mount Stromlo and\\ Siding Spring Observatories
Cotter Road, Weston Creek, ACT 2611, Australia\\
$^{19}$ Anglo-Australian Observatory, PO Box 296, Epping, NSW 1710, Australia\\
$^{20}$ Department of Physics, Stockholm University, AlbaNova University Center,
SE-10691 Stockholm, Sweden\\
$^{21}$ Department of Experimental Physics \& Astronomical Observatory, University 
of Szeged, H-6720 Szeged, D\'om t\'er 9, Hungary\\
$^{22}$ Institut f\"{u}r Astro- und Teilchenphysik, Leopold-Franzens-Universit\"{a}t
Innsbruck, Technikerstr. 25, A-6020 Innsbruck, Austria\\
$^{23}$ CEA/DSM/DAPNIA, Service d'Astrophysique, Saclay, F-91191 Gif-sur-Yvette Cedex, France
}

 \date{Accepted
.....; Received ....; in original form ....}

\maketitle

\begin{abstract}
We present optical and infrared observations of the unusual Type Ia
supernova (SN) 2004eo. The light curves and spectra closely resemble
those of the prototypical \a, and the luminosity at maximum ($M_B =
-19.08$) is close to the average for a SN~Ia.  However, the ejected
$^{56}$Ni mass derived by modelling the bolometric light curve (about
0.45~M$_\odot$) lies near the lower limit of the $^{56}$Ni mass
distribution observed in normal SNe~Ia.  Accordingly, SN~2004eo shows a
relatively rapid post--maximum decline in the light curve
($\Delta m_{15}(B)_{\rm true} = 1.46$), small expansion velocities in the
ejecta, and a depth ratio Si~II $\lambda$5972 / Si~II $\lambda$6355
similar to that of SN~1992A.  The physical
properties of SN~2004eo cause it to fall very close to the boundary
between the faint, low velocity gradient,
and high velocity gradient subgroups proposed by Benetti et al. (2005). Similar
behaviour is seen in a few other SNe~Ia. Thus, there may in fact
exist a few SNe~Ia with intermediate physical properties.
\end{abstract}

\begin{keywords}
supernovae: general --- supernovae: individual: SN 2004eo --- supernovae:
individual: SN 1992A --- galaxies: NGC 6928 --- photometry --- spectroscopy
\end{keywords}

\section{Introduction}

Thermonuclear supernovae (SNe) are among the most
important cosmological distance indicators; see Filippenko (2005) for
a recent review. Consequently, over the
course of the last few years, much effort has been invested in
obtaining extensive datasets. The modelling of such high--quality
data, particularly in three dimensions, should help improve our
understanding of the thermonuclear explosion mechanisms.

This work adds to the database of observations of nearby Type Ia
supernovae (SNe~Ia), obtained by the European Supernova Collaboration
(ESC)\footnotemark[1] \footnotetext[1]{\sl
http://www.mpa-garching.mpg.de/$\sim$rtn/} as part of a European
Research Training Network (RTN). The first supernova targeted by
the collaboration was SN~2002bo \cite{ben04}.  Since then, 15 other
nearby SNe~Ia (see Pastorello et al. 2007 and references therein) and
one peculiar SN~Ic (SN 2004aw, initially misclassified as a SN~Ia;
Taubenberger at al. \shortcite{tau05}) have been monitored by the ESC.
In addition, three papers based on ESC data and discussing systematic
properties of SNe~Ia have been published.  Benetti et
al. \shortcite{ben05} found evidence that thermonuclear SNe cluster in
three different subgroups on the basis of the observed
spectrophotometric properties [faint, low velocity gradient (LVG), and
high velocity gradient (HVG) SNe~Ia]. In an effort to find clues to explain
this diversity, Hachinger, Mazzali, $\&$ Benetti \shortcite{hach06}
explored the behaviour of a number of other spectroscopic parameters
of SNe~Ia.  They
found that the equivalent width ratios of some spectral lines (mainly
Si~II $\lambda$5972 and $\lambda$6355, S~II, and Fe~II) correlate with 
$\Delta m_{15}(B)$ (as defined by Phillips 1993). Also, using the spectra of a
number of SNe followed by the ESC, Mazzali et al.  \shortcite{maz05}
showed that high--velocity features, whose existence was first proposed
by Hatano et al. \shortcite{hata99} in early--time spectra of SN 1994D,
are actually a common characteristic of SNe~Ia.

The discovery of SN 2004eo provided an ideal opportunity for
increasing the sample of well--monitored, apparently normal SNe~Ia. It
fulfilled the selection criteria specified by the ESC --- it was a
relatively nearby ($v_{\rm rec} < 6000$ km s$^{-1}$) SN discovered well
before maximum light. Moreover, because of its proximity to the
celestial equator, SN 2004eo was observable from both hemispheres, 
allowing us to exploit all telescopes accessible to the various nodes of the
collaboration.  Finally, the location of SN 2004eo at the outskirts of
its host galaxy allowed for accurate photometric measurements and
suggested minimal internal extinction.

In this paper, we present the spectroscopic and photometric
observations of SN~2004eo, ranging from 11~d before maximum light to more 
than one year after maximum.  Most of the data have been collected by the
ESC, with additional contributions (generally photometry) from the SN
groups of the University of California (Berkeley) and of the 
Osservatorio Astronomico di Collurania (Italy).

The paper is organised as follows. In Sect. \ref{obs} we describe the
observations, including instrument details and data reduction
techniques.  In Sect. \ref{lc} we show the optical and infrared
(IR) light curves of \eo, present the colour evolution and bolometric
light curve, and analyse the parameters derived from the
photometry.  Sect. \ref{spec} is devoted to the study of the optical
and IR spectroscopic evolution of \eo.  A discussion and a short
summary follow in Sect. \ref{disc} and Sect. \ref{conc}, respectively.

\section{Observations} \label{obs}

\subsection{SN 2004eo and NGC 6928}
\begin{figure}
\includegraphics[width=8.5cm]{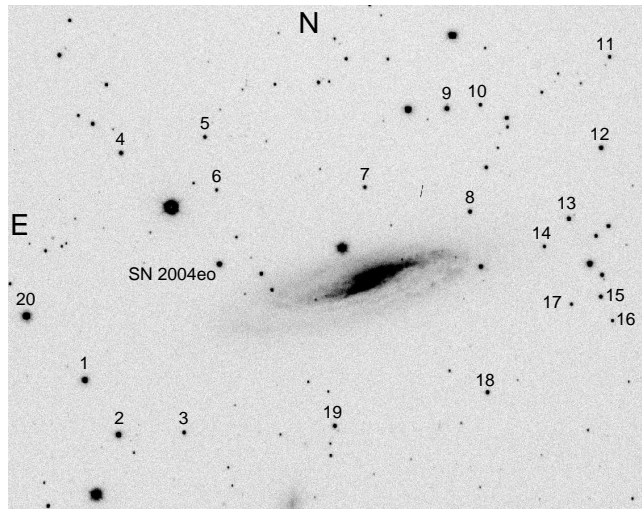}
\caption{SN 2004eo and a local sequence of stars in the field of NGC 6928. 
($V$-band image, obtained on 2004 October 2 with the Northern Optical
Telescope equipped with ALFOSC).  \label{field04eo}} 
\end{figure}

\begin{table}
\begin{center}
\caption{Main Parameters of  NGC 6928 and SN 2004eo. 
\label{gal_param}}
\footnotesize
\begin{tabular}{ccc}\hline \hline
\multicolumn{3}{c}{NGC 6928} \\ \hline 
Galaxy type & SB(s)ab & 1 \\
$\alpha$ (J2000.0) & 20$^{h}$32$^{m}$50$\fs$22 & 1 \\
$\delta$ (J2000.0) & $+09\degr55\arcmin35\farcs1$ & 1 \\ 
B$_{tot}$ & 13.40 & 2 \\
Diameter & $2\arcmin0 \times 0\arcmin6$ & 1 \\
$v_{\rm rad}$ & 4707 km s$^{-1}$ & 3 \\
$v_{\rm Vir}$ & 4809 km s$^{-1}$ & 2 \\
$\mu$$^{\ddag}$& $34.12 \pm 0.10$ &  2 \\ \hline \hline
\multicolumn{3}{c}{SN 2004eo} \\ \hline 
$\alpha$ (J2000.0) & 20$^{h}$32$^{m}$54\fs19 &5  \\
$\delta$ (J2000.0) & +09$\degr$55$\arcmin$42$\farcs$.7 & 5 \\
Offset SN--Galaxy & 59$^{\prime\prime}$.1E, 6$^{\prime\prime}$.5N &5 \\
SN Type & Ia & 6 \\
$E(B-V)_{\rm host}$ & 0 & 7 \\
$E(B-V)_{\rm Gal}$ & 0.109 & 4 \\
Discovery date (UT)& 2004 Sep. 17.5 & 5 \\
Discovery mag. & 17.8 & 5 \\
Predisc. limit date (UT) & 2004 Sep. 12 &5 \\
Limit mag. & 18.5 & 5 \\
$B_{\rm max}$ epoch (UT) & 2004 Sep. 30.7 & 7 \\
$B_{\rm max}$ epoch (JD) & 2,453,279.2 & 7 \\
$B_{\rm max}$ & 15.51 & 7 \\
$M_{B,max}$ & $-19.08$ & 7 \\
$\Delta m_{15}(B)_{\rm obs}$ & 1.45 & 7 \\ 
$\Delta m_{15}(B)_{\rm true}$ & 1.46 & 7 \\ 
$s^{-1}$ & 1.12 & 7 \\ 
$\Delta$$C_{12}$ & 0.49 & 7 \\
$t_r$ & 17.7 & 7 \\ 
$M_{Ni}$ (M$_\odot$) & 0.45 & 7 \\ \hline
\end{tabular}

$^\ddag$ computed with H$_{0}$ = 72 km
s$^{-1}$ Mpc$^{-1}$\\

1 = NED\footnotemark[2]; 
2 = LEDA\footnotemark[3];\\
3 = Theureau et al. \protect\shortcite{theu98};\\
4 = Schlegel et al. \protect\shortcite{schl98};\\
5 = Arbour et al. \protect\shortcite{arb04};\\
6 = Gonzales et al. \protect\shortcite{fol04};\\
7 = this paper.
\end{center}
\end{table}
\footnotetext[2]{\sl http://nedwww.ipac.caltech.edu}
\footnotetext[3]{\sl http://leda.univ-lyon1.fr}

SN 2004eo was discovered by K. Itagaki \cite{arb04} on 17.5 September 
2004 (UT dates are used throughout this paper)
at a magnitude of 17.8.  The SN coordinates are $\alpha
= 20^{h}32^{m}54\fs19$, $\delta = +09\degr55\arcmin42\farcs7$. It lies
$59\farcs1$ E and $6\farcs5$ N of the nucleus of the host galaxy, NGC
6928 (cf. Tab. \ref{gal_param}).  The location of the SN 
(Fig. \ref{field04eo}) suggests small
host-galaxy extinction.  This is supported by spectroscopic evidence
(see Sect. \ref{spec}).

The SN 2004eo early--time spectrum is dominated by Si~II and S~II lines,
showing that the SN was a Type Ia event \cite{fol04}.  The
relatively high velocities in the classification spectrum indicate
that the SN was discovered soon after explosion.  Indeed, the SN was
not detected on 12 September (limiting magnitude 18.5), only a few days
before discovery \cite{arb04}.  This, together with subsequent
observations, means that the light curve of SN~2004eo is one of the
most complete of any known SN.

\subsection{ESC and KAIT Observations}

SN 2004eo was monitored extensively, especially in $B$, $V$, $R$, and $I$
photometry and optical spectroscopy.  Observations cover a period from
about 11~d before $B$-band maximum to about 3 months
after maximum. Henceforth, we adopt as reference the epoch of the $B$-band 
maximum (see below). The $U$ and IR bands were less densely sampled, with the
observations only starting after maximum light. Due to the seasonal
gap, the SN was not observed for three months, between about 100 and
190~d past maximum.  In view of the slow evolution at late (nebular) phases,
subsequent observations were less frequent than near maximum.  On day +228 a 
high-quality nebular spectrum of SN~2004eo was obtained with the VLT.  The
log of all optical and IR spectra is given in Tab. \ref{journal}. The
photometric data are presented in Sect. \ref{lc}.

During the optical monitoring of SN~2004eo, 12 different instruments
were employed, as follows:

\begin{itemize}
\item 2.3-m telescope of the Siding Spring Observatory (Australia),
MSSSO11 imager ($0.59''$ pixel$^{-1}$); for spectroscopy the
Double Beam Spectrograph was used ($2048 \times 512$ pixel E2V CCD4210
detector in the blue arm; Site $1752 \times 532$ pixel detector in the red
arm).
\item 0.76-m Katzman Automatic Imaging Telescope (KAIT; Filippenko et al.
2001) of the Lick
Observatory (Mt. Hamilton, California, USA) (SITe AP7 CCD, $0.8''$ pixel$^{-1}$);
\item Liverpool Telescope (La Palma, Canary Islands, Spain), 
RATCAM (optical CCD, $0.27''$ pixel$^{-1}$);
\item Nordic Optical Telescope (La Palma), ALFOSC (E2V
$2048 \times 2048$ pixel CCD42-40 detector, $0.19''$ pixel$^{-1}$);
\item 1-m telescope of the Siding Spring Observatory (Australia), Wide
Field Camera (Eight $2048 \times 4096$ pixel CCDs, $0.375''$ pixel$^{-1}$);
\item Calar Alto 2.2-m telescope (Spain), CAFOS (Site CCD, $0.53''$ pixel$^{-1}$);
\item Tenagra II 0.81-m telescope (Arizona, US) (Site CCD, $0.87''$ pixel$^{-1}$);
\item Copernico 1.82-m telescope of Mt. Ekar (Asiago, Italy),
AFOSC (TEK $1024 \times 1024$ pixel thinned CCD, $0.473''$ pixel$^{-1}$);
\item ESO/MPI 2.2-m telescope (La Silla, Chile), Wide Field
Imager (mosaic of eight $2048 \times 4096$ pixel CCDs, $0.24''$ pixel$^{-1}$);
\item VLT-Antu telescope (Cerro Paranal, Chile), FORS1 
($2048 \times 2046$ pixel CCD detector, $0.2''$ pixel$^{-1}$);
\item Telescopio Nazionale Galileo (TNG) (La Palma), DOLORES ($2048 \times 4096$ pixel Loral CCD,
$0.275''$ pixel$^{-1}$);
\item William Herschel Telescope (WHT) (La Palma), ISIS
($2048 \times 4096$ pixel EEV12 CCD, $0.19''$ pixel$^{-1}$ in the blue arm;
$2047 \times 4611$ pixel Marconi2 CCD, $0.20''$ pixel$^{-1}$ in the red
arm); and
\item Shane 3-m telescope (Lick Observatory), Kast Double Spectrograph (two
UV-flooded Reticon $1200 \times 400$ pixel devices (one per arm), $0.8''$
pixel$^{-1}$).
\end{itemize}

In addition, six different instrumental configurations were used to collect
the IR data, as follows:

\begin{itemize}
\item AZT--24 1.08-m telescope of Campo Imperatore (Italy)
SWIRCAM ($1.04''$ pixel$^{-1}$);
\item 3.58-m Telescopio Nazionale Galileo (TNG) (La Palma), NICS
(HgCdTe Hawaii $1024 \times 1024$ pixel array, $0.25''$ pixel$^{-1}$);
\item Anglo--Australian Telescope (AAT) of the Siding Spring Observatory, 
IRIS2 ($1024 \times 1024$ pixel HgCd Te array, $0.45''$ pixel$^{-1}$);
\item WHT (La Palma), LIRIS (Hawaii $1024 \times 1024$ pixel HgCdTe array, $0.25''$
pixel$^{-1}$);
\item Calar Alto 3.5-m telescope, Omega2000 ($2048 \times 2048$ pixel
HAWAII-2 CCD, $0.45''$ pixel$^{-1}$); and
\item the VLT-Antu module, ISAAC ($1024 \times 1024$ pixel InSb Aladdin array,
  $0.148''$ pixel$^{-1}$).
\end{itemize}

\begin{table*}
\caption{Optical and IR Spectroscopic Observations of \eo. \label{journal}}
\tiny
\begin{tabular}{ccccccccc}\hline \hline
Date & JD-- & Phase$^\dag$ & Instrumental & Grism or & Resolution & Exptime &
 Range & Standard \\
 & 2,400,000& (days) & configuration & grating & (\AA) & (s) & (\AA) & star \\ \hline
19/09/04 & 53268.05 & -11.3 & SSO2.3m+DBS & B+R & 4.8,4.8 & 1200(x2)+1200(x2) & 3500--9200 & Feige110 \\
24/09/04 & 53272.74 & -6.6 & Lick-Shane3m+Kast & $^\ddag$&6.2,9.6 &1000+1000 & 3340--10700 &BD+284211,BD+174708 \\
27/09/04 & 53276.39 & -2.9 & WHT+ISIS & R300B+R158R & 3.6,6.5 & 900(x2)+900(x2) & 3070--10670 &BD+284211 \\
02/10/04 & 53281.45 & +2.2 & NOT+ALFOSC  & gm4+5 & 21,20 &1500+1500 & 3360--9810 &BD+284211 \\
07/10/04 & 53286.36 & +7.1 & CAHA2.2m+CAFOS & B+R200 & 12,11 &1800+1800 & 3180--10240 &Feige110 \\
11/10/04 & 53290.35 & +11.1& CAHA2.2m+CAFOS & B+R200 & 12,11 &1800+1800 & 3500--10280 &Feige110 \\
13/10/04 & 53291.79 & +12.5& Lick-Shane3m+Kast & $\ddag$&4.9,9.6 &1500+1500 & 3330--10400 &BD+284211,BD+174708 \\
14/10/04 & 53293.40 & +14.1& CAHA2.2m+CAFOS & B+R200 & 12,11 &1800+1800 & 3420--10300 &Feige110 \\
21/10/04 & 53300.37 & +21.1& NOT+ALFOSC  & gm4 &14.5 &1200(x2) & 3290--9230 &BD+174708\\
22/10/04 & 53300.95 & +21.7& SSO2.3m+DBS & B+R &4.8,4.8 &1200+1200 & 3390--9200 &Feige110 \\
24/10/04 & 53302.95 & +23.7& SSO2.3m+DBS & B+R &4.8,4.8 &1200(x2)+1200(x2) & 3380--9200 & Feige110\\
30/10/04 & 53309.38 & +30.1& NOT+ALFOSC  & gm4 &14.5 &1200x2 & 3460--9230 &BD+174708 \\
15/11/04 & 53325.32 & +46.0& CAHA2.2m+CAFOS & B200 & 12 & 2700& 3380--8760 &Feige110 \\
18/11/04 & 53328.32 & +49.0& Ekar1.82m+AFOSC & gm4 & 24 & 3600& 3710--7800 &-- \\
19/11/04 & 53329.35 & +50.1& NOT+ALFOSC  & gm4 & 14.5 &1800 & 3480--8910 &-- \\ 
23/11/04 & 53333.31 & +54.0& CAHA2.2m+CAFOS & B+R200 &12,11 &2700+2700 & 3170--9810 &Feige110 \\
07/12/04 & 53347.24 & +67.9& Ekar1.82m+AFOSC & gm4 & 24 &3600 & 3730--7800 &BD+284211 \\ 
11/12/04 & 53351.37 & +72.1& NOT+ALFOSC  & gm4 & 19 & 3600 & 3440--8900 & BD+174708 \\
16/05/05 & 53506.89 & +227.6& VLT-UT2+FORS1 & 300V+GG375&11.5 & 2280+2280& 3630--8900 &LTT7987 \\   \hline \hline
Date & MJD & Phase & Instrumental & Grism or & Resolution & DITxNDITxN(ABBA) &
 Range & Standard \\
 & & (days) & configuration & grating & (\AA) & Tot. exptime (s) &  (\AA) & star \\ \hline
02/10/04 & 53280.91 & +2.1 & TNG+NICS & Amici prism& -- & 2400 & 7500--25000 & BD+472802 \\
22/10/04 & 53300.45 & +21.7& AAT+IRIS2 & gmJs+Hs+K & -- & 1440+1440+1440 & 10430--24800 & BS7793\\  
05/11/04 & 53314.84 & +36.0& WHT+LIRIS & gmzJ+HK & -- & 2400+960 & 8900--23950 & HD194012\\ \hline
\end{tabular}

$^\dag$ The phase is with respect to the B-band maximum (JD=2,453,279.2);
$^\ddag$ Grt300/7500+Gm600/4310
\end{table*}

The use of a large number of different instruments required careful
homogenisation of our dataset. This is discussed in Sect. \ref{lc}

\subsection{Data Reduction}

All optical photometric images were pre--reduced (i.e., overscan, bias,
flatfield corrected, and trimmed) using standard IRAF\footnotemark[4]
\footnotetext[4]{IRAF is distributed by the National Optical Astronomy
Observatories, which are operated by the Association of Universities
for Research in Astronomy, Inc, under contract to the National Science
Foundation.} procedures.  Because of the high brightness of the night
sky in the IR, both imaging and spectroscopy required some additional
data processing, including
the creation of clean sky images and their subtraction from
the target frames.  Finally, the resulting images were spatially
registered and combined in order to improve the signal--to--noise
ratio (S/N).

The SN lies in a region only marginally contaminated by the 
host-galaxy light, as well as being clear of stars in the field.
Consequently, point--spread function (PSF) fitting is well suited to
provide excellent estimates of the SN magnitudes. 
A detailed description of the photometric data
reduction techniques can be found in Pastorello et
al. \shortcite{pasto05}. The PSF measurements were 
performed using SNOoPY\footnotemark[5].
\footnotetext[5]{SNOoPY, originally presented in Patat \shortcite{nando96o}, 
has been implemented in IRAF by E. Cappellaro. The
package is based on DAOPHOT, but optimised for SN magnitude
measurements.}  The SN magnitudes were then measured with
reference to a local sequence of 20 stars in the field of NGC 6928
(Fig. \ref{field04eo}). These, in turn, were calibrated by comparison
with photometric standard stars from the Landolt
\shortcite{land92} catalogue.  The $U$, $B$, $V$, $R$, and $I$ magnitudes 
of the local sequence stars, obtained averaging the measurements
of $\sim$20 photometric nights, are in Tab. \ref{stand_seq}.

The IR photometry was calibrated using 2MASS magnitudes for a number of
local standards in the field of NGC 6928, together with additional
calibrations from standard fields \cite{hunt98} observed on the
same nights as the SN.  The $K'$ magnitudes of NICS and Omega2000 were
converted to $K$ magnitudes using the relations derived by Wainscoat
$\&$ Cowie \shortcite{wain92}.

The first steps of the spectroscopic data reduction (overscan and bias
corrections, flatfielding, and trimming) are the same as for the
photometry. For the IR spectroscopy, the contribution of the night
sky was subtracted from the 2D IR spectrum using as reference a
spectrum from a different position along the slit.  Further data
processing was performed using standard IRAF procedures, in particular
some tasks from the package CTIOSLIT. After the optimised extraction
performed with the task APALL, the 1D spectra were wavelength--calibrated 
by comparison with spectra of arc lamps obtained
during the same night and with the same instrumental configuration. In 
addition, the wavelength calibration was checked
against bright night--sky emission lines.  

The SN spectra were then flux--calibrated using response 
curves derived from the spectra of standard stars \cite{hamu94,mass88} 
possibly observed during the same night. The spectra of the standard 
stars were also used to remove telluric features.  The flux
calibration of the spectra was finally checked against the photometry and, in
case of discrepancy, the spectral fluxes were rescaled to match the
photometry. This step is particularly important for the IR spectra,
where flux tables for spectroscopic standard stars are usually not 
available. The typical deviation from the photometric
values was less than 10$\%$ (but it can be up to 20$\%$ in the IR
spectra).

\begin{table*}
\caption{$U$, $B$, $V$, $R$, $I$ magnitudes of the local standards in the field
of NGC 6928 (see Fig. \ref{field04eo}). The uncertainties in brackets are the
root mean square of the individual exposures used to estimate the average 
magnitudes. 
\label{stand_seq}}
\begin{tabular}{cccccc}\hline \hline
ID star & $U$ & $B$ & $V$ & $R$ & $I$ \\ \hline 
1 & 16.506 (0.011) &16.505 (0.007)&15.873 (0.006)  &15.472 (0.007)&15.132	(0.008)\\
2 & 16.907 (0.006) &16.729 (0.016)&16.051 (0.005)  &15.624 (0.005)&15.266	(0.006)\\
3 & 18.903 (0.018) &18.491 (0.008)&17.690 (0.007)  &17.197 (0.008)&16.770	(0.008)\\
4 & 18.244 (0.019) &18.009 (0.008)&17.283 (0.009)  &16.829 (0.008)&16.437	(0.011)\\
5 & 18.751 (0.009) &18.581 (0.007)&17.877 (0.007)  &17.433 (0.010)&17.056	(0.011)\\
6 & 18.812 (0.018) &18.998 (0.016)&18.473 (0.009)  &18.061 (0.011)&17.716	(0.010)\\
7 & 18.899 (0.016) &18.647 (0.010)&17.957 (0.011)  &17.513 (0.012)&17.150	(0.010)\\
8 & 18.885 (0.008) &18.278 (0.016)&17.427 (0.012)  &16.840 (0.010)&16.349	(0.009)\\
9 & 18.896 (0.016) &17.720 (0.010)&16.622 (0.009)  &15.918 (0.010)&15.312	(0.008)\\
10& 18.828 (0.021) &18.637 (0.007)&17.920 (0.011)  &17.501 (0.013)&17.103	(0.011)\\
11& 18.663 (0.025) &18.734 (0.017)&18.038 (0.009)  &17.682 (0.015)&17.340	(0.017) \\
12& 17.907 (0.013) &17.819 (0.012)&17.095 (0.011)  &16.684 (0.012)&16.301	(0.010) \\
13& 18.809 (0.030) &18.115 (0.011)&17.196 (0.010)  &16.695 (0.011)&16.236	(0.011)\\
14& 19.410 (0.029) &19.096 (0.012)&18.311 (0.008)  &17.800 (0.014)&17.373	(0.011) \\
15& 18.232 (0.013) &18.274 (0.006)&17.650 (0.011)  &17.235 (0.014)&16.829	(0.012) \\
16& 19.666 (0.035) &19.591 (0.020)&18.802 (0.012)  &18.334 (0.014)&17.916	(0.016) \\
17& 19.757 (0.022) &19.241 (0.016)&18.348 (0.012)  &17.780 (0.013)&17.289	(0.009) \\
18& 18.618 (0.008) &18.350 (0.007)&17.562 (0.011)  &17.101 (0.011)&16.684	(0.008)\\
19& 18.408 (0.019) &18.248 (0.009)&17.553 (0.008)  &17.084 (0.009)&16.699	(0.007)\\
20& 15.597 (0.014) &15.562 (0.007)&14.923 (0.005)  &14.544 (0.005)&14.215 	(0.010)\\ \hline

\end{tabular}
\end{table*}

\section{Photometry} \label{lc}

\subsection{S--Correction} \label{scor}

Owing to the large number of different instruments used during the
optical follow--up observations of SN 2004eo, it was prudent to reduce the
photometry to a standard system applying a well-tested procedure called 
``S--correction,'' a technique which has already been applied 
to a number of SNe~Ia (see, e.g., Stritzinger et
al. \shortcite{max02}, Pignata et al. \shortcite{pig04}, Krisciunas et
al \shortcite{kris03,kris04}, Elias--Rosa et al. \shortcite{nancy06}, 
Pastorello et al. \shortcite{pasto05}, Stanishev et al. \shortcite{sta05}). 

Actually, despite the large
number of instrumental configurations used, the $U$, $B$, $V$ photometry of
SN~2004eo appeared to be fairly homogeneous even without S--correction.  In
contrast, the original $R$-band and, in particular, $I$-band light curves
show large scatter owing to differences in the filter transmission
curves between the various instruments, especially for data obtained with
CAFOS and ALFOSC.  For details of the filter transmission curves, see
Pastorello et al. \shortcite{pasto05} and Li et
al. \shortcite{li01}.

Application of the S--correction to the $R$ and $I$-band data reduced
significantly the scatter in the light curves. Only S--corrected $B$, 
$V$, $R$, and $I$ light curves are considered in our analysis. However, 
for the $U$ and IR $J$, $H$, and $K$ bands, S--corrections were not
feasible because of the incomplete spectroscopic wavelength coverage 
and/or sparse temporal sampling. The original optical magnitudes of 
SN~2004eo are listed in
Tab. \ref{opt_photometry}, while the S--corrections applied to the $B$,
$V$, $R$, and $I$ magnitudes of Tab. \ref{opt_photometry} are in
Tab. \ref{S_corr}.
 
\begin{figure*}
\includegraphics[width=8.8cm,angle=0]{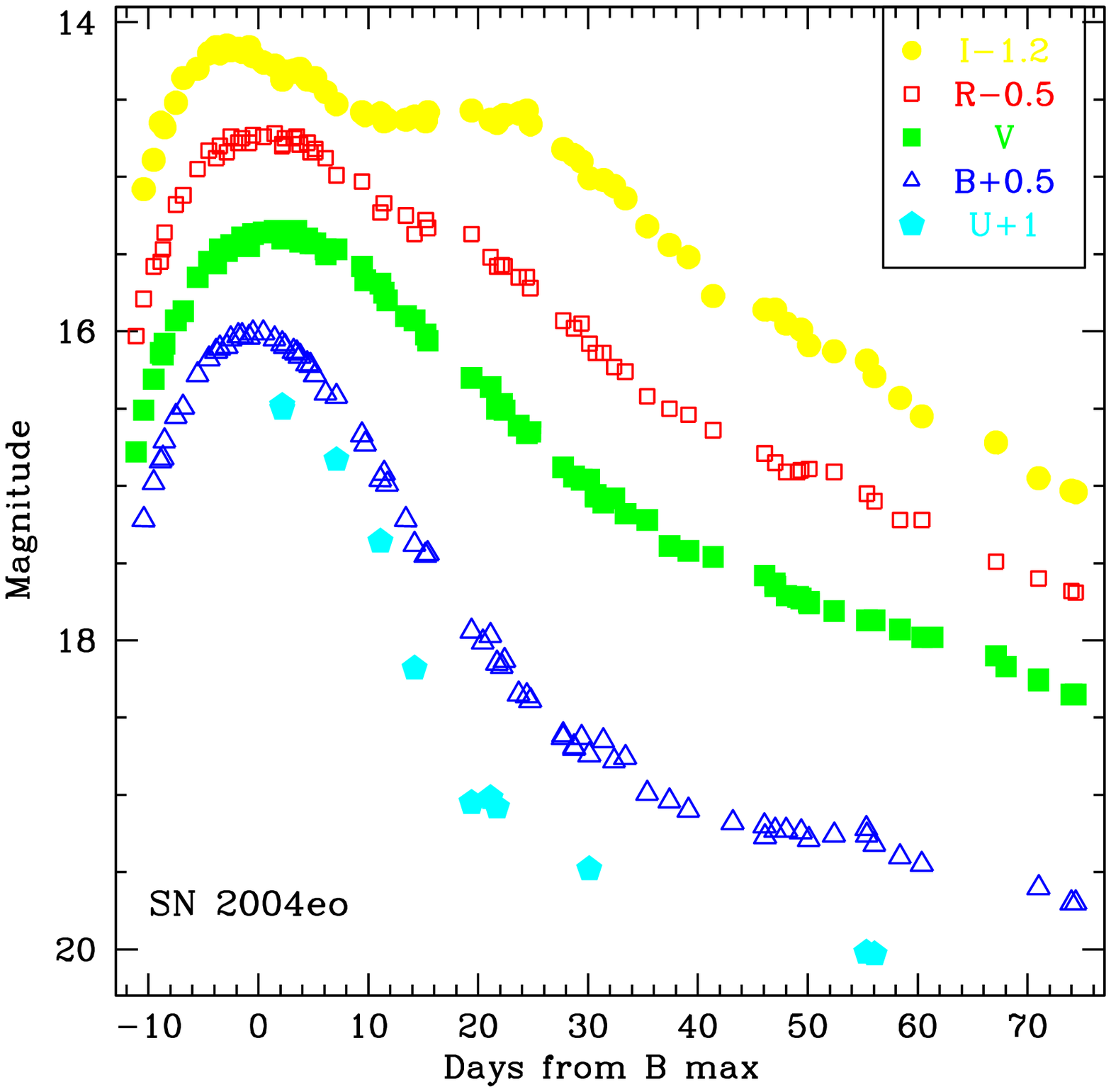}
\includegraphics[width=8.8cm,angle=0]{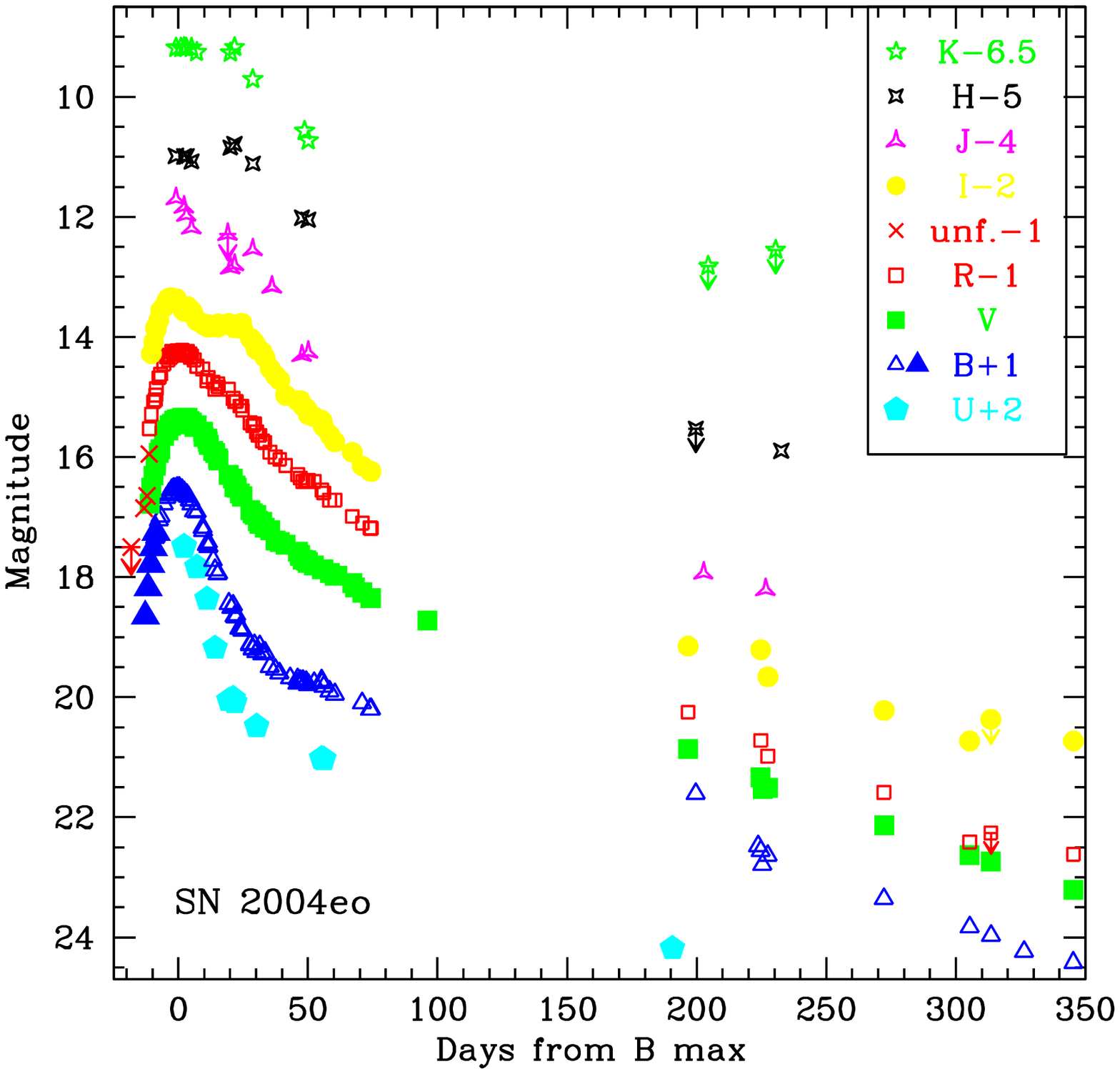}
\caption{Left: Early--time ($\sim$3 months) $UBVRI$ light curves of
SN~2004eo. Right: Complete $UBVRIJHK$ light curves of SN~2004eo, up to
almost one year past explosion. The very early unfiltered magnitudes of
Itagaki \protect\cite{arb04} (crosses) and the early $B$-band
observations of Gonzales et
al. \protect\shortcite{fol04} (filled triangles) are also included. A few 
detection limits are shown. S--correction (see
Sect. \ref{scor}) and K--correction (Sect. \ref{kcor}) have been
applied to the optical light curves (for the $U$-band photometry,
K--correction only).
\label{light_cur} }
\end{figure*}

\subsection{K--Correction} \label{kcor}

The redshift of the host galaxy of SN~2004eo, $z = 0.016$, is sufficiently
large to produce a significant effect on the observed magnitudes.
In order to compare the intrinsic luminosities and colours of
different SNe~Ia, it is necessary to correct the photometry to the rest frame. 
This can be done comparing the
fluxes of unredshifted and redshifted spectra of the same event
through standard spectral bandpasses. 
In the literature this technique is usually referred to as ``K correction,''
and detailed descriptions of the application to SNe~Ia can be found in 
Leibundgut \shortcite{bruno90}, Hamuy et al. \shortcite{ham93}, Kim et
al. \shortcite{kim96}, and Nugent et al. \shortcite{nug02}.
 
\onecolumn 
\begin{longtable}{cccccccc}
\caption{\label{opt_photometry} Original optical photometry of SN 2004eo.} \\ \hline \hline
Date & JD--2,400,000 & $U$ & $B$ & $V$ & $R$ & $I$ & Instrument \\ \hline
19/09/04&  53268.07 & ~~~~~~~~~~~~~~~~~~~~ &  & 16.751 (0.010) & 16.484 (0.013) & ~~~~~~~~~~~~~ &  0 \\
20/09/04&  53268.76 & & 16.689 (0.021) & 16.485 (0.023) & 16.238 (0.033) & 16.226 (0.037)  &  1\\
21/09/04&  53269.68 & & 16.447 (0.012) & 16.282 (0.010) & 16.025 (0.013) & 16.034 (0.013)  &  1\\
21/09/04&  53270.33 & & 16.309 (0.016) & 16.133 (0.015) & 15.978 (0.012) & 15.833 (0.024)  &  2\\
22/09/04&  53270.52 & & 16.286 (0.014) & 16.141 (0.010) & 15.902 (0.014) & 15.839 (0.010)  &  2\\
22/09/04&  53270.67 & & 16.188 (0.042) & 16.050 (0.016) & 15.807 (0.016) & 15.825 (0.020)  &  1\\
23/09/04&  53271.69 & & 16.029 (0.014) & 15.897 (0.012) & 15.621 (0.014) & 15.676 (0.016)  &  1\\
23/09/04&  53272.36 & & 15.966 (0.013) & 15.858 (0.009) & 15.556 (0.013) & 15.549 (0.011)  &  2\\
25/09/04&  53273.66 & & 15.759 (0.016) & 15.625 (0.018) & 15.392 (0.018) & 15.465 (0.031)  &  1\\
26/09/04&  53274.69 & & 15.662 (0.018) & 15.532 (0.016) & 15.272 (0.019) & 15.384 (0.032)  &  1\\
26/09/04&  53275.37 & & 15.616 (0.012) & 15.552 (0.011) & 15.301 (0.014) & 15.402 (0.011)  &  2\\
27/09/04&  53275.70 & & 15.597 (0.019) & 15.451 (0.012) & 15.237 (0.015) & 15.398 (0.017)  &  1\\
27/09/04&  53276.33 & & 15.588 (0.013) & 15.476 (0.010) & 15.264 (0.013) & 15.403 (0.010)  &  2\\
28/09/04&  53276.70 & & 15.536 (0.024) & 15.426 (0.014) & 15.179 (0.014) & 15.383 (0.019)  &  1\\
28/09/04&  53277.37 & & 15.514 (0.015) & 15.445 (0.009) & 15.198 (0.017) & 15.424 (0.009)  &  2\\
29/09/04&  53277.70 & & 15.519 (0.015) & 15.370 (0.015) & 15.185 (0.010) & 15.398 (0.017)  &  1\\
29/09/04&  53278.36 & & 15.534 (0.012) & 15.453 (0.010) & 15.193 (0.012) & 15.430 (0.010) &  2\\
30/09/04&  53278.71 & & 15.504 (0.017) & 15.351 (0.020) & 15.160 (0.013) & 15.426 (0.012) &  1\\
01/10/04&  53279.66 & & 15.506 (0.021) & 15.339 (0.014) & 15.167 (0.014) & 15.473 (0.015) &  1\\
02/10/04&  53280.69 & & 15.545 (0.015) & 15.329 (0.014) & 15.147 (0.025) & 15.492 (0.020) &   1\\
02/10/04&  53281.37 & & & 15.387 (0.010) & & &  3\\
02/10/04&  53281.38 & 15.425 (0.018) &  & 15.385 (0.008) & 15.217 (0.010) & 15.613 (0.009) &   3\\
02/10/04&  53281.39 & 15.439 (0.018) & 15.569 (0.010) & 15.391 (0.009) & 15.205 (0.012) & 15.617 (0.012) &3\\
03/10/04&  53281.63 &  & 15.594 (0.018) & 15.338 (0.012) & 15.171 (0.016) & 15.534 (0.021) &   1\\
03/10/04&  53282.43 &  & 15.624 (0.011) & 15.358 (0.013) & 15.145 (0.013) & 15.587 (0.009) &   2\\
04/10/04&  53282.68 &  & 15.634 (0.055) & 15.337 (0.012) & 15.160 (0.015) & 15.527 (0.011) &   1\\
04/10/04&  53282.98 &  & 15.663 (0.019) & 15.404 (0.013) & 15.222 (0.015) & 15.509 (0.035) &   4\\
05/10/04&  53283.66 &  & 15.712 (0.011) & 15.386 (0.009) & 15.204 (0.011) & 15.569 (0.016) &   1\\
05/10/04&  53283.94 &  & 15.721 (0.011) & 15.421 (0.009) & 15.273 (0.011) & 15.565 (0.011) &   4\\
05/10/04&  53284.36 &  & 15.780 (0.018) &  & 15.224 (0.013) & 15.626 (0.018) &   2\\
05/10/04&  53284.39 &  & 15.777 (0.011) & 15.434 (0.009) & 15.211 (0.010) & 15.653 (0.009) &   2\\
06/10/04&  53285.33 &  & 15.895 (0.011) & 15.473 (0.010) & 15.267 (0.013) & 15.722 (0.012) &   2\\
06/10/04&  53285.34 &  &  & 15.512 (0.130) &  &  & 5$^\ddag$\\
07/10/04&  53286.30 & 15.946 (0.018) & 15.925 (0.011) & 15.490 (0.009) & 15.420 (0.014) & 15.570 (0.013) & 6\\
10/10/04&  53288.63 &  & 16.182 (0.013) & 15.577 (0.013) & 15.459 (0.020) & 15.766 (0.035) &   1\\
10/10/04&  53288.92 &  & 16.243 (0.016) & 15.675 (0.009) &  & 15.777 (0.012) & 4\\
11/10/04&  53290.30 & 16.492 (0.018) & 16.492 (0.016) & 15.726 (0.047) & 15.677 (0.038) & 15.648 (0.022) & 6\\
12/10/04&  53290.63 &  & 16.446 (0.012) & 15.747 (0.013) & 15.611 (0.012) & 15.825 (0.011) &  1\\
12/10/04&  53290.92 &  & 16.509 (0.011) & 15.804 (0.011) &  & 15.811 (0.011) &  4\\
14/10/04&  53292.63 &  & 16.738 (0.015) & 15.902 (0.013) & 15.702 (0.011) & 15.817 (0.016) &   1\\
14/10/04&  53293.42 & 17.282 (0.019) & 16.906 (0.016) & 15.959 (0.011) & 15.832 (0.022) & 15.683 (0.013) & 6\\
15/10/04&  53294.42 &  & 16.962 (0.011) & 16.036 (0.013) & 15.665 (0.011) & 15.919 (0.011) & 2\\
16/10/04&  53294.62 &  & 16.961 (0.027) & 16.057 (0.014) & 15.782 (0.015) & 15.766 (0.015) & 1\\
20/10/04&  53298.61 & 18.125 (0.074) & 17.488 (0.029) & 16.334 (0.014) & 15.826 (0.017) & 15.768 (0.016) & 7\\
21/10/04&  53299.63 &  & 17.563 (0.235) &  &  &  &  1\\
21/10/04&  53300.32 & 18.092 (0.021) & 17.529 (0.011) & 16.406 (0.011) & 15.984 (0.013) & 15.853 (0.011) &3\\
22/10/04&  53300.90 &  &  & 16.525 (0.008) &  &  & 0 \\
22/10/04&  53300.92 & 18.165 (0.068) & 17.681 (0.014) & 16.530 (0.009) & 16.040 (0.011) & 15.827 (0.016) &0\\
22/10/04&  53301.37 &  & 17.710 (0.041) & 16.511 (0.024) & 15.936 (0.012) & 15.907 (0.015) & 2\\
23/10/04&  53301.60 &  & 17.681 (0.070) & 16.522 (0.022) & 16.031 (0.031) & 15.785 (0.035) & 1\\
24/10/04&  53302.90 &  & 17.869 (0.031) & 16.654 (0.025) & 16.102 (0.067) & 15.782 (0.088) & 0\\
25/10/04&  53303.64 &  & 17.912 (0.066) & 16.675 (0.022) & 16.090 (0.016) & 15.766 (0.026) & 1\\
25/10/04&  53303.96 &  & 17.917 (0.041) & 16.689 (0.022) & 16.176 (0.015) & 15.846 (0.013) & 4\\
28/10/04&  53306.92 &  & 18.159 (0.072) &  &  &  &  4\\
28/10/04&  53306.94 &  & 18.145 (0.059) & 16.918 (0.018) & 16.392 (0.021) & 16.001 (0.013) &  4\\
29/10/04&  53307.93 &  & 18.224 (0.096) & 16.973 (0.024) & 16.447 (0.031) & 16.042 (0.018) &4\\
29/10/04&  53307.95 &  & 18.240 (0.105) &  &  &  &4\\ \hline
\\
\caption{continued.}\\
\hline\hline
Date & JD--2,400,000 & $U$ & $B$ & $V$ & $R$ & $I$ & Instrument \\ \hline
30/10/04&  53308.61 &  & 18.200 (0.032) & 16.977 (0.018) & 16.415 (0.016) & 16.092 (0.017) &  1\\ 
30/10/04&  53309.33 & 18.623 (0.023) & 18.308 (0.011) & 17.012 (0.009)	&16.548 (0.010) & 16.228 (0.011) & 3\\
31/10/04&  53309.90 &  &  & 17.091 (0.050) & &  & 4\\
31/10/04&  53309.91 &  &  & 17.104 (0.017) & 16.616 (0.026) &  & 4\\
01/11/04&  53310.60 &  & 18.218 (0.023) &17.127 (0.017)& 16.601 (0.019)& 16.207 (0.030) &  1\\
02/11/04&  53311.59 &  & 18.328 (0.110) &17.115 (0.060)& 16.694 (0.044)& 16.259 (0.115) &  7\\
03/11/04&  53312.60 &  & 18.327 (0.025) &17.194 (0.047)& 16.722 (0.021)& 16.316 (0.034) &  1\\
05/11/04&  53314.59 &  & 18.544 (0.029) &17.256 (0.015)& 16.876 (0.059)& 16.507 (0.028) &  7\\
07/11/04&  53316.60 &  & 18.608 (0.060) &17.411 (0.028)& 16.965 (0.026)& 16.603 (0.028) &  1\\
08/11/04&  53318.33 &  & 18.649 (0.070) &17.455 (0.029)& 17.024 (0.036)& 16.651 (0.035) &  8\\
11/11/04&  53320.58 &  &  & 17.497 (0.020)& 17.095 (0.021)& 16.935 (0.042) & 7\\
12/11/04&  53322.36 &  & 18.729 (0.210) &  &  &  & 2\\
15/11/04&  53325.27 &  & 18.810 (0.050) & 17.659 (0.019) & 17.243 (0.016) & 17.005 (0.033) & 6\\ 
15/11/04&  53325.32 &  & 18.813 (0.048) &  &  &  & 2\\
16/11/04&  53326.21 &  &  & 17.656 (0.029) &  &  &  8\\
16/11/04&  53326.23 &  & 18.778 (0.054) & 17.676 (0.045) & 17.320 (0.050) & 16.972 (0.110) & 8\\
17/11/04&  53327.24 &  & 18.786 (0.240) & 17.732 (0.045) & 17.375 (0.017) & 17.072 (0.016) & 8\\
18/11/04&  53328.32 &  &  & 17.738 (0.135) & 17.364 (0.155) &  &8\\ 
19/11/04&  53328.59 &  & 18.794 (0.100) & 17.740 (0.050) & 17.341 (0.020) & 17.145 (0.036) &  1\\
19/11/04&  53329.31 &  & 18.814 (0.240) & 17.766 (0.220) &  &  & 8\\
19/11/04&  53329.32 &  &  & 17.775 (0.105) & 17.363 (0.130) &17.217 (0.130) & 8\\
22/11/04&  53331.59 &  & 18.809 (0.500) & 17.826 (0.300) & 17.370 (0.220) & 17.263 (0.180) & 1\\
25/11/04&  53334.53 & 19.109 (0.260) & 18.833 (0.275) & & & & 9 \\
25/11/04&  53334.59 &  & 18.812 (0.215) & 17.889 (0.065) & 17.505 (0.070) & 17.309 (0.200) & 1\\
25/11/04&  53335.27 & 19.130 (0.043) & 18.913 (0.017) & 17.943 (0.016) & 17.560 (0.013) & 17.452 (0.022) & 6\\
28/11/04&  53337.59 &  & 18.944 (0.082) & 17.946 (0.027) & 17.666 (0.024) & 17.540 (0.037) & 1\\
30/11/04&  53339.58 &  & 18.987 (0.145) & 18.014 (0.031) & 17.653 (0.054) & 17.673 (0.130) & 7\\
01/12/04&  53340.56 &  & & 18.023 (0.046) &  &  & 7\\
06/12/04&  53346.32 &  & & 18.117 (0.023)& 17.937 (0.014) &17.793 (0.040) & 8\\
07/12/04&  53347.24 &  & & 18.184 (0.016) &  &  & 8\\
10/12/04&  53350.19 &  & & 18.264 (0.057) &  &  & 8\\
10/12/04&  53350.20 &  & 19.133 (0.048)& 18.272 (0.023) &18.064 (0.032)& 18.014 (0.033) &  8\\
13/12/04&  53353.20 &  & 19.239 (0.080)& 18.369 (0.014) &18.156 (0.017)& 18.087 (0.018) &  8\\
14/12/04&  53353.59 &  & 19.238 (0.280)& 18.359 (0.080) &18.161 (0.125)& 18.086 (0.215) &  1\\
04/01/05&  53375.22 &  &  &18.717 (0.550) &  &  & 8\\
09/04/05&  53469.89 & 22.181 (0.210) &  &  &  &  & 9\\
15/04/05&  53475.88 &  &  &20.702 (0.025) &21.207 (0.048) &20.914 (0.135) & 9\\ 
18/04/05&  53478.89 &  &20.478 (0.018) &  &  &  & 9\\
12/05/05&  53502.90 & & 21.354 (0.013)  &  &  &  & 9\\
13/05/05&  53503.91 & & 21.662 (0.014) & 21.340 (0.011) & 21.730 (0.018) & 21.063 (0.023) & 10\\
14/05/05&  53504.65 &  &21.688 (0.300) &21.354 (0.460) & $\geq$21.60 & $\geq$ 21.16 & 6\\
16/05/05&  53506.64 &  & 21.543 (0.050)& 21.335 (0.060)& 22.011 (0.087)& 21.448 (0.285)&6\\
29/06/05&  53551.49 & & 22.260 (0.100) & 21.957 (0.092) & 22.619 (0.135)& 22.009 (0.197)&6\\
02/08/05&  53584.52 & $\geq$24.24& 22.793 (0.190) & 22.368 (0.044) &23.219 (0.120) & 22.518 (0.180) & 11\\
10/08/05&  53592.69 & & 22.842 (0.093) & 22.582 (0.124) &$\geq$23.23 & $\geq$22.14 & 9 \\ 
23/08/05&  53605.53 &  & 23.337 (0.059)  &  &  &  & 10\\
11/09/05&  53624.55 &  & 23.521 (0.045)  & 23.217 (0.044) & 23.633 (0.080) & 22.578 (0.075) & 10\\ \hline
\end{longtable}
\begin{centering}
0 = SSO2.3-m + imager, 1 = KAIT + CCD, 2 = LT, 3 = NOT+ALFOSC, 4 =
SSO1-m + WFI,
5 = Nw41-cm + CCD,
6 = CAHA2.2-m + CAFOS, 7 = Ten0.81-m +
CCD, 8 = Ekar1.82-m + AFOSC,
9 = ESO/MPI2.2-m + WFI, 10 = VLT-Antu + FORS1, 11 = TNG + DOLORES
\end{centering}

$^\ddag$ This $V$-band observation was kindly provided by M. Fiaschi,
and obtained using the 0.41-m Newton/Cassegrain Telescope of the 
Gruppo Astrofili di Padova, Italy, equipped with a $1024 \times 1024$ pixel 
E2V CCD. 
\twocolumn

\onecolumn 
\begin{longtable}{ccccccc}
\caption{S--correction to be added to the data of SN 2004eo in
Tab. \ref{opt_photometry}. \label{S_corr}}  \\ \hline  \hline 
Date & JD--2,400,000 & $B$ & $V$ & $R$ & $I$ & Instrument \\ \hline
19/09/04&   53268.07 &         &   0.005  & -0.004 &        &  0\\
20/09/04&   53268.76 &  -0.015 &   0      &  0.006 & -0.001 &  1\\
21/09/04&   53269.68 &  -0.014 &   0      &  0.006 & -0.002 &  1\\
21/09/04&   53270.33 &  -0.006 &  -0.014  &  0.017 & -0.035 &  2\\
22/09/04&   53270.52 &  -0.006 &  -0.014  &  0.017 & -0.036 &  2\\
22/09/04&   53270.67 &  -0.013 &   0      &  0.006 & -0.004 &  1\\
23/09/04&   53271.69 &  -0.012 &   0      &  0.007 & -0.005 &  1\\
23/09/04&   53272.36 &  -0.007 &  -0.015  &  0.019 & -0.044 &  2\\
25/09/04&   53273.66 &  -0.011 &  -0.001  &  0.008 & -0.008 &  1\\
26/09/04&   53274.69 &  -0.011 &  -0.001  &  0.008 & -0.009 &  1\\
26/09/04&   53275.37 &  -0.007 &  -0.017  &  0.024 & -0.059 &  2\\
27/09/04&   53275.70 &  -0.010 &  -0.002  &  0.008 & -0.011 &  1\\
27/09/04&   53276.33 &  -0.006 &  -0.018  &  0.025 & -0.062 &  2\\
28/09/04&   53276.70 &  -0.010 &  -0.002  &  0.008 & -0.012 &  1\\
28/09/04&   53277.37 &  -0.006 &  -0.018  &  0.026 & -0.067 &  2\\
29/09/04&   53277.70 &  -0.009 &  -0.002  &  0.008 & -0.013 &  1\\
29/09/04&   53278.36 &  -0.005 &  -0.018  &  0.027 & -0.072 &  2\\
30/09/04&   53278.71 &  -0.008 &  -0.003  &  0.009 & -0.014 &  1\\
01/10/04&   53279.66 &  -0.008 &  -0.003  &  0.009 & -0.015 &  1\\
02/10/04&   53280.69 &  -0.007 &  -0.004  &  0.010 & -0.016 &  1\\
02/10/04&   53281.38 &  -0.001 &  -0.011  &  0.014 & -0.047 &  3\\
03/10/04&   53281.63 &  -0.007 &  -0.004  &  0.010 & -0.017 &  1\\
03/10/04&   53282.43 &  -0.002 &  -0.017  &  0.037 & -0.085 &  2\\
04/10/04&   53282.68 &  -0.006 &  -0.004  &  0.010 & -0.017 &  1\\
04/10/04&   53282.98 &  -0.012 &  -0.004  & -0.006 & -0.020 &  4\\
05/10/04&   53283.66 &  -0.005 &  -0.004  &  0.011 & -0.017 &  1\\
05/10/04&   53283.94 &  -0.010 &  -0.006  & -0.006 & -0.019 &  4\\
05/10/04&   53284.36 &  -0.001 &  -0.017  &  0.043 & -0.094 &  2\\
06/10/04&   53285.33 &    0    &  -0.017  &  0.045 & -0.099 &  2\\
06/10/04&   53285.34 &         &  -0.028  &        &        &  5\\
07/10/04&   53286.30 &  -0.006 &  -0.032  &  0.005 &  0.117 &  6\\
10/10/04&   53288.63 &  -0.001 &  -0.003  &  0.010 & -0.016 &  1\\
10/10/04&   53288.92 &  -0.002 &  -0.009  &        & -0.015 &  4\\
11/10/04&   53290.30 &  -0.008 &  -0.034  &  0.006 &  0.111 &  6\\
12/10/04&   53290.63 &    0    &  -0.001  &  0.008 & -0.015 &  1\\
12/10/04&   53290.92 &   0.002 &  -0.009  &        & -0.011 &  4\\
14/10/04&   53292.63 &  -0.001 &   0.002  &  0.007 & -0.013 &  1\\
14/10/04&   53293.42 &  -0.018 &  -0.034  &  0.007 &  0.096 &  6\\
15/10/04&   53294.42 &   0.002 &  -0.013  &  0.071 & -0.109 &  2\\
16/10/04&   53294.62 &  -0.004 &   0.007  &  0.005 & -0.011 &  1\\
20/10/04&   53298.61 &  -0.004 &  -0.001  &  0.003 & -0.025 &  7\\
21/10/04&   53299.63 &  -0.012 &          &        &        &  1\\
21/10/04&   53300.32 &  -0.013 &  -0.009  &  0.001 & -0.044 &  3\\
22/10/04&   53300.91 &   0.009 &  -0.001  & -0.010 &  0.001 &  0\\
22/10/04&   53301.37 &   0.002 &  -0.005  &  0.086 & -0.112 &  2\\
23/10/04&   53301.60 &  -0.015 &   0.020  & -0.001 & -0.006 &  1\\
24/10/04&   53302.90 &   0.008 &   0.001  & -0.012 &  0.002 &  0\\
25/10/04&   53303.64 &  -0.020 &   0.023  & -0.002 & -0.005 &  1\\
25/10/04&   53303.96 &   0.007 &   0.002  & -0.013 &  0.003 &  4\\
28/10/04&   53306.93 &   0.005 &   0.004  & -0.014 &  0.005 &  4\\
29/10/04&   53307.93 &   0.004 &   0.005  & -0.014 &  0.006 &  4\\
30/10/04&   53308.61 &  -0.027 &   0.025  & -0.004 & -0.002 &  1\\
30/10/04&   53309.33 &  -0.026 &  -0.008  & -0.004 & -0.026 &  3\\
31/10/04&   53309.91 &         &   0.006  & -0.015 &        &  4\\
01/11/04&   53310.60 &  -0.028 &   0.025  & -0.005 & 0      &  1\\
02/11/04&   53311.59 &  -0.010 &   0.005  & -0.003 & -0.013 &  7\\
03/11/04&   53312.60 &  -0.028 &   0.024  & -0.005 &  0.001 &  1\\
05/11/04&   53314.59 &  -0.011 &   0.007  & -0.001 & -0.011 &  7\\ \hline
\\
\caption{continued.}\\
\hline \hline
Date &  JD--2,400,000 & $B$ & $V$ & $R$ & $I$ & Instrument \\ \hline
07/11/04&   53316.60 &  -0.030 &   0.023  & -0.005 &  0.004 &  1\\
08/11/04&   53318.33 &  -0.009 &   0.009  & -0.027 &  0.032 &  8\\
11/11/04&   53320.58 &         &   0.005  &  0.002 & -0.006 &  7\\
12/11/04&   53322.36 &  -0.008 &          &        &        &  2\\
15/11/04&   53325.27 &  -0.074 &  -0.039  &  0.005 & -0.005 &  6\\
15/11/04&   53325.32 &  -0.006 &          &        &        &  2\\
16/11/04&   53325.61 &         &          & -0.001 &  0.011 &  1\\
16/11/04&   53326.22 &  -0.007 &   0.009  & -0.020 &  0.037 &  8\\
17/11/04&   53327.24 &  -0.006 &   0.009  & -0.019 &  0.037 &  8\\
18/11/04&   53328.32 &         &   0.008  & -0.018 &        &  8\\
19/11/04&   53328.59 &  -0.016 &   0.016  &  0.001 &  0.013 &  1\\
19/11/04&   53329.32 &  -0.005 &   0.008  & -0.017 &  0.039 &  8\\
22/11/04&   53331.59 &  -0.012 &   0.014  &  0.002 &  0.014 &  1\\
25/11/04&   53334.53 &  -0.061 &          &        &        &  9\\   
25/11/04&   53334.59 &  -0.007 &   0.013  &  0.004 &  0.015 &  1\\
25/11/04&   53335.27 &  -0.041 &  -0.045  &  0.002 & -0.036 &  6\\
28/11/04&   53337.59 &  -0.003 &   0.012  &  0.004 &  0.016 &  1\\
30/11/04&   53339.58 &   0.001 &  -0.018  &  0.015 &  0.001 &  7\\
01/12/04&   53340.56 &         &  -0.018  &        &        &  7\\
06/12/04&   53346.32 &         &  -0.003  & -0.006 &  0.039 &  8\\
07/12/04&   53347.24 &         &  -0.004  &        &        &  8\\
10/12/04&   53350.20 &   0.002 &  -0.005  & -0.004 &  0.047 &  8\\
13/12/04&   53353.20 &   0.003 &  -0.006  & -0.001 &  0.057 &  8\\
14/12/04&   53353.59 &   0.003 &  -0.006  & -0.001 &  0.059 &  8\\
04/01/05&   53375.22 &         &  -0.007  &        &        &  8\\
15/04/05&   53475.88 &         &   0.055  &  0.059 &  0.126 &  9\\
18/04/05&   53478.89 &   0.198 &          &        &        &  9\\
12/05/05&   53502.90 &   0.198 &          &        &        &  9\\
13/05/05&   53503.91 &  -0.028 &  -0.110  &   0.007&   0.037& 10\\
14/05/05&   53504.65 &   0.168 &   0.072  &  -0.013&   0.103&  6\\
16/05/05&   53506.64 &   0.168 &   0.072  &  -0.013&   0.103&  6\\
29/06/05&   53551.49 &   0.168 &   0.072  &  -0.013&   0.103&  6\\
02/08/05&   53584.52 &   0.111 &   0.163  &   0.214&   0.100& 11\\
10/08/05&   53592.69 &   0.198 &   0.055  &   0.059&   0.126&  9\\ 
23/08/05&   53605.53 & -0.028 &  & &  & 10\\
11/09/05&   53624.55 & -0.028 & -0.110 & 0.007 & 0.037 & 10\\ \hline
\end{longtable}
\begin{centering}
0 = SSO2.3-m + imager, 1 = KAIT + CCD, 2 = LT + RATCAM, 3 = NOT+ALFOSC, 4 =
SSO1-m + WFI,
5 = Nw41-cm GAP + CCD, 6 = CAHA2.2-m + CAFOS, 7 = Ten0.81-m +
CCD, 8 = Ekar1.82-m + AFOSC,
9 = ESO/MPI2.2-m + WFI, 10 = VLT-Antu + FORS1, 11 = TNG + DOLORES
\end{centering}
\twocolumn

\onecolumn 
\begin{longtable}{cccccccc}
\caption{K--correction to be added to the data of SN 2004eo in
Tab. \ref{opt_photometry}. \label{K_corr}}  \\ \hline  \hline 
Date & JD--2,400,000 & $U$ & $B$ & $V$ & $R$ & $I$ & Instrument \\ \hline
19/09/04 &  53268.07 & ~~~~~~~~~~~~~~~~~~~~~  &  ~~~~~~~~~~~~~ &   0.020  &  0.050  &   & 0 \\
20/09/04 &  53268.76 &   &   0.047 &   0.022  &  0.050  &  0.058 & 1 \\
21/09/04 &  53269.68 &   &   0.043 &   0.023  &  0.050  &  0.057 & 1 \\
21/09/04 &  53270.33 &   &   0.040 &   0.025  &  0.050  &  0.055 & 2 \\
22/09/04 &  53270.52 &   &   0.039 &   0.025  &  0.050  &  0.055 & 2 \\
22/09/04 &  53270.67 &   &   0.039 &   0.026  &  0.050  &  0.054 & 1\\
23/09/04 &  53271.69 &   &   0.034 &   0.028  &  0.050  &  0.052 & 1 \\
23/09/04 &  53272.36 &   &   0.032 &   0.029  &  0.050  &  0.050 & 2 \\
25/09/04 &  53273.66 &   &   0.027 &   0.027  &  0.050  &  0.040 & 1 \\
26/09/04 &  53274.69 &   &   0.025 &   0.025  &  0.050  &  0.029 & 1 \\
26/09/04 &  53275.37 &   &   0.023 &   0.023  &  0.050  &  0.021 & 2 \\
27/09/04 &  53275.70 &   &   0.022 &   0.022  &  0.050  &  0.018 & 1 \\
27/09/04 &  53276.33 &   &   0.020 &   0.020  &  0.050  &  0.011 & 2 \\
28/09/04 &  53276.70 &   &   0.019 &   0.020   &  0.051 &  0.009 & 1 \\
28/09/04 &  53277.37 &   &   0.018 &   0.020   &  0.054 &  0.008 & 2 \\
29/09/04 &  53277.70 &   &   0.017 &   0.020   &  0.055 &  0.007 & 1 \\
29/09/04 &  53278.36 &   &   0.016 &   0.020   &  0.058 &  0.006 & 2 \\
30/09/04 &  53278.71 &   &   0.015 &   0.020   &  0.059 &  0.005 & 1 \\
01/10/04 &  53279.66 &   &   0.014 &   0.020   &  0.063 &  0.004 & 1 \\
02/10/04 &  53280.69 &   &   0.012 &   0.020   &  0.067 &  0.002 & 1 \\
02/10/04 &  53281.37 &   &   &   0.020   &   &    & 3 \\
02/10/04 &  53281.38 &   0.059 &   &   0.020   &  0.070 &  0     & 3 \\
02/10/04 &  53281.39 &   0.059 &   0.010 &   0.020   &  0.070 &  0     & 3 \\
03/10/04 &  53281.63 &   &   0.010 &   0.020   &  0.070 &  0.001 & 1 \\
03/10/04 &  53282.43 &   &   0.008 &   0.018  &  0.070  &  0.008 & 2 \\
04/10/04 &  53282.68 &   &   0.008 &   0.017  &  0.070  &  0.010 & 1 \\
04/10/04 &  53282.98 &   &   0.007 &   0.017  &  0.070  &  0.012 & 4 \\
05/10/04 &  53283.66 &   &   0.006 &   0.015  &  0.070  &  0.018 & 1 \\
05/10/04 &  53283.94 &   &   0.005 &   0.015  &  0.070  &  0.020 & 4 \\
05/10/04 &  53284.36 &   &   0.003 &    &  0.070  &  0.024 & 2 \\
05/10/04 &  53284.39 &   &   0.003 &   0.014  &  0.070  &  0.024 & 2 \\
06/10/04 &  53285.33 &   &   0.001 &   0.012  &  0.070  &  0.032 & 2 \\
06/10/04 &  53285.34 &   &    &   0.012  &    &   & 5 \\
07/10/04 &  53286.30 &  -0.118 &  -0.002 &   0.010  &  0.070  &  0.040 & 6 \\
10/10/04 &  53288.63 &   &  -0.007 &   0.004  &  0.059 &  0.034 & 1 \\
10/10/04 &  53288.92 &   &  -0.009 &   0.004  &   &  0.034 & 4 \\
11/10/04 &  53290.30 &  -0.135 &  -0.020 &   0     &  0.050  &  0.030 & 6 \\
12/10/04 &  53290.63 &   &  -0.022 &   0     &  0.050  &  0.028 & 1 \\
12/10/04 &  53290.92 &   &  -0.024 &   0     &    &  0.026 & 4 \\
14/10/04 &  53292.63 &   &  -0.020 &   0     &  0.045  &  0.025 & 1 \\
14/10/04 &  53293.42 &  -0.100 &  -0.010 &   0     &  0.040  &  0.030 & 6 \\
15/10/04 &  53294.42 &   &  -0.016 &  -0.006  &  0.040  &  0.029 &2  \\
16/10/04 &  53294.62 &   &  -0.017 &  -0.007  &  0.040  &  0.028 &1  \\
20/10/04 &  53298.61 &  -0.078 &  -0.040 &  -0.030  &  0.040  &  0.023 &7  \\
21/10/04 &  53299.63 &   &  -0.046 &    &    &   &1  \\
21/10/04 &  53300.32 &  -0.070 &  -0.050 &  -0.040  &  0.040  &  0.020 &3  \\
22/10/04 &  53300.90 &   &   &  -0.031  &  &  & 0 \\
22/10/04 &  53300.92 &  -0.089 &  -0.041 &  -0.031  &  0.049 &  0.020 & 0 \\
22/10/04 &  53301.37 &   &  -0.038 &  -0.032  &  0.052 &  0.018 & 2 \\
23/10/04 &  53301.60 &   &  -0.037 &  -0.033  &  0.053 &  0.017 & 1 \\
24/10/04 &  53302.90 &   &  -0.030 &  -0.040  &  0.060 &  0.010 & 0 \\
25/10/04 &  53303.64 &   &  -0.031 &  -0.040   &  0.058 &  0.010  &1  \\
25/10/04 &  53303.96 &   &  -0.032 &  -0.040   &  0.057 &  0.010  &4  \\
28/10/04 &  53306.92 &   &  -0.036 &     &   &   &4  \\
28/10/04 &  53306.94 &   &  -0.036 &  -0.040   &  0.048 &  0.010  &4  \\
29/10/04 &  53307.93 &   &  -0.038 &  -0.040   &  0.045 &  0.010  &4  \\ 
29/10/04 &  53307.95 &   &  -0.038 &     &  &   &4  \\ 
30/10/04 &  53308.61 &   &  -0.039 &  -0.040   &  0.042 &  0.010  &1  \\ \hline
\\
\caption{continued.}\\
\hline \hline
Date &  JD--2,400,000 & $U$ & $B$ & $V$ & $R$ & $I$ & Instrument \\ \hline
30/10/04 &  53309.33 & ~~ -0.140 &  -0.040 &  -0.040   &  0.040 &  0.010  &3  \\
31/10/04 &  53309.90 &   &    &  -0.040   &    &   &4  \\
31/10/04 &  53309.91 &   &    &  -0.040   &  0.040  &   &4  \\
01/11/04 &  53310.60 &   &  -0.040  &  -0.040   &  0.040  &  0.014 &1  \\
02/11/04 &  53311.59 &   &  -0.040  &  -0.040   &  0.040  &  0.017 &7  \\
03/11/04 &  53312.60 &   &  -0.040  &  -0.040   &  0.040  &  0.020 &1  \\
05/11/04 &  53314.59 &   &  -0.040  &  -0.040   &  0.040  &  0.026 &7  \\
07/11/04 &  53316.60 &   &  -0.040  &  -0.040   &  0.040  &  0.033 &1  \\
08/11/04 &  53318.33 &   &  -0.040  &  -0.040   &  0.040  &  0.038 &8  \\
11/11/04 &  53320.58 &   &    &  -0.040   &  0.040  &  0.045 &7  \\
12/11/04 &  53322.36 &   &  -0.040  &     &    &   &2  \\
15/11/04 &  53325.27 &   &  -0.040  &  -0.040   &  0.040  &  0.060 &6  \\
15/11/04 &  53325.32 &   &  -0.040  &     &     &  &2  \\
16/11/04 &  53326.21 &   &   &  -0.037  &   &  &  8\\
16/11/04 &  53326.23 &   &  -0.043 &  -0.037  &  0.047 &  0.053 &  8\\
17/11/04 &  53327.24 &   &  -0.046 &  -0.034  &  0.053 &  0.046 &  8\\
18/11/04 &  53328.32 &   &   &  -0.030  &  0.060 &   &  8\\
19/11/04 &  53328.59 &   &  -0.042 &  -0.027  &  0.055 &  0.036 &  1\\
19/11/04 &  53329.31 &   &  -0.021 &  -0.020  &   &   &  8\\
19/11/04 &  53329.32 &   &   &  -0.020  &  0.041 &  0.030 &  8\\
22/11/04 &  53331.59 &   &  -0.037 &  -0.026  &  0.04  &  0.053 &  1\\
25/11/04 &  53334.53 &  -0.093 &  -0.048 &    &   &  &  9\\
25/11/04 &  53334.59 &   &  -0.048 &  -0.028  &  0.042 &  0.071 &  1\\
25/11/04 &  53335.27 &   &  -0.047 &  -0.027  &  0.043 &  0.072 &  6\\
28/11/04 &  53337.59 &   &  -0.044 &  -0.024  &  0.046 &  0.075 &  1\\
30/11/04 &  53339.58 &   &  -0.041 &  -0.021  &  0.049 &  0.077 &  7\\
01/12/04 &  53340.56 &   &   &  -0.020  &   &   &  7\\
06/12/04 &  53346.32 &   &   &  -0.011  &  0.059 &  0.084 & 8 \\
07/12/04 &  53347.24 &   &   &  -0.010  &   &   & 8 \\
10/12/04 &  53350.19 &   &   &  -0.010  &   &   & 8 \\
10/12/04 &  53350.20 &   &  -0.038 &  -0.010  &  0.038 &  0.089 & 8 \\
13/12/04 &  53353.20 &   &  -0.040 &  -0.009  &  0.029 &  0.090 & 8 \\
14/12/04 &  53353.59 &   &  -0.040 &  -0.008  &  0.029 &  0.090 & 1 \\
04/01/05 &  53375.22 &   &   &   0.007  &   &   & 8 \\
09/04/05 &  53469.89 &   N &   &  &  &   & 9 \\
15/04/05 &  53475.88 &      &   &   0.078  & -0.010 &  0.106 & 9 \\
18/04/05 &  53478.89 &      &  -0.070 &     &  &   & 9 \\
12/05/05 &  53502.90 &      &  -0.070 &     &  &   & 9 \\
13/05/05 &  53503.91 &      &  -0.070 &   0.100  & -0.020 &  0.110 &10  \\
14/05/05 &  53504.65 &      &  -0.070 &   0.100  & -0.020 &  0.110 & 6 \\
16/05/05 &  53506.64 &      &  -0.070 &   0.100  & -0.020 &  0.110 & 6 \\
29/06/05 &  53551.49 &      &  -0.070 &   0.100  & -0.020 &  0.110 & 6 \\
02/08/05 &  53584.52 &  N  &  -0.070 &   0.100  & -0.020 &  0.110 &11  \\
10/08/05 &  53592.69 &      &  -0.070 &   0.100  & -0.020 &  0.110 & 9 \\
23/08/05 &  53605.53 &      &  -0.070 &     &  &  &10  \\
11/09/05 &  53624.55 &      &  -0.070 &   0.100  & -0.020 &  0.110 &10  \\ \hline
\end{longtable}
\begin{centering}
N = the K--correction in the $U$ band was not computed for these two epochs.\\

0 = SSO2.3-m + imager, 1 = KAIT + CCD, 2 = LT + RATCAM, 3 = NOT+ALFOSC, 4 =
SSO1-m + WFI,
5 = Nw41-cm GAP + CCD, 6 = CAHA2.2-m + CAFOS, 7 = Ten0.81-m +
CCD, 8 = Ekar1.82-m + AFOSC,
9 = ESO/MPI2.2-m + WFI, 10 = VLT-Antu + FORS1, 11 = TNG + DOLORES.
\end{centering}
\twocolumn

\onecolumn 
\begin{longtable}{cccccccc}
\caption{\label{SK_opt_photometry} S-- and K--corrected optical photometry of SN 2004eo.} \\ \hline \hline
Date & JD--2,400,000 & $U$ & $B$ & $V$ & $R$ & $I$ & Instrument \\ \hline
19/09/04&  53268.07 & ~~~~~~~~~~~~~~~~~~~~~~~ & ~~~~~~~~~~ & 16.776 (0.014) & 16.530 (0.014) &  &  0 \\
20/09/04&  53268.76 & & 16.721 (0.024) & 16.507 (0.025) & 16.293 (0.034) & 16.283 (0.039)  &  1\\
21/09/04&  53269.68 & & 16.476 (0.016) & 16.305 (0.013) & 16.081 (0.015) & 16.089 (0.019)  &  1\\
21/09/04&  53270.33 & & 16.342 (0.020) & 16.144 (0.018) & 16.045 (0.019) & 15.853 (0.033)  &  2\\
22/09/04&  53270.52 & & 16.319 (0.018) & 16.152 (0.014) & 15.969 (0.020) & 15.858 (0.025)  &  2\\
22/09/04&  53270.67 & & 16.214 (0.043) & 16.076 (0.018) & 15.863 (0.017) & 15.875 (0.024)  &  1\\
23/09/04&  53271.69 & & 16.051 (0.017) & 15.925 (0.015) & 15.678 (0.016) & 15.723 (0.021)  &  1\\
23/09/04&  53272.36 & & 15.991 (0.018) & 15.874 (0.012) & 15.625 (0.019) & 15.555 (0.025)  &  2\\
25/09/04&  53273.66 & & 15.775 (0.020) & 15.651 (0.020) & 15.450 (0.019) & 15.497 (0.034)  &  1\\
26/09/04&  53274.69 & & 15.677 (0.021) & 15.555 (0.018) & 15.330 (0.020) & 15.404 (0.035)  &  1\\
26/09/04&  53275.37 & & 15.632 (0.017) & 15.558 (0.015) & 15.375 (0.020) & 15.364 (0.026)  &  2\\
27/09/04&  53275.70 & & 15.609 (0.022) & 15.471 (0.015) & 15.295 (0.016) & 15.405 (0.021)  &  1\\
27/09/04&  53276.33 & & 15.602 (0.017) & 15.478 (0.013) & 15.339 (0.019) & 15.352 (0.025)  &  2\\
28/09/04&  53276.70 & & 15.545 (0.026) & 15.444 (0.016) & 15.238 (0.015) & 15.380 (0.023)  &  1\\
28/09/04&  53277.37 & & 15.526 (0.019) & 15.447 (0.013) & 15.278 (0.022) & 15.365 (0.025)  &  2\\
29/09/04&  53277.70 & & 15.527 (0.019) & 15.388 (0.017) & 15.248 (0.012) & 15.392 (0.021)  &  1\\
29/09/04&  53278.36 & & 15.544 (0.017) & 15.454 (0.013) & 15.278 (0.019) & 15.364 (0.025) &  2\\
30/09/04&  53278.71 & & 15.511 (0.020) & 15.368 (0.022) & 15.228 (0.015) & 15.417 (0.018) &  1\\
01/10/04&  53279.66 & & 15.512 (0.024) & 15.356 (0.016) & 15.239 (0.015) & 15.462 (0.021) &  1\\
02/10/04&  53280.69 & & 15.550 (0.019) & 15.345 (0.016) & 15.224 (0.026) & 15.478 (0.024) &   1\\
02/10/04&  53281.37 & & & 15.396 (0.014) & & &  3\\
02/10/04&  53281.38 & 15.484 (0.039) &  & 15.394 (0.013) & 15.301 (0.013) & 15.567 (0.024) &   3\\
02/10/04&  53281.39 & 15.498 (0.039) & 15.578 (0.015) & 15.400 (0.013) & 15.289 (0.015) & 15.571 (0.025) &3\\
03/10/04&  53281.63 &  & 15.597 (0.021) & 15.353 (0.015) & 15.251 (0.017) & 15.519 (0.025) &   1\\
03/10/04&  53282.43 &  & 15.630 (0.016) & 15.359 (0.016) & 15.252 (0.019) & 15.510 (0.025) &   2\\
04/10/04&  53282.68 &  & 15.635 (0.056) & 15.350 (0.015) & 15.240 (0.016) & 15.520 (0.017) &   1\\
04/10/04&  53282.98 &  & 15.657 (0.022) & 15.417 (0.016) & 15.286 (0.017) & 15.501 (0.038) &   4\\
05/10/04&  53283.66 &  & 15.712 (0.016) & 15.397 (0.012) & 15.285 (0.013) & 15.570 (0.021) &   1\\
05/10/04&  53283.94 &  & 15.715 (0.016) & 15.431 (0.012) & 15.337 (0.013) & 15.566 (0.017) &   4\\
05/10/04&  53284.36 &  & 15.782 (0.021) &  & 15.337 (0.019) & 15.556 (0.029) &   2\\
05/10/04&  53284.39 &  & 15.779 (0.016) & 15.433 (0.013) & 15.323 (0.017) & 15.583 (0.025) &   2\\
06/10/04&  53285.33 &  & 15.896 (0.016) & 15.468 (0.014) & 15.382 (0.019) & 15.655 (0.026) &   2\\
06/10/04&  53285.34 &  &  & 15.496 (0.132) &  &  & 5\\
07/10/04&  53286.30 & 15.828 (0.039) & 15.917 (0.023) & 15.468 (0.014) & 15.495 (0.015) & 15.727 (0.039) & 6\\
10/10/04&  53288.63 &  & 16.174 (0.017) & 15.578 (0.016) & 15.528 (0.021) & 15.784 (0.038) &   1\\
10/10/04&  53288.92 &  & 16.232 (0.020) & 15.671 (0.012) &  & 15.796 (0.018) & 4\\
11/10/04&  53290.30 & 16.357 (0.039) & 16.464 (0.025) & 15.693 (0.048) & 15.733 (0.039) & 15.789 (0.043) & 6\\
12/10/04&  53290.63 &  & 16.424 (0.016) & 15.746 (0.016) & 15.669 (0.014) & 15.838 (0.017) &  1\\
12/10/04&  53290.92 &  & 16.487 (0.016) & 15.795 (0.014) &  & 15.826 (0.017) &  4\\
14/10/04&  53292.63 &  & 16.717 (0.019) & 15.904 (0.016) & 15.754 (0.013) & 15.829 (0.021) &   1\\
14/10/04&  53293.42 & 17.182 (0.040) & 16.878 (0.025) & 15.925 (0.016) & 15.874 (0.023) & 15.809 (0.039) & 6\\
15/10/04&  53294.42 &  & 16.948 (0.016) & 16.017 (0.016) & 15.776 (0.018) & 15.839 (0.026) & 2\\
16/10/04&  53294.62 &  & 16.940 (0.029) & 16.056 (0.016) & 15.827 (0.016) & 15.783 (0.020) & 1\\
20/10/04&  53298.61 & 18.047 (0.082) & 17.444 (0.032) & 16.303 (0.018) & 15.869 (0.019) & 15.766 (0.022) & 7\\
21/10/04&  53299.63 &  & 17.505 (0.235) &  &  &  &  1\\
21/10/04&  53300.32 & 18.022 (0.041) & 17.466 (0.016) & 16.357 (0.015) & 16.025 (0.016) & 15.829 (0.025) &3\\
22/10/04&  53300.90 &  &  & 16.493 (0.012) &  &  & 0 \\
22/10/04&  53300.92 & 18.076 (0.076) & 17.649 (0.018) & 16.498 (0.012) & 16.079 (0.013) & 15.848 (0.021) &0\\
22/10/04&  53301.37 &  & 17.674 (0.043) & 16.474 (0.026) & 16.074 (0.019) & 15.813 (0.027) & 2\\
23/10/04&  53301.60 &  & 17.631 (0.071) & 16.509 (0.024) & 16.083 (0.032) & 15.796 (0.038) & 1\\
24/10/04&  53302.90 &  & 17.847 (0.033) & 16.615 (0.025) & 16.151 (0.067) & 15.794 (0.089) & 0\\
25/10/04&  53303.64 &  & 17.861 (0.067) & 16.657 (0.024) & 16.146 (0.018) & 15.771 (0.029) & 1\\
25/10/04&  53303.96 &  & 17.892 (0.043) & 16.651 (0.024) & 16.220 (0.016) & 15.859 (0.019) & 4\\
28/10/04&  53306.92 &  & 18.128 (0.073) &  &  &  &  4\\
28/10/04&  53306.94 &  & 18.114 (0.060) & 16.882 (0.020) & 16.426 (0.022) & 16.016 (0.019) &  4\\
29/10/04&  53307.93 &  & 18.190 (0.097) & 16.937 (0.025) & 16.478 (0.032) & 16.058 (0.023) &4\\
29/10/04&  53307.95 &  & 18.202 (0.106) &  &  &  &4\\
30/10/04&  53308.61 &  & 18.134 (0.034) & 16.962 (0.020) & 16.453 (0.017) & 16.100 (0.022) &  1\\ \hline 
\\
\caption{continued.}\\
\hline\hline
Date & JD--2,400,000 & $U$ & $B$ & $V$ & $R$ & $I$ & Instrument \\ \hline
30/10/04&  53309.33 & 18.483 (0.042) & 18.242 (0.016) & 16.964 (0.013)	&16.584 (0.013) & 16.212 (0.025) & 3\\
31/10/04&  53309.90 &  &  & 17.057 (0.051) & &  & 4\\
31/10/04&  53309.91 &  &  & 17.070 (0.019) & 16.642 (0.027) &  & 4\\
01/11/04&  53310.60 &  & 18.150 (0.026) &17.112 (0.019)& 16.636 (0.020)& 16.221 (0.033) &  1\\
02/11/04&  53311.59 &  & 18.278 (0.111) &17.080 (0.061)& 16.731 (0.045)& 16.263 (0.116) &  7\\
03/11/04&  53312.60 &  & 18.259 (0.027) &17.178 (0.048)& 16.757 (0.022)& 16.337 (0.037) &  1\\
05/11/04&  53314.59 &  & 18.493 (0.032) &17.223 (0.019)& 16.915 (0.060)& 16.522 (0.032) &  7\\
07/11/04&  53316.60 &  & 18.538 (0.061) &17.393 (0.029)& 17.001 (0.027)& 16.639 (0.031) &  1\\
08/11/04&  53318.33 &  & 18.599 (0.071) &17.424 (0.031)& 17.037 (0.037)& 16.720 (0.038) &  8\\
11/11/04&  53320.58 &  &  & 17.461 (0.023)& 17.137 (0.023)& 16.974 (0.045) & 7\\
12/11/04&  53322.36 &  & 18.681 (0.210) &  &  &  & 2\\
15/11/04&  53325.27 &  & 18.696 (0.054) & 17.580 (0.022) & 17.288 (0.017) & 17.060 (0.049) & 6\\ 
15/11/04&  53325.32 &  & 18.767 (0.049) &  &  &  & 2\\
16/11/04&  53326.21 &  &  & 17.628 (0.031) &  &  &  8\\
16/11/04&  53326.23 &  & 18.728 (0.055) & 17.648 (0.047) & 17.347 (0.050) & 17.062 (0.110) & 8\\
17/11/04&  53327.24 &  & 18.734 (0.240) & 17.707 (0.047) & 17.409 (0.019) & 17.152 (0.021) & 8\\
18/11/04&  53328.32 &  &  & 17.716 (0.135) & 17.406 (0.155) &  &8\\ 
19/11/04&  53328.59 &  & 18.736 (0.100) & 17.729 (0.050) & 17.397 (0.021) & 17.194 (0.038) &  1\\
19/11/04&  53329.31 &  & 18.788 (0.240) & 17.754 (0.220) &  &  & 8\\
19/11/04&  53329.32 &  &  & 17.763 (0.105) & 17.387 (0.130) &17.286 (0.130) & 8\\
22/11/04&  53331.59 &  & 18.760 (0.500) & 17.814 (0.300) & 17.412 (0.220) & 17.330 (0.180) & 1\\
25/11/04&  53334.53 & 19.016 (0.262) & 18.724 (0.275) & & & & 9 \\
25/11/04&  53334.59 &  & 18.763 (0.215) & 17.874 (0.065) & 17.551 (0.070) & 17.395 (0.200) & 1\\
25/11/04&  53335.27 & 19.037 (0.055) & 18.825 (0.026) & 17.871 (0.020) & 17.605 (0.014) & 17.488 (0.043) & 6\\
28/11/04&  53337.59 &  & 18.897 (0.083) & 17.934 (0.028) & 17.716 (0.025) & 17.631 (0.039) & 1\\
30/11/04&  53339.58 &  & 18.947 (0.145) & 17.975 (0.033) & 17.717 (0.055) & 17.751 (0.131) & 7\\
01/12/04&  53340.56 &  & & 17.985 (0.047) &  &  & 7\\
06/12/04&  53346.32 &  & & 18.103 (0.025)& 17.990 (0.016) &17.916 (0.042) & 8\\
07/12/04&  53347.24 &  & & 18.170 (0.019) &  &  & 8\\
10/12/04&  53350.19 &  & & 18.249 (0.058) &  &  & 8\\
10/12/04&  53350.20 &  & 19.097 (0.049)& 18.257 (0.025) &18.098 (0.033)& 18.150 (0.036) &  8\\
13/12/04&  53353.20 &  & 19.201 (0.080)& 18.354 (0.018) &18.184 (0.019)& 18.234 (0.023) &  8\\
14/12/04&  53353.59 &  & 19.201 (0.280)& 18.345 (0.080) &18.189 (0.125)& 18.235 (0.215) &  1\\
04/01/05&  53375.22 &  &  &18.717 (0.550) &  &  & 8\\
09/04/05&  53469.89 & 22.181 (0.210) &  &  &  &  & 9\\
15/04/05&  53475.88 &  &  &20.835 (0.028) &21.256 (0.049) &21.146 (0.136) & 9\\ 
18/04/05&  53478.89 &  &20.611 (0.021) &  &  &  & 9\\
12/05/05&  53502.90 & & 21.483 (0.017)  &  &  &  & 9\\
13/05/05&  53503.91 & & 21.565 (0.018) & 21.328 (0.014) & 21.718 (0.020) & 21.210 (0.027) & 10\\
14/05/05&  53504.65 &  &21.786 (0.300) &21.524 (0.460) & $\geq$21.57 & $\geq$ 21.37 & 6\\
16/05/05&  53506.64 &  & 21.641 (0.054)& 21.507 (0.061)& 21.978 (0.087)& 21.661 (0.288)&6\\
29/06/05&  53551.49 & & 22.358 (0.102) & 22.129 (0.093) & 22.586 (0.135)& 22.222 (0.200)&6\\
02/08/05&  53584.52 & $\geq$24.24& 22.834 (0.190) & 22.630 (0.045) &23.413 (0.120) & 22.728 (0.181) & 11\\
10/08/05&  53592.69 & & 22.970 (0.094) & 22.737 (0.125) &$\geq$23.27 & $\geq$22.37 & 9 \\ 
23/08/05&  53605.53 &  & 23.239 (0.060)  &  &  &  & 10\\
11/09/05&  53624.55 &  & 23.423 (0.046)  & 23.207 (0.045) & 23.620 (0.080) & 22.725 (0.076) & 10\\ \hline
\end{longtable}
\begin{centering}
0 = SSO2.3-m + imager, 1 = KAIT + CCD, 2 = LT, 3 = NOT+ALFOSC, 4 =
SSO1-m + WFI,
5 = Nw41-cm + CCD,
6 = CAHA2.2-m + CAFOS, 7 = Ten0.81-m +
CCD, 8 = Ekar1.82-m + AFOSC,
9 = ESO/MPI2.2-m + WFI, 10 = VLT-Antu + FORS1, 11 = TNG + DOLORES
\end{centering}

\twocolumn
\begin{table*}
\caption{Infrared Photometry of SN 2004eo.}  
\label{IR_photometry}
\begin{tabular}{llcccc} \\ \hline
Date & JD--2,400,000 & $J$ & $H$ & $K$ & Instrument \\ \hline\hline
29/09/04&  53278.29 &  15.695 (0.073) & 15.985 (0.097) & 15.681 (0.177) & A \\
01/10/04&  53280.27 &  -- & -- & 15.680 (0.191) & A \\
02/10/04&  53281.43 &  15.834 (0.065) & 15.991 (0.085) & 15.677 (0.150) & B \\
03/10/04&  53282.33 &  15.973 (0.077) & 15.991 (0.098) & 15.682 (0.337) & A \\
05/10/04&  53284.31 &  16.185 (0.101) & 16.077 (0.134) & 15.685 (0.523) & A \\
07/10/04&  53286.26 &  -- & -- & 15.757 (0.322) &A\\
19/10/04&  53298.28 &  $\geq$16.29 & -- & -- & A \\
20/10/04&  53299.30 &  16.851 (0.145) &15.847 (0.119) & 15.764 (0.182) & A \\
22/10/04&  53300.88 &  16.802 (0.087) &15.788 (0.072) & 15.674 (0.092) & C \\
29/10/04&  53307.93 &  16.545 (0.062) &16.114 (0.085) & 16.209	(0.177) & C\\
05/11/04&  53315.29 &  17.165 (0.192) & -- & -- & D \\
05/11/04&  53315.35 &  17.173 (0.097) & -- & -- & D \\
17/11/04&  53326.93 & 18.313 (0.076) & 17.015 (0.099) & -- & C \\
18/11/04&  53327.90 &  -- & -- & 17.065 (0.161) & C \\
19/11/04&  53329.32 &  18.254 (0.103) & 17.050 (0.091) & 17.226 (0.198) & B\\
16/04/05&  53476.89 &  -- & $\geq$19.73 & -- & E\\ 
18/04/05&  53478.87 &  -- & $\geq$20.53 & -- & E\\ 
21/04/05&  53481.86 & 21.935 (0.102) & -- & -- & E\\ 
22/04/05&  53483.68 &  $\geq$21.04 & $\geq$20.05 & $\geq$19.32 & F \\
15/05/05&  53505.83 & 22.206 (0.282) & -- & -- & E\\ 
19/05/05&  53509.64 & -- & -- &$\geq$19.05  & F \\
21/05/05&  53511.83 &-- & 20.898 (0.312) & -- & E\\ \hline
\end{tabular}
\\
A = CI + SWIRCAM, B = TNG + NICS, C = AAT + IRIS2, \\
D = WHT + LIRIS, E = VLT-Antu + ISAAC,
F = CAHA3.5-m + OMEGA2000
\end{table*}

Following the prescription of Hamuy et al. \shortcite{ham93}, and
using the spectra of SN~2004eo presented in Sect. \ref{spec}, we
computed the K--correction to be applied to the photometric data of
SN~2004eo.  In order to cover the epochs for which no spectra were
available, the corrections were linearly interpolated.  

The K--corrections to be added to the $UBVRI$ SN magnitudes of
Tab. \ref{opt_photometry} are reported in Tab. \ref{K_corr}. They are
typically below $\sim$0.1 mag and are, in general, larger at
late phases. Since the nebular spectra evolve very slowly, the
K--correction was assumed to be constant at phases later than 200~d.

Due to the lack of adequate spectral coverage in the IR region, no
K--correction was applied to the $J$, $H$, and $K$ photometry
(Tab. \ref{IR_photometry}).

\subsection{Optical and Infrared Light Curves}

The S-- and K--corrected optical magnitudes are reported in 
Tab. \ref{SK_opt_photometry}.
Early--time (up to $\sim$3 months post-explosion) S-- and K--corrected
optical light curves of SN~2004eo are shown in Fig. \ref{light_cur}
(left). Phase 0 corresponds to the epoch of maximum in the $B$-band
light curve, which was estimated to have occurred on September 30.7
(JD = 2,453,279.2, see Sect. \ref{param}).

Our early--time photometric coverage adequately samples all the phases
of the early light curve (i.e., pre--maximum rise, $B$-band maximum
phase, $\Delta m_{15}(B)$ decline, $I$-band secondary peak).  In
addition, some photometry was obtained during the nebular phase.  In
Fig. \ref{light_cur}(right) the complete $U$ through $K$ light curves are
shown, including some additional data from the literature
\cite{arb04,fol04}.  Unfortunately, the $U$ band and IR observations
only commenced near maximum light, and so the coverage for these wavelengths
is incomplete.

As is characteristic of SNe~Ia, the $V$- and $R$-band peaks are delayed with
respect to the $B$-band maximum, while the $I$-band maximum occurs few
days earlier (Sect. \ref{param}).  A clear secondary peak is visible
in the $I$-band light curve, though less prominent than usual (see
Fig. \ref{light_cur}--left).  A hint of a plateau--like feature is
detectable in the $R$-band curve at the time of the $I$-band secondary
maximum.

\begin{figure}
\includegraphics[width=9.05cm,angle=0]{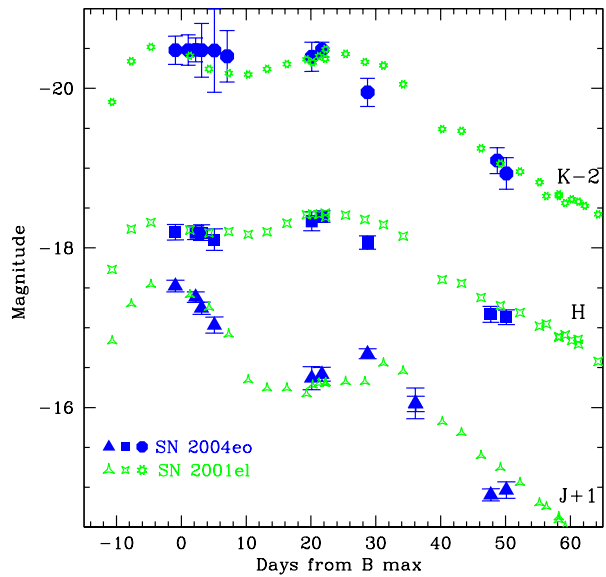}
\caption{Comparison between the $JHK$ absolute light curves of SN~2004eo 
and SN~2001el. Only measurement and photometric calibration errors
have been taken into account for SN~2004eo. For SN~2001el we adopted 
$\mu = 31.29$ mag, $E(B-V) = 0.22$ mag, and a $B$-band maximum epoch of 
JD = 2,452,182.5 (Krisciunas et al. 2003). 
\label{IRlc} }
\end{figure}

The evolution of the $J$-band light curve is similar to that of the $I$
band, although the post-maximum decline is steeper ($\Delta J \approx
1.1$ mag to the minimum). At about 4 weeks, the $J$-band
light curve shows evidence of a secondary
maximum analogous to that observed in the $I$ band.  After that, the 
$J$-band light curve declines very rapidly, fading by about 1.5 mag
in 3 weeks.  Elias et al. \shortcite{eli81}, Meikle \shortcite{peter00}, and
Phillips et al. \shortcite{phil03} noted that the $H$ and $K$-band light 
curves of SNe~Ia show a secondary maximum with brightness similar to 
that of the early maximum,
resulting in relatively flat $H$ and $K$ light curves during the
first month.  Thereafter, the IR light curves decline steeply.
A few late--time IR observations were also obtained, and the SN was
recovered at $\sim$230~d, at $J \approx 22.2$ mag
and $H \approx 20.9$ mag.  The SN was not detected in the $K$ band at
$\sim$205~d to a limiting magnitude of $\sim$19.3.

In Fig. \ref{IRlc} the absolute IR light curves of SN~2004eo, 
computed using the
distance to the host galaxy discussed in Sect. \ref{color_uvoir}, 
are compared with those of the well-observed SN~2001el \cite{kris03}.  
 Despite different values of $\Delta m_{15}(B)$ for the two SNe (1.15 for
SN~2001el and 1.46 for SN~2004eo, see Sect. \ref{param}), the $J$, $H$, and
$K$ light curves turn out to be quite similar.  

\begin{figure*}
\includegraphics[width=10.2cm,angle=270]{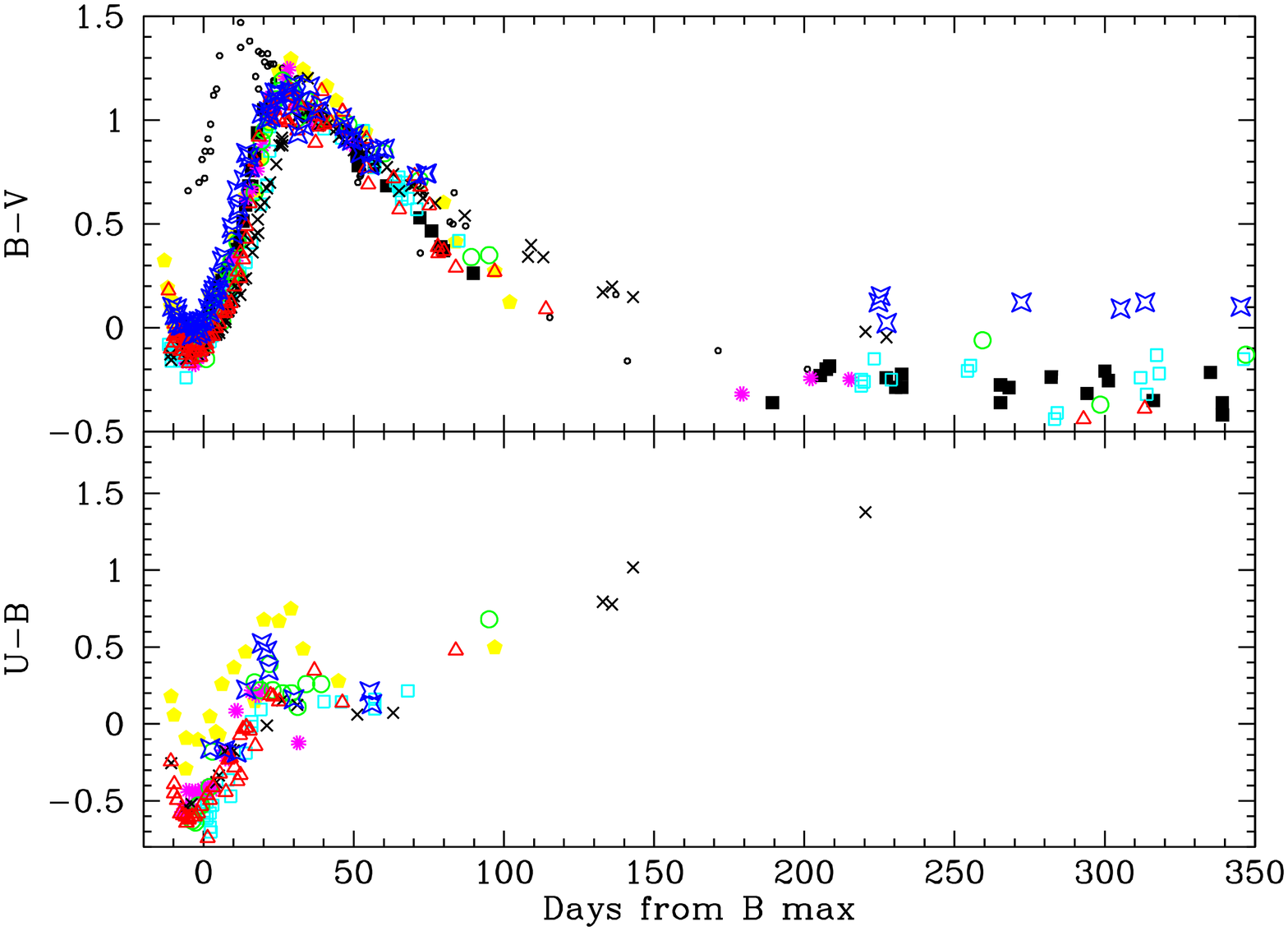}
\includegraphics[width=10.2cm,angle=270]{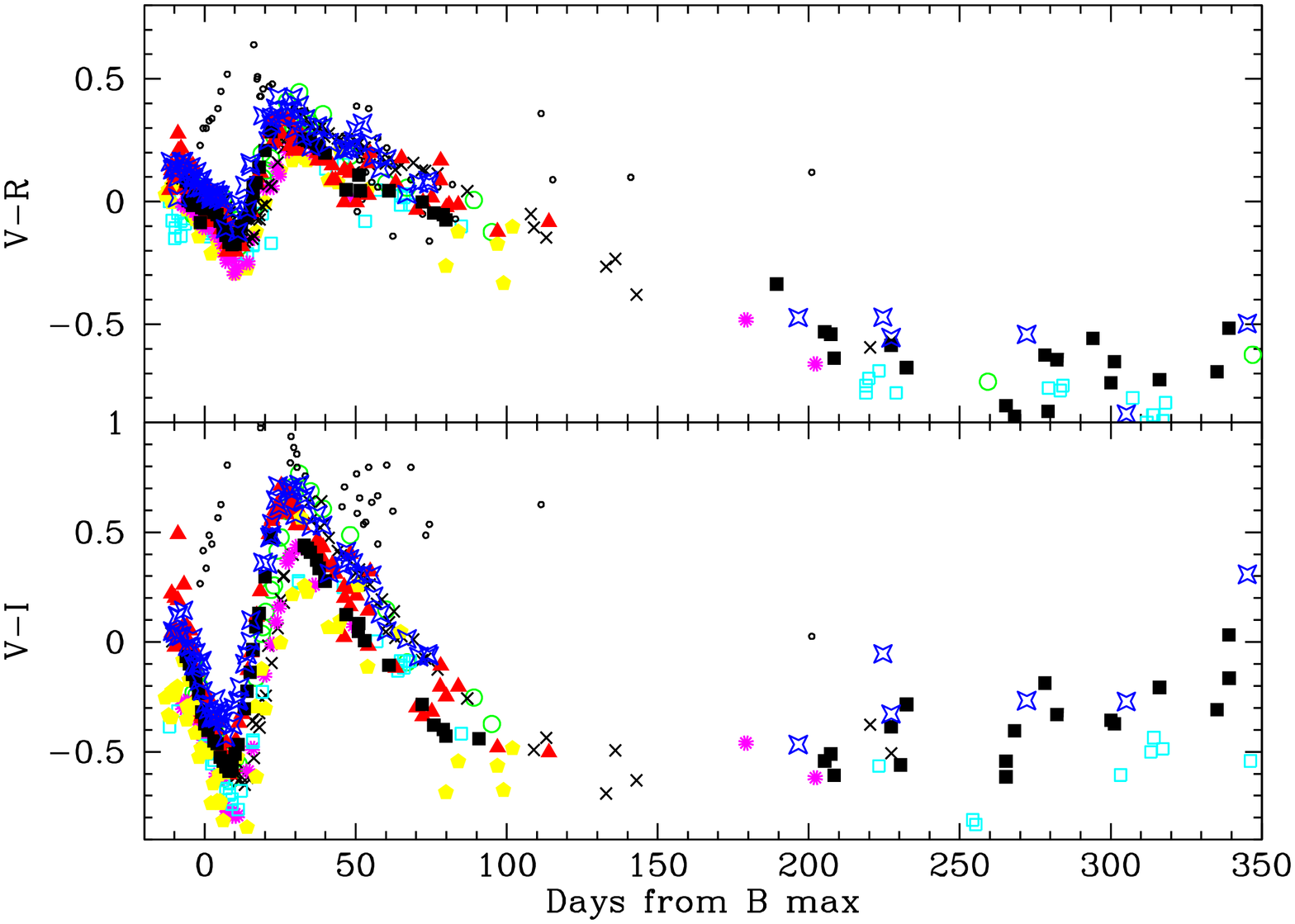}
\caption{Colour evolution of SN~2004eo compared with other
SNe~Ia. Upper figure: $B-V$ (top) and $U-B$ (bottom) colour curves.
Lower figure: $V-R$ (top) and $V-I$ (bottom) colour curves.
S--correction and K--correction have been applied to the data of SN~2004eo.
\label{colour_ev} }
\end{figure*}

\subsection{Colour Evolution and {\it uvoir} Light Curve} \label{color_uvoir}

The excellent photometric coverage of SN 2004eo allows a detailed
comparison of the colour and luminosity evolution with those of other
well--studied SNe~Ia.  The distance of NGC 6928 was computed from
the recession velocity corrected for Local Group infall into the Virgo
cluster (v$_{Vir}$ = 4810 km s$^{-1}$). Assuming $H_0$ = 72 km
s$^{-1}$ Mpc$^{-1}$, we obtain a distance $d \approx 67$ Mpc,
or $\mu = 34.12 \pm 0.10$ mag (LEDA).

The total reddening was estimated by taking into account only the
Galactic contribution, $E(B-V) = 0.109$ mag \cite{schl98}.  The
Cardelli, Clayton, $\&$ Mathis \shortcite{card89} law was used to
estimate the extinction in the different bands.  The SN
has a peripheral location in the host galaxy and narrow interstellar
Na~I~D lines are not detected.  This suggests that the light of
SN~2004eo was not significantly extinguished in the host galaxy.

We compared the colour and {\it uvoir} light curves of SN~2004eo with
those of other normal SNe~Ia (SNe 1992A, 1996X, 1994D, 2002bo, 2002er,
and 2003du) spanning a range of different $\Delta m_{15}(B)$ values.
Comparisons were also made with the more peculiar SN 1991T (Filippenko
et al. 1992b; Phillips et al. 1992) and SN 1991bg (Filippenko et al. 
1992a; Leibundgut et al. 1993). In 
Tab. \ref{main_parI} we list the JD of the $B$-band maximum,
distance modulus, and extinction for each SN.

\begin{figure*}
\includegraphics[width=13cm,angle=270]{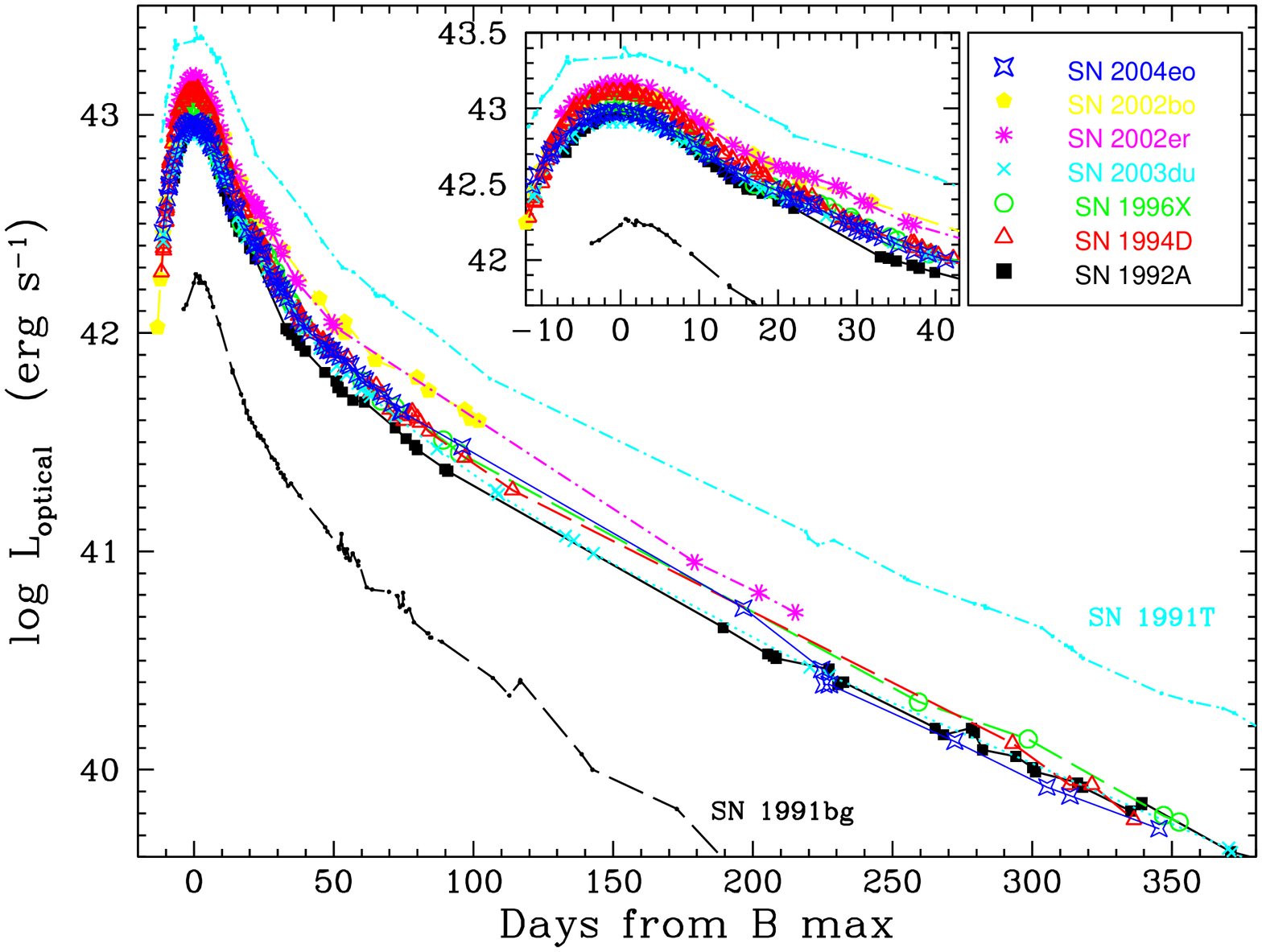}
\caption{The integrated {\it UBVRI} ({\sl uvoir}) light curve of
SN~2004eo compared with those of typical SNe~Ia plus the peculiar Type
Ia SNe 1991T and 1991bg.  The SN sample is the same as in
Fig. \ref{colour_ev}.  An enlarged detail of the pseudo--bolometric
light curves between $-12$~d and +43~d is also shown. Lacking 
$U$-band observations for SNe~1991bg, 1992A, and 2002bo, the contribution
of this band to their total luminosity was estimated following
Contardo et al. \protect\shortcite{con00}.
\label{bolo_cur} }
\end{figure*}

\begin{table*}
\begin{center}
\caption{Main parameters adopted for our sample of SNe~Ia. \label{main_parI}}
\footnotesize
\begin{tabular}{cccccc}\hline \hline
SN  & JD($B_{max}$) & $\Delta m_{15}(B)$ & $\mu$ & $E(B-V)$ & sources \\ \hline
2004eo & 2,453,279.2 & 1.46 & 34.12 & 0.109 & 0,1 \\
1991T & 2,548,374.5 & 0.94 & 30.74 & 0.22 & 2,3,4,5,6,7  \\
1991bg & 2,448,605.5 & 1.94 & 31.32 & 0.04 & 5,8,9,10,11 \\ 
1992A & 2,448,640.5 & 1.45 & 31.41$^\ddag$ &0.06 & 6,12,1 \\
1994D & 2,449,432.5 & 1.31 & 31.14 & 0.03 & 6,11,13,14,15,16 \\
1996X & 2,450,191.5 & 1.32 & 32.17 & 0.07 &  17,18,1\\
2002bo & 2,452,356.5 & 1.17 & 31.45 & 0.38 & 5,19,20,1 \\
2002er & 2,452,524.16 & 1.33 & 32.87 & 0.36 & 21,1 \\
2003du & 2,452,766.38 & 1.02 & 32.42 & 0.01 & 22,23,1 \\ \hline
\end{tabular}
\end{center}

0 = This paper; 1 = LEDA;
2 = Schmidt et al. \protect\shortcite{brian94}; 
3 = Lira et al. \protect\shortcite{lira98};\\
4 = Cappellaro et al. \protect\shortcite{capp97};
5 = Krisciunas et al. \protect\shortcite{kris04};
6 = Altavilla et al. \protect\shortcite{alt04};\\ 
7 = Saha et al. \protect\shortcite{saha01};
8 = Filippenko et al. \protect\shortcite{alex92a}; 
9 = Leibundgut et al. \protect\shortcite{bruno93};\\
10 = Turatto et al. \protect\shortcite{tura96}; 
11 = Tonry et al. \protect\shortcite{ton01};
12 = Suntzeff \protect\shortcite{nick96}; \\
13 = Richmond et al. \protect\shortcite{rich95};
14 = Tsvetkov $\&$ Pavlyuk \protect\shortcite{tsve95}; 
15 = Patat et al. \protect\shortcite{nando96};\\
16 = Meikle et al. \protect\shortcite{peter96}; 
17 = Riess et al. \protect\shortcite{riess99};
18 = Salvo et al. \protect\shortcite{mary01}; \\
19 = Benetti et al. \protect\shortcite{ben04};
20 = Stehle et al.  \protect\shortcite{ste05};
21 = Pignata et al. \protect\shortcite{pig04};\\ 
22 = Stanishev et al. \protect\shortcite{sta05};
23 = Leonard et al. \protect\shortcite{leo05};
$^\ddag$ = average $\mu$ from different sources.
\end{table*}

As can be seen, in Fig. \ref{colour_ev} the colour curves of SN 2004eo
show an evolution similar to that of normal SNe~Ia, and different from
those of SN~1991bg which are redder until about +3 weeks past maximum.  After
the initial shift to the blue, the $B-V$ colour of SN~2004eo reddens
from 0 at +5~d to 1.2 at about +1 month.  Later on (at $\ge$
+30~d, Fig. \ref{colour_ev}, top), the $B-V$ curve becomes bluer
again. At late times ($\ge$ +200~d) the $B-V$ colour of SN~2004eo is
unusually red ($\sim$0.1 mag). This colour is 0.3--0.4 mag redder
than in other SNe~Ia at comparable phases.  Some of the
difference may be attributed to the fact that most of the reference
SNe are not S--corrected. S--correction can be fairly large at late
phases because of the presence of strong emission lines. However,
other more physical reasons may account for this
anomalous late--time $B-V$ colour (see Sect. \ref{opt_spec}).
Instead, there is no corresponding red excess of the late--epoch $V-R$
and $V-I$ colour curves (see below).

The $U-B$ colour of SN~2004eo (Fig. \ref{colour_ev}, bottom) increases
from $-0.3$ at a few days after maximum, to 0.5 at about +20~d.
During the subsequent month, it remains almost constant, $U-B
\approx 0.2$ mag.  The $V-R$ colour curve of SN~2004eo
(Fig. \ref{colour_ev}, top) evolves from 0.3 to $-$0.1 mag during the period
$-1$~week to +1~week. It then reddens from $-$0.1 to 0.6 mag between +10
and +20~d, before turning blueward again ($V-R \approx -0.5$ in
the late nebular phase). A similar evolution is observed in the $V-I$
colour (Fig. \ref{colour_ev} bottom).

Fig. \ref{bolo_cur} shows the pseudo--bolometric ({\it uvoir}) 
light curve of SN~2004eo, compared with those
of other SNe~Ia, including the peculiar SNe~1991T and 1991bg.
The {\it uvoir} light curve was obtained by integrating the fluxes in
the optical region from the $U$ band to the $I$ band.
The J, H and K--band data were not included because most of the 
comparison SNe do not have well--sampled NIR light curves.
The light curve is similar to those of typical SNe~Ia, fainter than
that of SN~1991T, and definitely brighter than that of SN~1991bg.  In particular,
SN~2004eo appears to have a luminosity evolution similar to that of
SN~1992A (at least up to +1 month).

Finally, using available $J$, $H$, and $K$ data, we estimate the IR
contribution to the SN bolometric luminosity to be negligible
near maximum ($\sim$2--3$\%$), increasing to over 20$\%$ by about 
+50~d.
 
\subsection{Main Parameters of SN~2004eo from the Photometry} \label{param}

\begin{figure}
\includegraphics[width=7.6cm,angle=270]{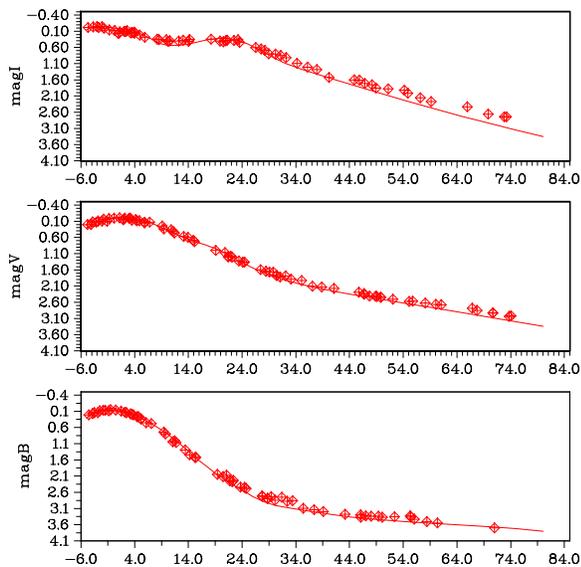}
\caption{Comparison of the $I$ (top), $V$ (middle), and $B$ (bottom) 
light curves of SN~2004eo with those of the template Type Ia SN 1992A
\protect\cite{ham96}. The light curves have been offset vertically
by various amounts for clarity.
\label{BVI_fit} }
\end{figure}

\begin{table*}
\begin{center}
\caption{Main parameters describing the $BVRI$ light curves of SN
2004eo, as derived from polynomial fits. The Galactic absorptions in
the different bands are those reported by Schlegel et
al. \protect\shortcite{schl98}. The distance modulus adopted is
34.12 mag.}
\label{SN04eo_param}
\footnotesize
\begin{tabular}{ccccc}\hline \hline
Parameters & $B$ & $V$ & $R$ & $I$ \\ \hline
JD$_{max}$ & 2,453,279.2$\pm$0.5 & 2,453,280.8$\pm$0.4 & 2,453,280.1$\pm$0.7& 2,453,276.3$\pm$1.0 \\
JD($\lambda$)$_{max} -$ JD(B)$_{max}$ & -- & $+1.6^d$ & $+1.0^d$ & $-3.2^d$ \\
A$_\lambda$ & 0.468 & 0.360 & 0.290 & 0.211 \\
m$_{\lambda,max}$ & 15.51$\pm$0.02 & 15.35$\pm$0.02 & 15.22$\pm$0.03 & 15.36$\pm$0.04 \\
M$_{\lambda,max}$ & --19.08$\pm$0.10 & --19.13$\pm$0.10 & --19.19$\pm$0.10 & --18.97$\pm$0.11 \\ 
$\Delta$m$_{15}(\lambda)_{obs}$ & 1.45$\pm$0.04 & 0.74$\pm$0.04 & 0.57$\pm$0.07 & 0.45$\pm$0.10\\
$\Delta$m(neb) (mag/100$^{d}$)& 1.53$\pm$0.11& 1.48$\pm$0.06& 1.60$\pm$0.26& 1.25$\pm$0.29\\ \hline
\end{tabular}
\end{center}
\end{table*}

A $\chi^2$ test shows that the best match to the SN~2004eo light
curves in all bands is given by SN 1992A \cite{ham96}. As shown in
Fig. \ref{BVI_fit}, the fit is excellent in the $B$ and $V$ bands, while
some difference is seen in the $I$ band, such as in the depth of the
minimum.  This is not surprising, since differences in the light
curves of SNe~Ia are usually more evident in this band \cite{nick96}.

The epoch and apparent magnitude of the $B$-band maximum (JD =
2,453,279.2 $\pm$ 0.5, $B=15.51 \pm 0.02$ mag) were estimated by fitting the
observations of the first month with a fifth or sixth--order polynomial.  The
epochs of the $V$ and $R$-band maxima were similarly determined and occur,
respectively, at +1.6 and +0.9~d, while the $I$-band maximum is at
about $-3$~d.  Fitting the $I$-band light curve with a
higher--degree spline, we found the epoch of the $I$-band secondary
maximum to be at about +19~d (i.e., $\sim$22~d after the main
$I$-band maximum).

Given the peak magnitude and adopted distance and reddening (see Sec
3.4), the absolute $B$-band peak magnitude is $M_{B,max} =
-19.08 \pm 0.10$ mag. This is only marginally fainter than the average
for normal SNe~Ia \cite{gib00}. The absolute magnitudes of the $V$, $R$, 
and $I$-band maxima are, respectively, $M_{V,max} = -19.13 \pm 0.10$ mag, 
$M_{R,max} = -19.19 \pm 0.10$ mag, and $M_{I,max} = -18.97 \pm 0.11$ mag 
(the uncertainties include fitting errors plus uncertainty in the distance).

The observed $\Delta m_{15}(B)_{obs}$ was found to be 1.45 $\pm$ 0.04 mag.
In Tab. \ref{SN04eo_param} we show the corresponding $\Delta m_{15}$
values for the $V$, $R$, and $I$ bands, together with other light--curve
parameters.  The reddening--corrected \cite{phil99}
$\Delta m_{15}(B)_{true}$ = 1.46; because of the low reddening,
the correction was negligible.

It has been established that the luminosity of SNe~Ia correlates with
the decline rate after maximum light; the slower the SN declines, the
brighter the absolute peak magnitude \cite{psk77,phil93,ham95,ham96b,riess05}. 
More recently, other calibrations of the relations between absolute peak
magnitudes and light-curve shape have been
determined and refined using ever larger samples of well-studied SNe~Ia.

The application of these relations reduced the scatter in the Hubble
diagram of SNe~Ia and significantly improved their effectiveness as
distance indicators.  As detailed by Pastorello et al.
\shortcite{pasto05}, we make use of some of these relations to
determine the peak luminosity of SN~2004eo (see
Tab. \ref{sn04eo_M_B}). All relations have been rescaled to $H_0$ = 72
km s$^{-1}$ Mpc$^{-1}$.



Wang et al. \shortcite{wang05} recently introduced a new photometric
parameter, the intrinsic $B-V$ colour at 12~d after maximum light
($\Delta C_{12}$). Using the relations reported in their Tab. 1 and
Tab. 2 (see also Pastorello et al. 2007), we can derive the absolute
magnitudes at maximum for the $B$, $V$, and $I$ bands.  For SN~2004eo,
$\Delta C_{12}=0.49 \pm 0.07$ is found. The predicted magnitudes are given
in Tab. \ref{sn04eo_M_B}.

\begin{table}
\begin{center}
\caption{Decline--rate corrected absolute $B$, $V$, and $I$ magnitudes 
obtained by various methods (see text) and (last line) weighted averages 
of the different estimates. They can be compared with the absolute
magnitudes derived using the distance modulus
(Tab. \ref{SN04eo_param}).}
\label{sn04eo_M_B}
\footnotesize
\begin{tabular}{ccccc}\hline \hline
Method & $M_{B,max}$ & $M_{V,max}$ & $M_{I,max}$ \\ \hline
Phillips$^{(1)}$ & --18.83$\pm$0.18 &  --18.89$\pm$0.18 &--18.70$\pm$0.18 \\ 
Altavilla$^{(2)}$ & --18.99$\pm$0.08 & & \\
Reindl$^{(3)}$ &  --18.94$\pm$0.07 & --18.94$\pm$0.05 & --18.72$\pm$0.08 \\
Wang$^{(4)}$ & --19.01$\pm$0.16 & --19.00$\pm$0.13 & --18.72$\pm$0.12\\ \hline \hline
Weighted avg.& --18.95$\pm$0.07 & --18.94$\pm$0.04 & --18.72$\pm$0.01 \\  \hline
\end{tabular}
$^{(1)}$ Phillips et al. \protect\shortcite{phil99};
$^{(2)}$ Altavilla et al. \protect\shortcite{alt04};\\ 
$^{(3)}$ Reindl et al. \protect\shortcite{rein05};
$^{(4)}$ Wang et al. \protect\shortcite{wang05}.\\
\end{center}
\end{table}

The different methods appear to be in good agreement.
Only for the $B$ magnitude obtained applying the Phillips
et al. \shortcite{phil99} relation do we see a non--negligible deviation.  
Averaging the absolute
magnitudes, we obtain $M_{B,max} = -18.95 \pm 0.07$, $M_{V,max} =
-18.94 \pm 0.04$, and $M_{I,max} = -18.72 \pm 0.01$ mag.  Given the
uncertainties, these values are consistent with the directly
measured magnitudes (cf. Tab. \ref{SN04eo_param}), which are marginally
brighter.

Another useful photometric parameter is the stretch factor $s^{-1}$
\cite{perl97}.  The measured value for SN~2004eo is $s^{-1} = 1.12
\pm 0.04$.  This value is similar to that derived using a relation
between $s^{-1}$ and $\Delta m_{15}(B)_{true}$ \cite{alt04}: $s^{-1}
= 1.17 \pm 0.08$.

Finally, we can estimate the explosion epoch by applying the method
proposed by Riess et al. \shortcite{ries99b}.  Assuming that the SN
luminosity is proportional to the square of the time elapsed since
explosion and considering all photometric points starting around $-8.5$~d 
(including $B$-band measurements from Gonzalez et al. 2004),
we obtain a rise time $t_r = 17.7 \pm 0.6$~d.  This is a slightly short 
time for a relatively fast--declining SN~Ia, but surprisingly similar 
to the rising time determined by Garg et al. \shortcite{garg07} using 
$VR$ broad--band observations of a sample of 14 SNe`Ia at $z = 0.11$--0.35, 
behind the Large Magellanic Cloud ($t_r = 17.6 \pm 1.3$(stat) $\pm$ 
0.07(sys)). 

Consequently, the epoch of the explosion of SN~2004eo is estimated 
to be JD = 2,453,261.5 $\pm$ 0.8.

\subsection{Ejected Mass of $^{56}$Ni} \label{Ni_mass}

An important physical parameter of SNe~Ia, the mass of $^{56}$Ni
synthesized, can be estimated by modelling the bolometric light
curves.  The bolometric light curve of SN~2004eo was derived from the
observed {\sl uvoir} light curve (see Sect. \ref{color_uvoir} and Fig.
\ref{bolo_cur}), applying the UV and IR corrections of Suntzeff
\shortcite{nick96}.

The light--curve model was computed using a grey Monte Carlo code
developed by Mazzali et al. \shortcite{mazz01}. The code accounts for
the propagation and deposition of gamma--ray photons and positrons
emitted in the radioactive decay chain $^{56}$Ni to $^{56}$Co to
$^{56}$Fe, followed by diffusion through the ejecta of the photons
which ultimately constitute the observed SN light.

\begin{figure}
\includegraphics[width=8.5cm]{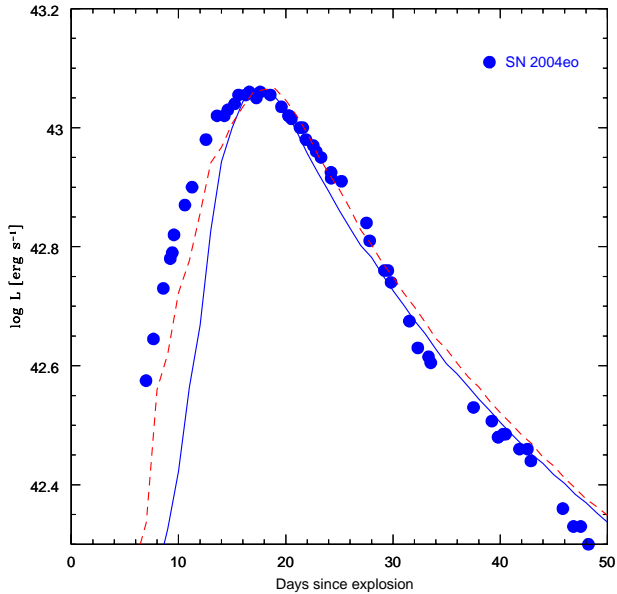}
\caption{Comparison between the bolometric light curves of SN~2004eo
(filled points), the model adopted for SN~2002bo \protect\cite{ste05}
(dashed line), and the W7--based model (solid line), both scaled to a
$^{56}$Ni mass of 0.45~M$_\odot$.
\label{LC_model} }
\end{figure}

In Fig. \ref{LC_model} we compare the bolometric light curve of
SN~2004eo with the model computed for SN~2002bo \cite{ste05}. This
originally invoked 0.50--0.55~M$_\odot$ of ejected $^{56}$Ni mass, but
here it has been scaled down to 0.45~M$_\odot$ in order to fit the
lower bolometric luminosity of SN~2004eo. The bolometric light curve of
SN~2004eo is slightly broader than the model, but is otherwise
consistent with it. This mass is close to the
lower limit of the $^{56}$Ni mass range observed in normal SNe~Ia,
about 0.4--1.1~M$_\odot$ \cite{capp97}.


For completeness, we also show in Fig. \ref{LC_model} a synthetic
light curve based on the W7 density distribution \cite{ken84} scaled
to a $^{56}$Ni mass of 0.45~M$_\odot$. This yields a rather poorer
reproduction of the observed light curve, especially in the rising
phase.  The better match of the Stehle et al. (2005) model at this
time is due to their inclusion of outward mixing of $^{56}$Ni.

\subsection{Decline Rate During the Nebular Phase}

The $B$, $V$, $R$, and $I$ decline rates of SN~2004eo during the 
nebular phase (between +190~d and +350~d) were also computed (see
Tab. \ref{SN04eo_param}, last line).  The slopes of the $B$, $V$, 
and $R$ light curves (about 1.5--1.6 mag/100~d) are in good agreement 
with the average slope ($\sim$1.4 mag/100~d) computed by Lair et al. 
\shortcite{lair06} for normal and luminous SNe~Ia (or high--velocity 
and low--velocity gradient SNe~Ia,
following the nomenclature introduced by Benetti et al. 2005; see
Sect. \ref{disc}).  During this phase, the ejecta become transparent
to the gamma rays and the luminosity is increasingly powered by
the energy deposition of the trapped positrons.  However, a fraction
of the positrons may escape, resulting in a steeper light-curve
decline than would be expected from the combined effects of just
$^{56}$Co decay and increasing gamma--ray escape.  
Indeed, this steeper decline in the $B$, $V$, and $R$ light curves 
is usually observed in SNe~Ia at late phases \cite{capp97,mil99}.  

The behaviour of the late-time $I$-band magnitude decline is 
expected to be slightly different. Lair et al. \shortcite{lair06} 
indeed found that a slower average decline rate in the
late--time $I$-band light curve (0.94 mag/100~d) could be a common
characteristic of normal SNe~Ia. This is true also for SN~2004eo, although 
the $I$-band light curve actually declined at a slightly higher rate 
(1.25 mag/100~d). 
Lair et al. suggested that the reason for the slow $I$-band decline could be 
a shift of the late--time flux from optical to IR wavelengths.  
This is consistent with an almost
constant IR luminosity during the nebular phase, as was found for
SN~2000cx at 1--1.5 years past the $B$-band maximum \cite{soll04}.

\begin{figure*}
\includegraphics[width=13.25cm]{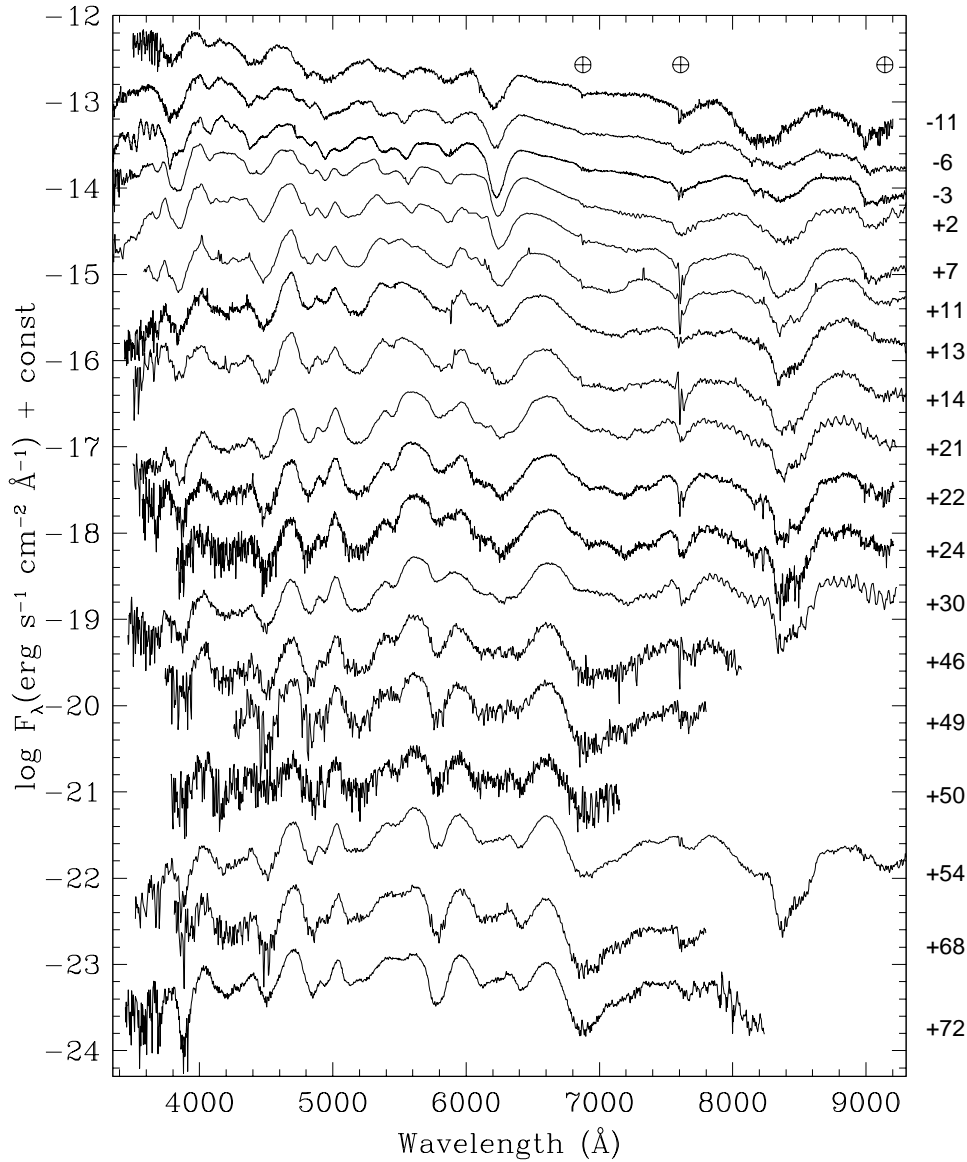}
\caption{Evolution of the optical spectra of SN~2004eo. 
The spectra have not been corrected for the host-galaxy redshift. 
The positions of the main telluric features are marked with 
the symbol $\oplus$. The phases are labeled on the right.
\label{spec_ev} }
\end{figure*}

\section{Spectroscopy} \label{spec}

Spectroscopic monitoring of SN~2004eo spanned the period $-$11~d to +72~d.
Three spectra were obtained before maximum brightness, and 15 after.  
In addition, a late-time VLT spectrum was obtained at about 8 months.  IR
spectroscopic observations were performed at +2, +22, and +36~d.
Unfortunately, the resolution and/or the signal--to--noise ratio (S/N)
of the IR spectra obtained for SN 2004eo are rather poor.

\subsection{Optical Spectra} \label{opt_spec}

\begin{figure}
\includegraphics[width=8.5cm]{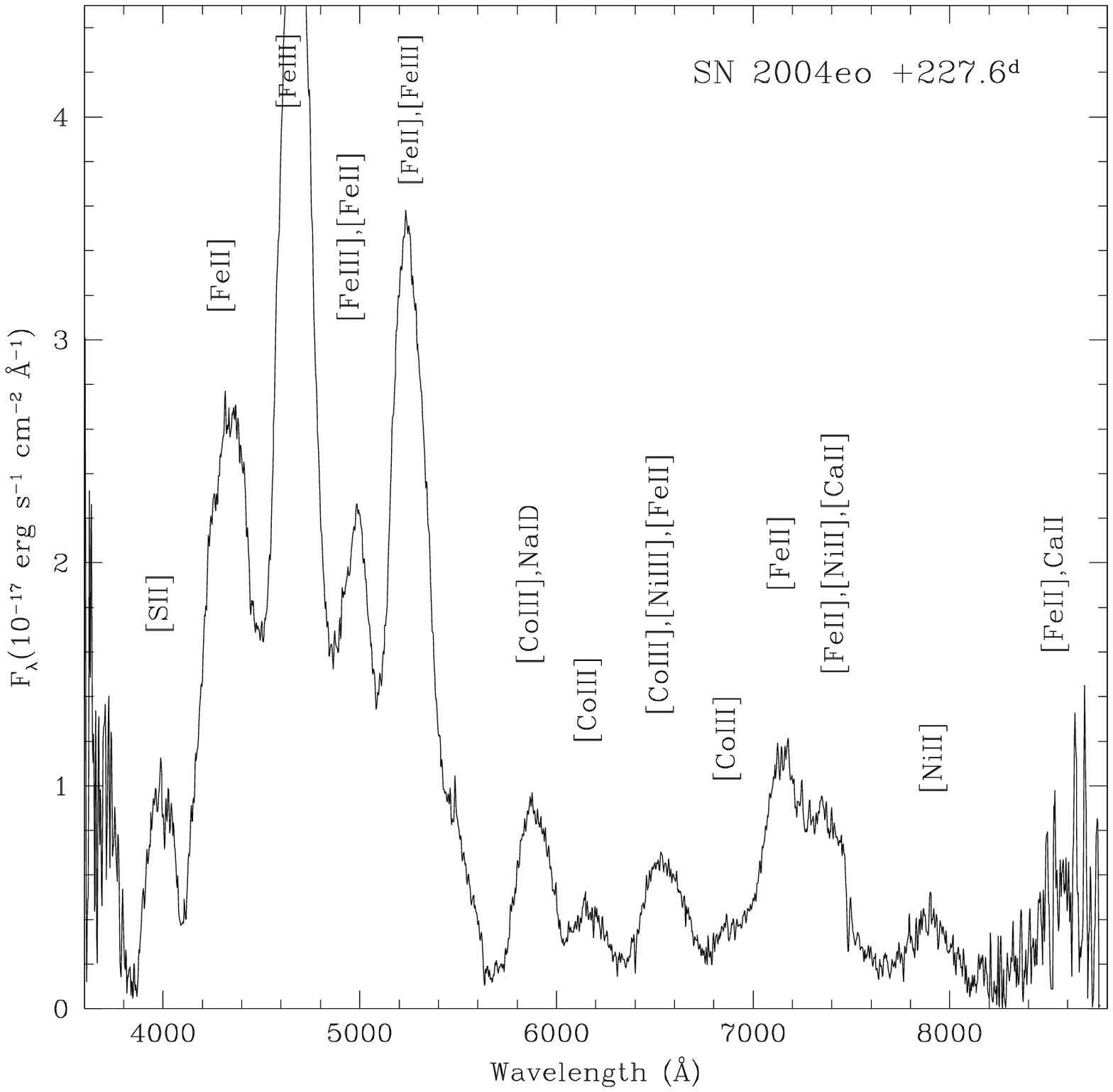}
\caption{Nebular (phase = 227.6~d) spectrum of SN~2004eo taken
with VLT + FORS1. Identifications of the main features are also shown.
\label{spec_neb} }
\end{figure}

The optical spectra of SN~2004eo obtained during the first season are
shown in Fig. \ref{spec_ev}.  The earliest spectra exhibit the broad
P--Cygni lines typical of SNe~Ia: the characteristic deep absorption 
near 6150~\AA\ due to Si~II $\lambda\lambda$6347, 6371 (hereafter Si~II
$\lambda$6355), the Si~II $\lambda\lambda$5958, 5979~\AA\ feature 
(hereafter Si~II $\lambda$5972), the W--shaped feature near 5400~\AA\ 
attributed to S~II $\lambda$5454 and S~II $\lambda$5606.  Other 
prominent features at pre--maximum epochs are Ca~II H$\&$K, Mg~II 
$\lambda$4481, and several blends due to Fe~II, Si~II, and Si~III.
Fe~III is also possibly detected below 5000\AA. At red wavelengths, 
particularly strong features are the Ca~II near-IR triplet, possibly 
with a high--velocity component \cite{maz05}, and Mg~II 
$\lambda\lambda$9218, 9244.  Despite contamination from the 7600~\AA\ 
telluric feature, O~I $\lambda$7774 is clearly visible. However, there 
is no evidence of C~II lines in the early-time spectra.

\begin{figure*}
\includegraphics[width=8.5cm]{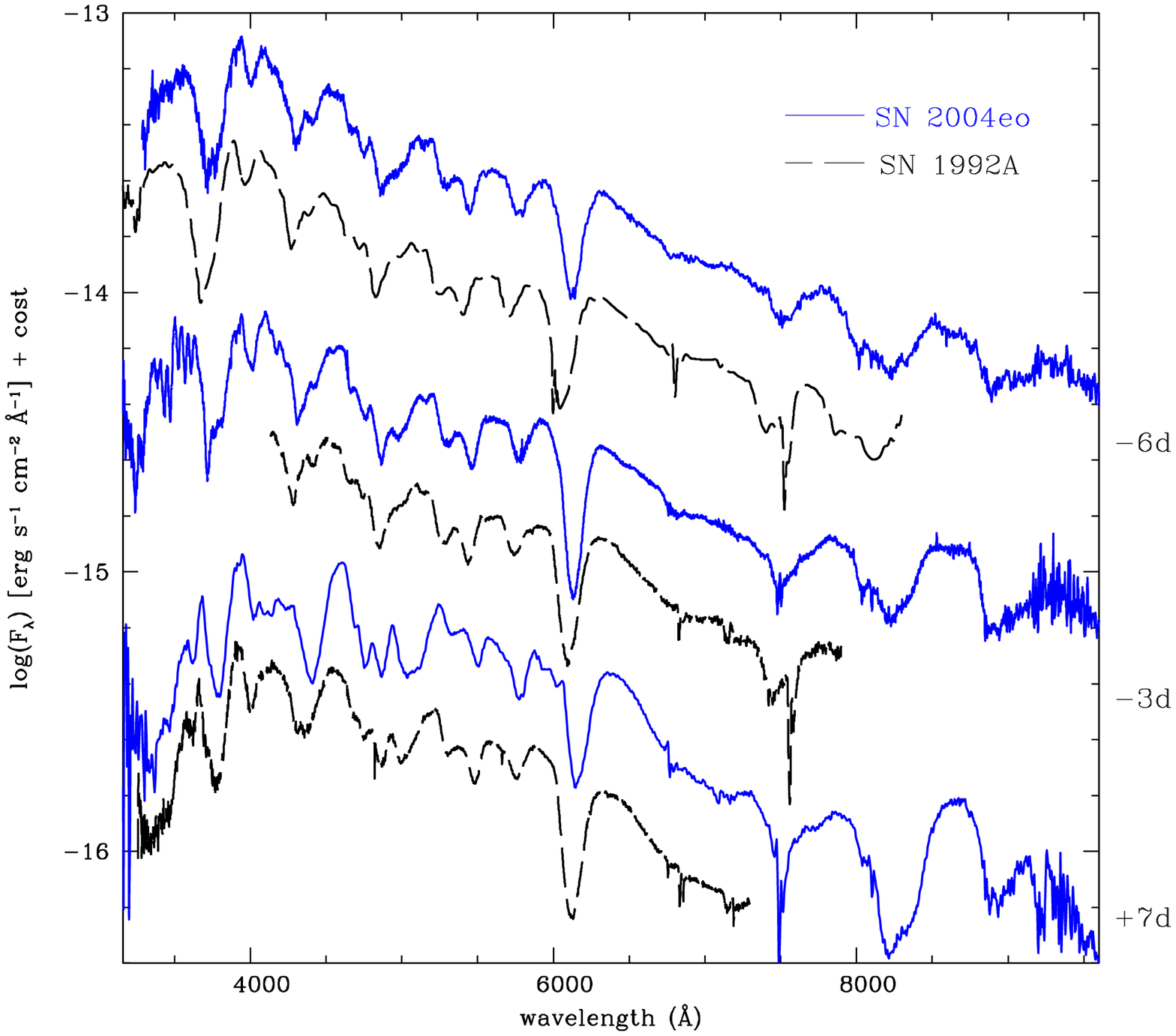}
\includegraphics[width=8.5cm]{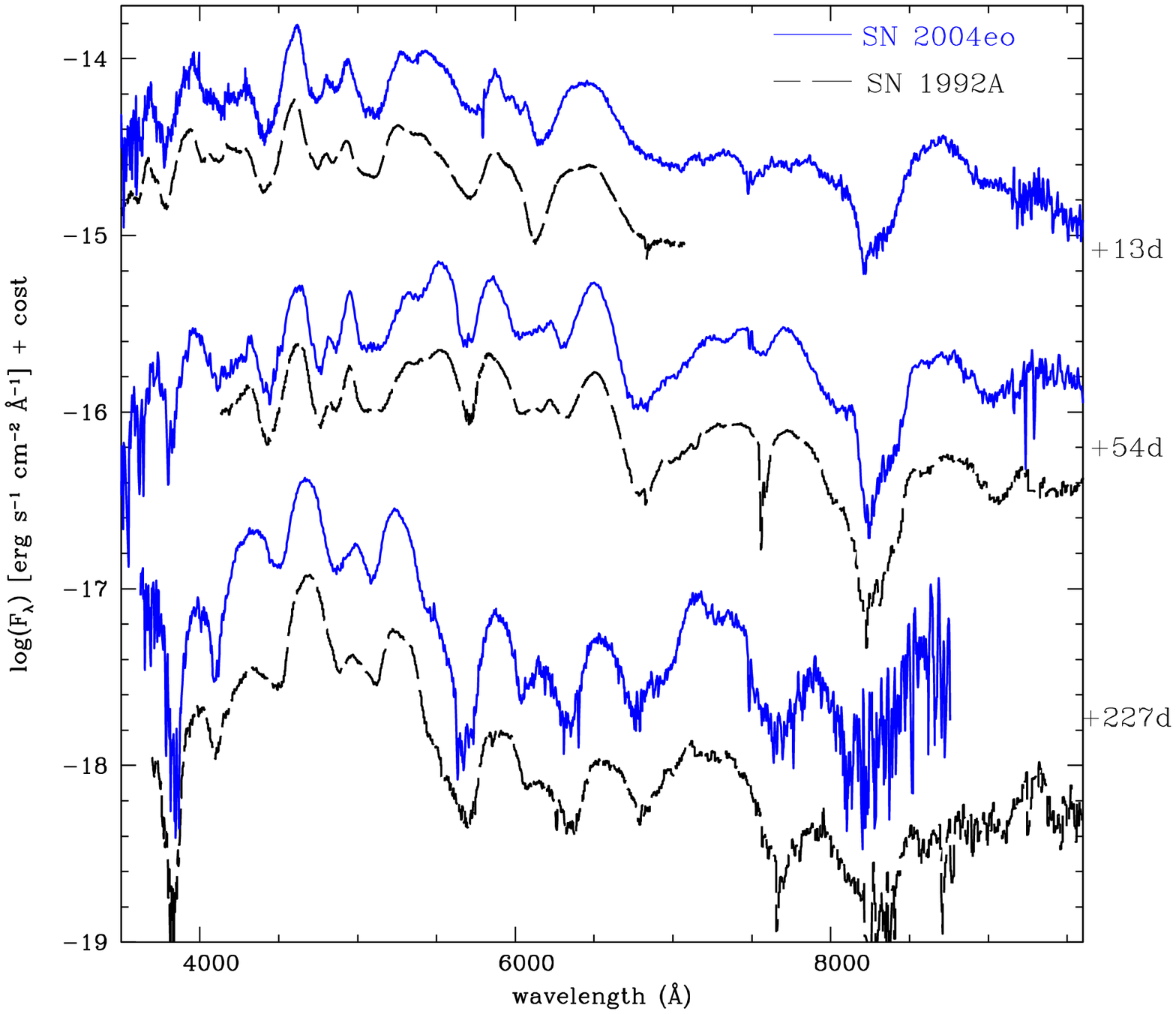}
\caption{Left: comparison between spectra of SN 2004eo and SN 1992A
around maximum brightness. Right: same as left, but for later phases. 
All spectra have been corrected for reddening and shifted to the 
host--galaxy rest frame.
\label{spec_04eo92A} }
\end{figure*}

\begin{figure*}
\includegraphics[width=8.5cm]{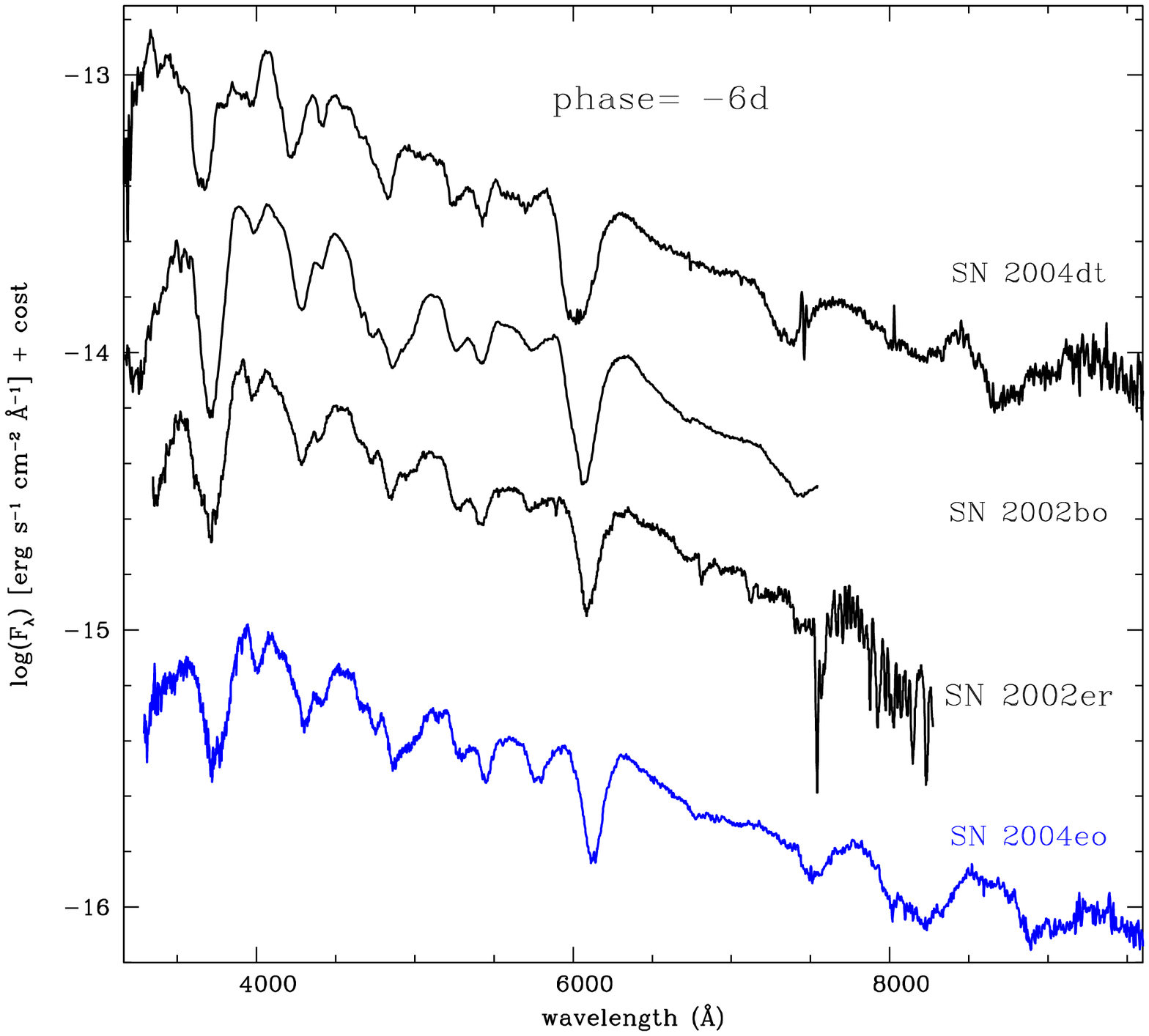}
\includegraphics[width=8.5cm]{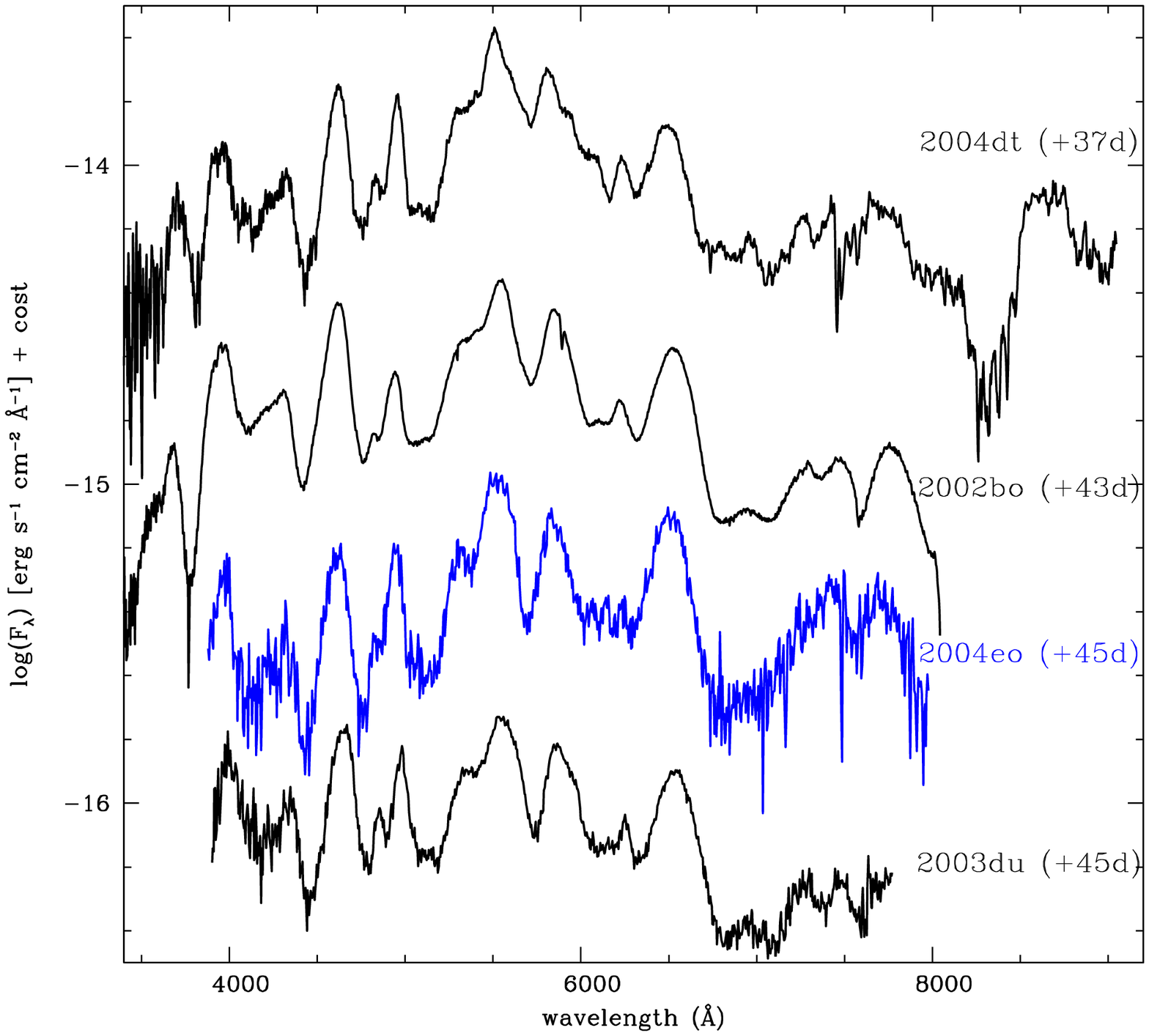}
\caption{Left: comparison of a pre--maximum spectrum ($\sim -6$~d)
of SN~2004eo with those of the normal SNe~Ia 2004dt, 2002bo, and 2002er
at similar phases (acquired by the ESC).  Right: comparison of a
post--max spectrum ($\sim$ +45~d) of SN~2004eo with those of the
normal SNe~Ia 2004dt, 2002bo, and 2003du at similar phases.  All
spectra have been corrected for reddening and shifted to the
host--galaxy rest frame.
\label{spec_norm_pre} }
\end{figure*}

After maximum light the Ca~II near-IR triplet is very prominent, while the
strengths of the Si~II and S~II lines rapidly decline.  In particular,
Si~II $\lambda$5972 is progressively replaced by the Na~I~D
$\lambda\lambda$5890, 5896 feature. However, Si~II $\lambda$6355 is
visible up to about 1 month past maximum.  Consistent with normal 
SN~Ia behaviour, the post--maximum spectra show strong line
blanketing at blue wavelengths due to Fe~II, Ni~II, Co~II, Ti~II and
Cr~II lines. Apart from the persistent Ca~II H\&K, Ca~II near-IR triplet, 
and Na~I~D features, the spectra at $\sim$50~d post--maximum are dominated
by iron--group lines \cite{bran05}.

A spectrum of SN~2004eo was obtained in the nebular phase 
at +227.6~d using the VLT UT2 equipped
with FORS1 (Fig. \ref{spec_neb}).  Strong forbidden lines of
iron--group elements ([Fe~II], [Fe~III], [Ni~II], [Ni~III], and
[Co~III]) dominate the spectrum, especially in the region 
4000--5500~\AA. Other well--developed features are also visible: 
Na~I~D, [Ca~II] $\lambda\lambda$7193, 7324, and the Ca~II near-IR triplet. 
Moreover, the prominent feature at 4000~\AA~can be tentatively 
identified as [S~II] \cite{bow97}.

\begin{figure}
\includegraphics[width=8.3cm,angle=0]{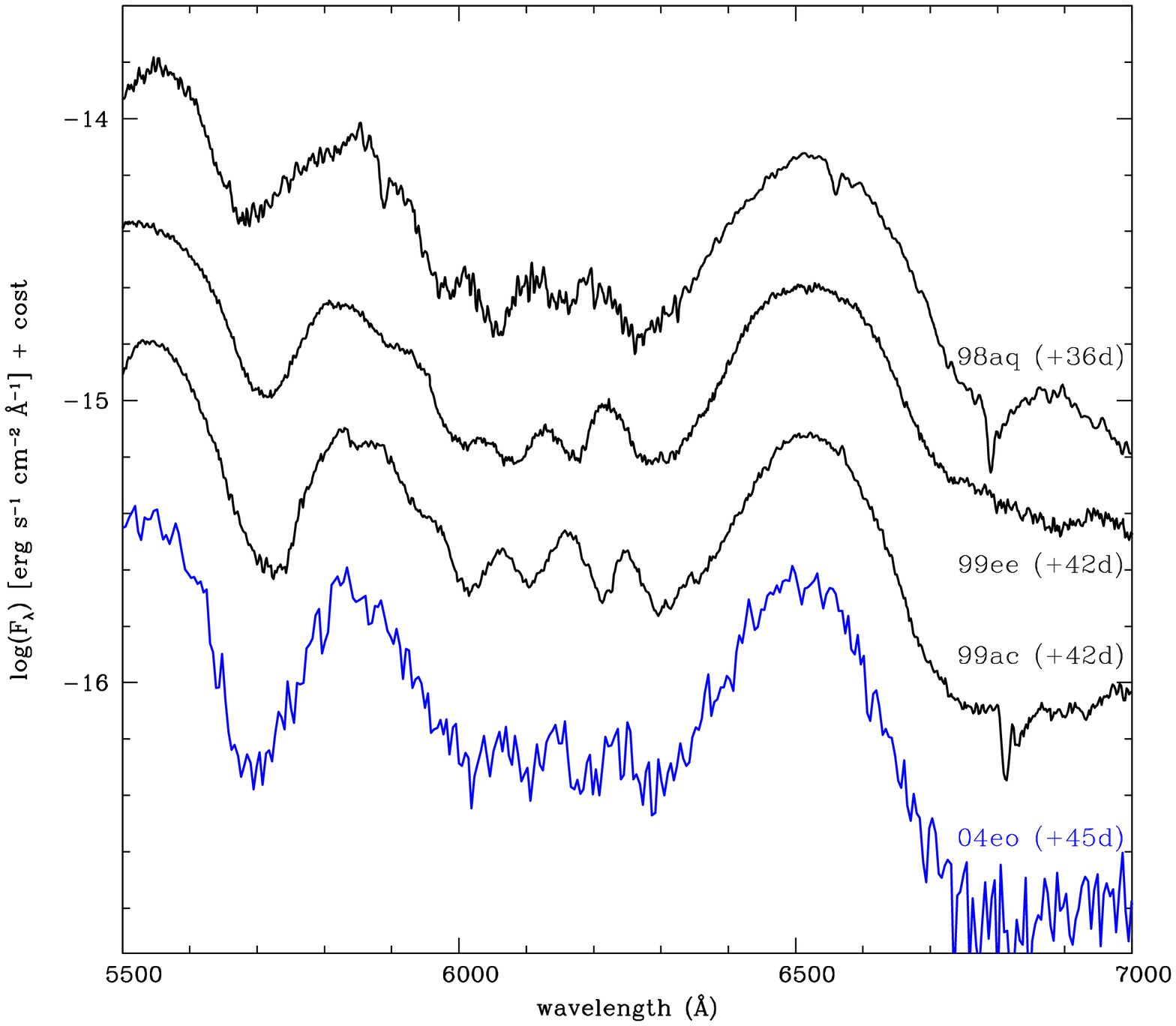}
\caption{Detail of spectra of SN~2004eo and other SNe~Ia at
$\sim$+40--45~d, illustrating the narrow, weak features in the 
wavelength region 6000--6300~\AA.
\label{zoom_noise} }
\end{figure}

\begin{figure*}
\includegraphics[width=8.5cm]{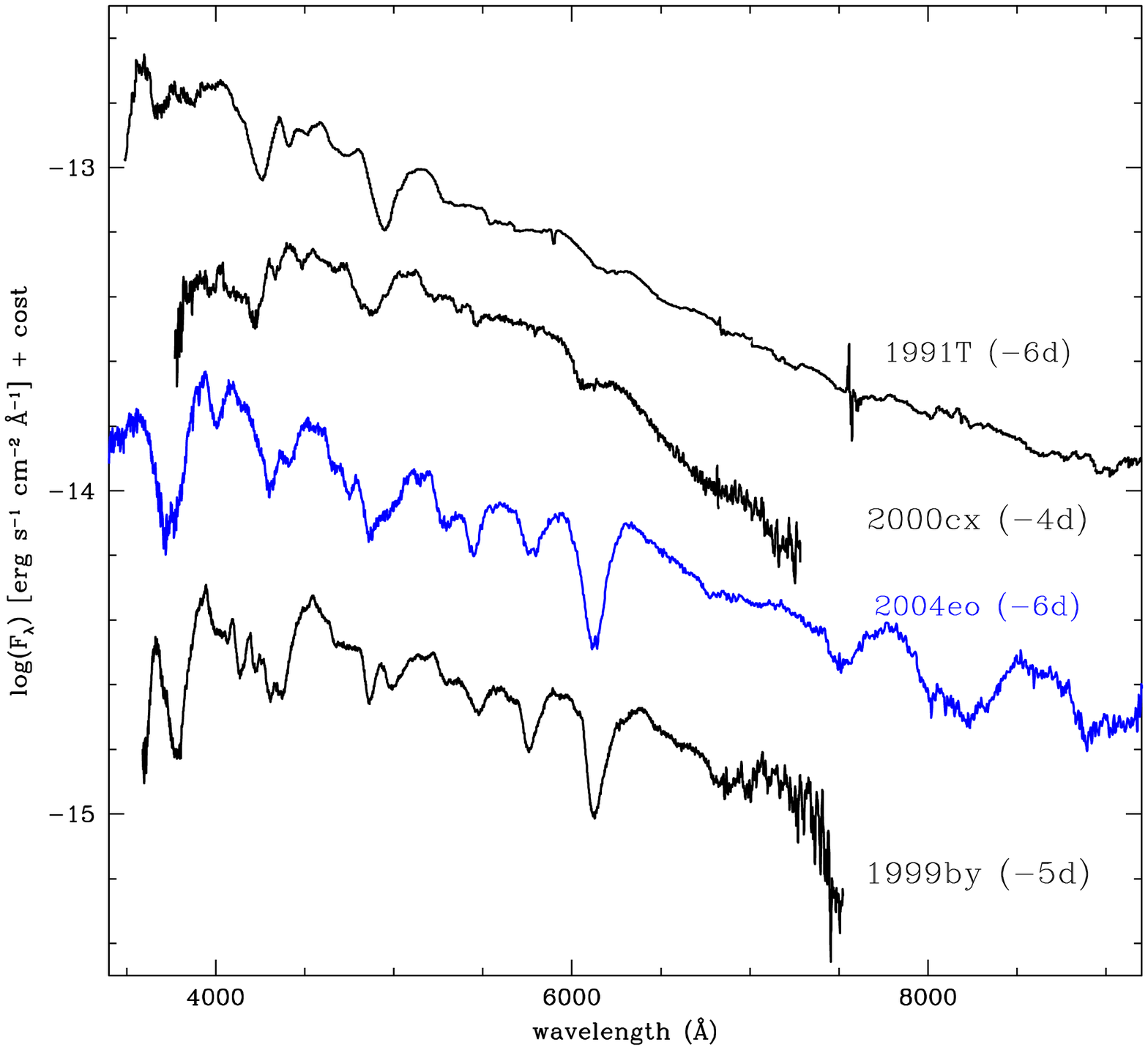}
\includegraphics[width=8.5cm]{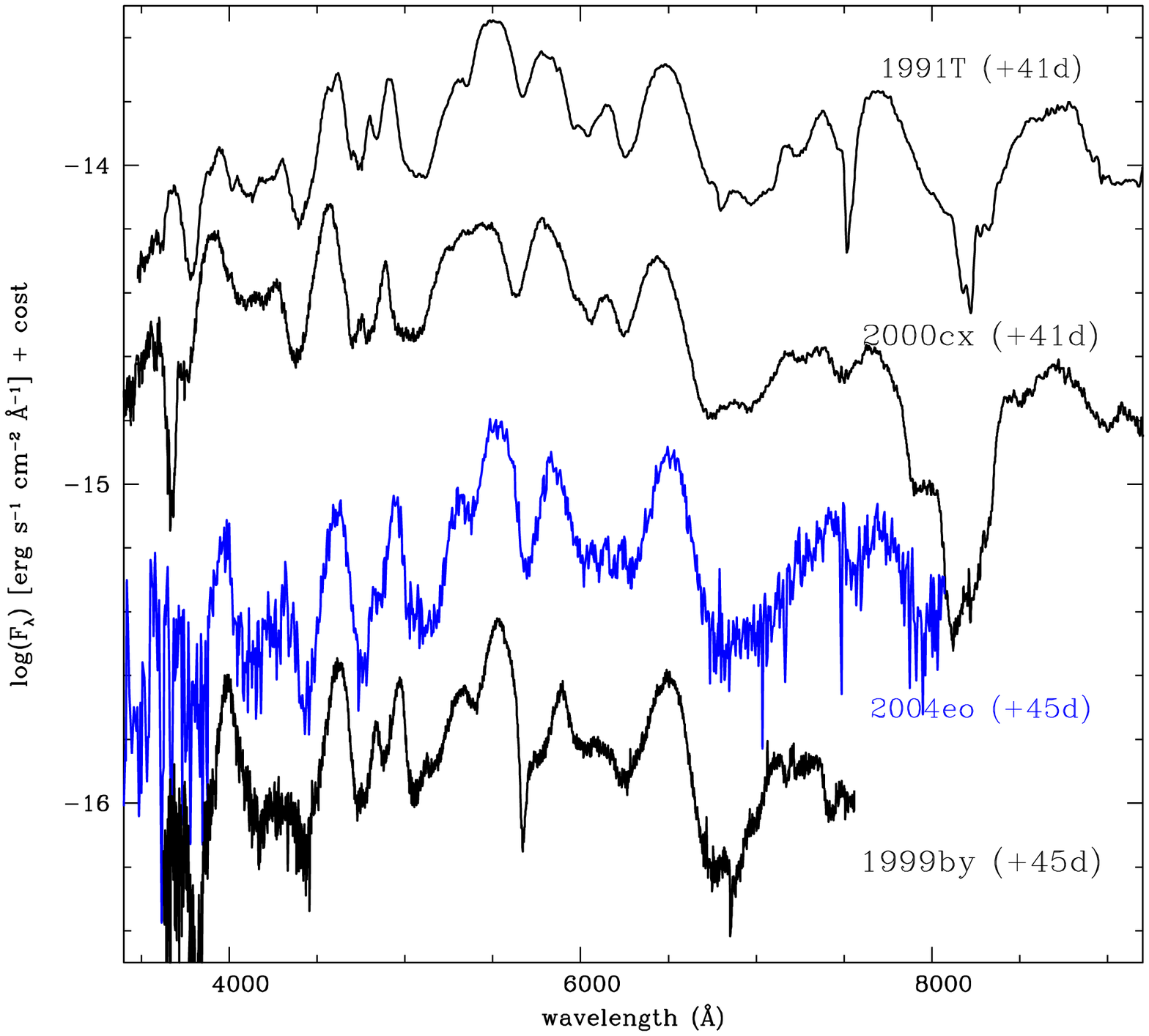}
\caption{Left: comparison of a pre--maximum ($\sim -$6~d) spectrum
of SN~2004eo with those of the peculiar SNe~Ia 1991T, 2000cx, and
1999by.  Right: as for the left--hand panel, but at $\sim$+40--45~d. 
All of the spectra, at both phases, have been corrected for 
reddening and shifted to the host--galaxy rest frame.
\label{spec_pec_pre} }
\end{figure*}

As mentioned in Sect. \ref{color_uvoir} and in Sect. \ref{param},
there are several photometric similarities between SN~2004eo and
SN~1992A (the shape of the light curve, the colour evolution, and the
quasi--bolometric luminosity).  In Fig. \ref{spec_04eo92A} (left)
photospheric spectra of SNe~2004eo and 1992A are compared, at $\sim
-$6, $-$3, and +7~d.  At all epochs the spectra of SNe~2004eo and
1992A show comparable line strengths, with some differences in
the line velocities, with those of SN~2004eo being significantly
lower.  In Fig. \ref{spec_04eo92A} (right) spectra of the same two SNe
are shown, at about +13, +54, and +227~d.  At +227~d, there is
some difference between the two events in the relative strength of
features in the $B$ and $V$ regions, which may help to explain the
anomalously red $B-V$ colour of SN~2004eo after +200~d.  Below
5000~\AA, the [Fe~III] $\lambda$4700 / [Fe~II] $\lambda$4300 intensity 
ratio \cite{liu97} appears to be lower in SN~2004eo with respect to 
SN~1992A. This is consistent with a slightly lower $^{56}$Ni mass
and hence lower temperature in SN~2004eo (Mazzali et al., in
preparation).  However, despite these minor differences, most of the
spectral features confirm the strong similarity of the two SNe,
already apparent in their photometric parameters (see,
e.g., Sect. \ref{param}).

\begin{figure}
\includegraphics[width=8.62cm,angle=0]{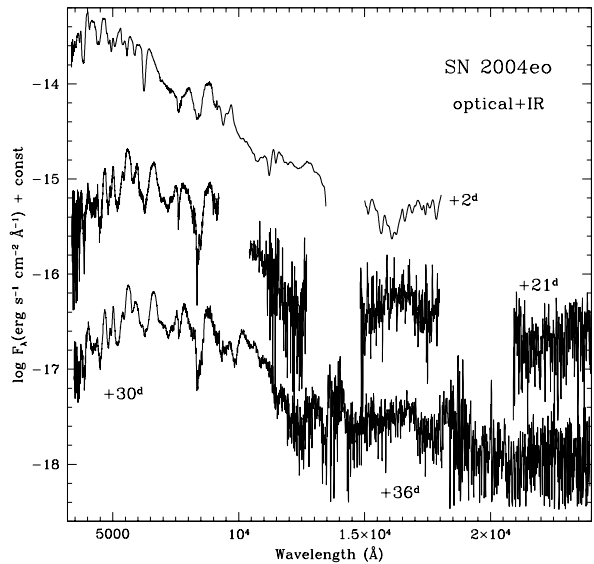}
\caption{Optical and IR spectra of SN~2004eo at comparable
epochs. The +2~d AMICI spectrum was truncated at 18,000~\AA\ because of its 
poor quality.  The
+22~d IRIS2 spectrum was smoothed using a boxcar of 8 pixels
(i.e., $\sim$20~\AA). No smoothing was applied to the +36~d LIRIS
spectrum. This spectrum is combined with an ALFOSC optical spectrum 
obtained 6~d earlier.
\label{spec_evIR} }
\end{figure}

In Fig. \ref{spec_norm_pre} pre--maximum spectra ($\sim -$6~d, left) 
and post--maximum spectra ($\sim$ +45~d, right) of SN~2004eo are
compared with those of normal SNe~Ia at similar phases 
\cite{alt07,ben04,kot05,sta05}.  
At the pre--maximum phase, of particular note is the
greater prominence of the Si~II $\lambda$5972 line in SN~2004eo, again
suggesting lower temperatures in this event.  At +45~d, SN~2004eo
shows a curious sequence of faint, narrow features in the
6000--6300~\AA~region.  This has occasionally been observed in other
SNe~Ia and is probably produced by Fe~II lines.  An enlargement of this
spectral region is shown in Fig. \ref{zoom_noise}, compared with the same
region of spectra of SNe 1998aq \cite{bran03}, 
1999ac \cite{gabri05}, and 1999ee
\cite{ham02b} at comparable phases, available in the SUSPECT \footnotemark[6]  
\footnotetext[6]{{\it http://bruford.nhn.ou.edu/$\sim$suspect/index1.html}}
archive of SN spectra.

Fig. \ref{spec_pec_pre} shows the same two spectra of SN~2004eo as
Fig. \ref{spec_norm_pre}, but now compared with pre-maximum (left
panel) and post-maximum (right panel) spectra of three peculiar SNe~Ia:
SN 1991T \cite{mazz95,gomez98}, SN 2000cx \cite{li01}, and the 
underluminous (SN~1991bg--like) SN~1999by \cite{gar04}.
Both the pre- and post-maximum spectra of SN~2004eo show some
similarity to those of SN 1999by, although the line velocities of
SN~2004eo are somewhat higher and its Si~II lines (especially that at
$\lambda$5972) are slightly less prominent.  But the most important
difference is the lack of clear evidence, in SN~2004eo, of Ti~II lines
which characterise the blue spectral region of SN~1999by and other
low--luminosity SNe~Ia (Filippenko et al. 1992a). 

\subsection{Infrared Spectra}

Fig. \ref{spec_evIR} shows the optical and IR spectra of SN~2004eo at
three epochs: +2~d, +3 weeks, and about +5 weeks. The spectrum at +2~d 
(see Tab.~2) shows a blue continuum. To the red of the Ca~II near-IR
triplet, the Mg~II doublet is visible at $\sim$9230~\AA\ (possibly
blended with Mn~II at 9440~\AA) and the Mg~II triplet at
$\sim$10920~\AA.  Moreover, a broad absorption feature near
12100~\AA\ may be attributed to a blend of Ca~II lines. Si~II
dominates the broad absorption near 16000~\AA, possibly blended with 
Mg~II lines.  Despite the low resolution, Fe~II and Co~II lines are
also possibly detected \cite{mar03}. As we progress from +2~d to 
+22~d (see Tab. \ref{journal}) the spectrum becomes redder.  
Despite the low S/N in the IR spectrum, the $J$-band shows a
prominent Fe~II absorption line at $\sim$12300~\AA, while the 
$H$-band region is dominated by a very broad emission feature 
produced by iron--group lines (Fe~II, Ni~II, Co~II). In the 
$\sim$+30~d spectrum (Tab. \ref{journal}), the IR region is 
characterised by a number of broad, prominent features mainly 
due to Fe~II, Ni~II, Co~II, and Si~II \cite{ham02b}. 

\section{A Bridge Connecting Different Subgroups of Type Ia Supernovae ?} \label{disc}

\begin{figure*}
\includegraphics[width=13.3cm]{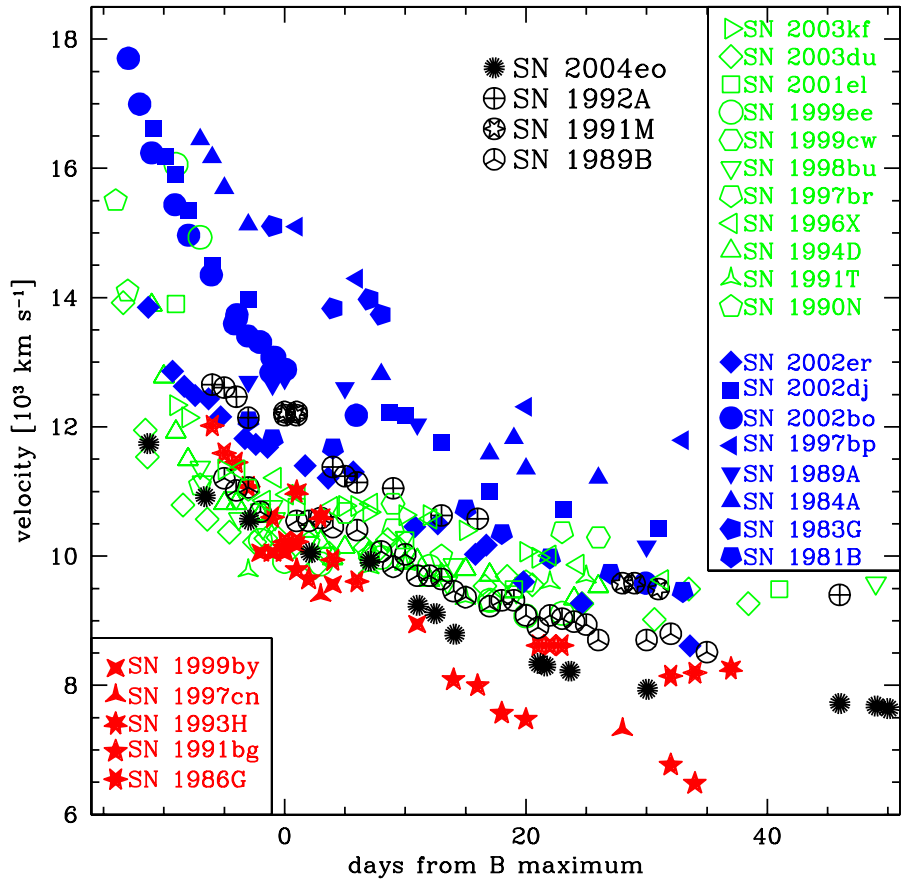}
\caption{Evolution of the Si~II $\lambda$6355 line velocity of SN~2004eo
and other SNe~Ia from Benetti et al. (2005). The colours adopted for
the different subclasses are the same as in Benetti et al.:
starred red symbols, {\it faint} SNe~Ia; filled blue symbols, {\it HVG}
SNe~Ia; open green symbols, {\it LVG} SNe~Ia.  The only exceptions
are for SNe~2004eo together with the ``nonstandard'' SNe 1989B, 1991M,
and 1992A, which are labelled with black symbols. For data source
references, see Benetti et al. (2005).
\label{spec_ev_vel} }
\end{figure*}

In Sect. \ref{param} we found that the best match to SN~2004eo is
SN~1992A \cite{nick96}.  Both these SNe show somewhat lower than
normal luminosity at maximum, $M_B \approx -19.1$ mag (see
Sect. \ref{param}), and rapidly declining light curves
($\Delta$m$_{15}(B)$ = 1.46 and 1.47, respectively).  Nevertheless,
there is some difference in the colours, SN~2004eo being redder than
SN~1992A at late phases.  There are also strong similarities in
the spectroscopic evolution (Sect. \ref{opt_spec}), but again with some
differences,  the most significant being in the line velocities
(lower in SN~2004eo) and the relative strengths of the most prominent
nebular features. This mixture of similarities and differences between
the two events supports the important conclusion of Benetti et
al. \shortcite{ben04} that a single-parameter characterisation of 
SNe~Ia does not specify the full diversity of their
observed behaviour (see also Hatano et al., 2000).

\begin{figure*}
\includegraphics[width=12.5cm,angle=0]{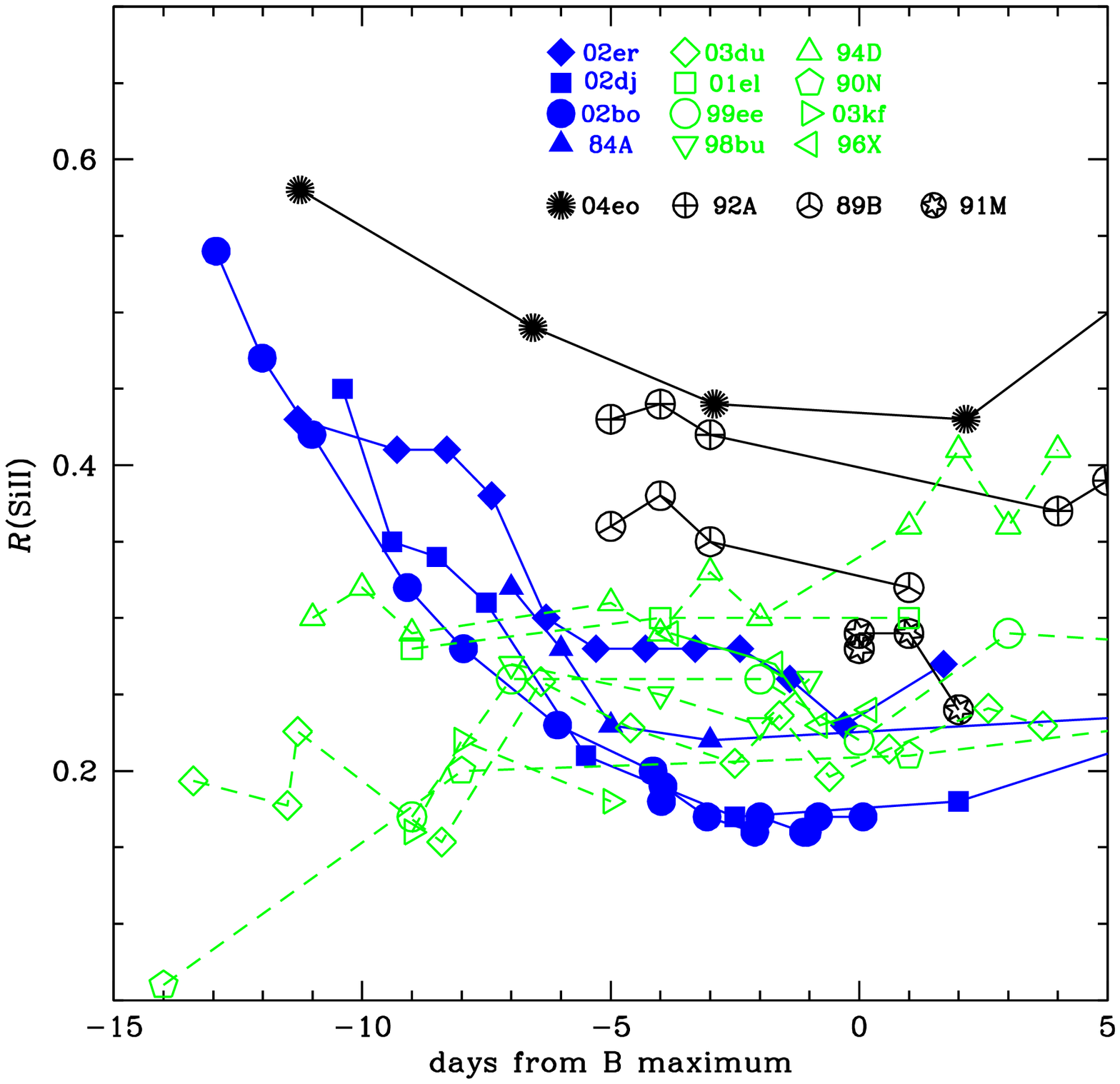}
\caption{Evolution of the depth ratio Si~II $\lambda$5972 / Si~II 
$\lambda$6355 in the spectra of SN~2004eo, together with those of 
other SNe~Ia from Benetti et al. (2005). Green open symbols, {\it LVG}
group; blue filled symbols, {\it HVG} group; black symbols, SN~2004eo
plus the nonstandard SNe 1989B, 1991M, and 1992A.
\label{spec_ev_RSi} }
\end{figure*}

In spite of the fact that SN 2004eo is not particularly underluminous
(see also the comparison with the bolometric light curve of the
low--luminosity SN~1991bg in Fig. \ref{bolo_cur},
Sect. \ref{color_uvoir}), it shows some characteristics of
low--luminosity SNe~Ia.  In general the spectroscopic evolution, the
strength of the Si~II lines, and the line velocities (as we will see
below) are not very different from those of SN~1999by.  
However, while the
$\Delta$m$_{15}(B)$ is higher and the colour redder than those of
normal SNe~Ia, they do not attain the extreme values of {\it faint}
SNe~Ia.  These findings suggest that SN~2004eo is a normal SN~Ia,
but with some properties (especially the line velocities) 
in common with low--luminosity SNe~Ia.

Benetti et al. \shortcite{ben05} analysed the photometric and
spectroscopic properties of a sample of well--observed SNe~Ia and
found that they can be divided into three different subgroups:
\begin{itemize}
\item {\it Faint} SNe~Ia similar to SN 1991bg, showing low expansion
velocities, rapid evolution of the Si~II $\lambda$6355 velocity, 
and low peak luminosity and temperature. The values of these 
parameters are the result of a low ejected $^{56}$Ni mass.
\item Normal--luminosity (mean absolute magnitude $M_B = -19.3$ mag) 
SNe~Ia, also showing a high velocity gradient ({\it HVG}), but 
having higher expansion velocities in the Si~II lines than do 
the {\it faint} SNe~Ia.
\item Normal and overluminous (SN 1991T-like) SNe~Ia, showing a low
velocity gradient ({\it LVG}) in the Si~II lines. These events
populate a narrow strip in the Si~II velocity evolution diagram
(cf. Fig. \ref{spec_ev_vel}).
\end{itemize}

\begin{figure*}
\includegraphics[width=8.8cm,angle=270]{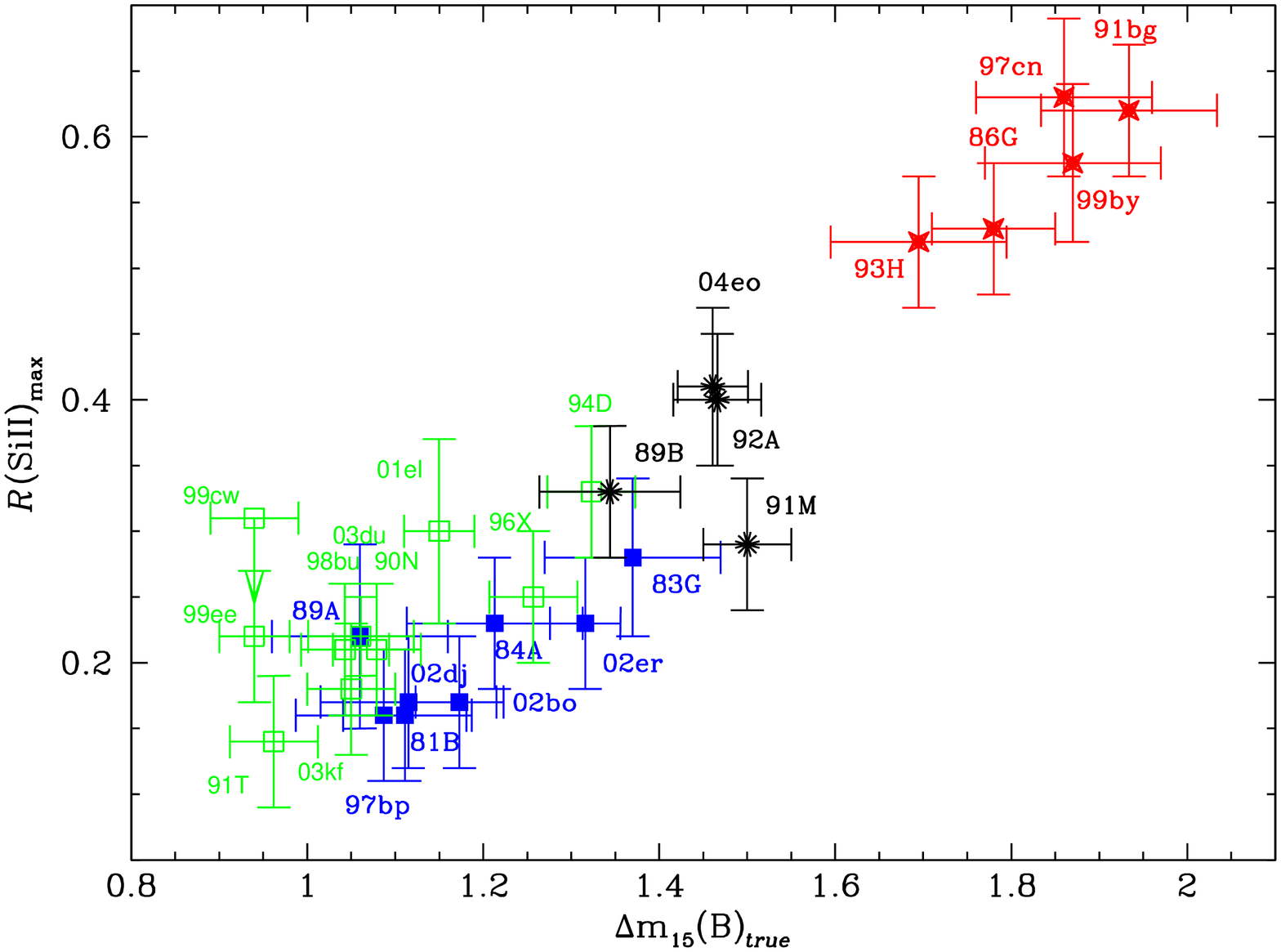}
\includegraphics[width=8.8cm,angle=270]{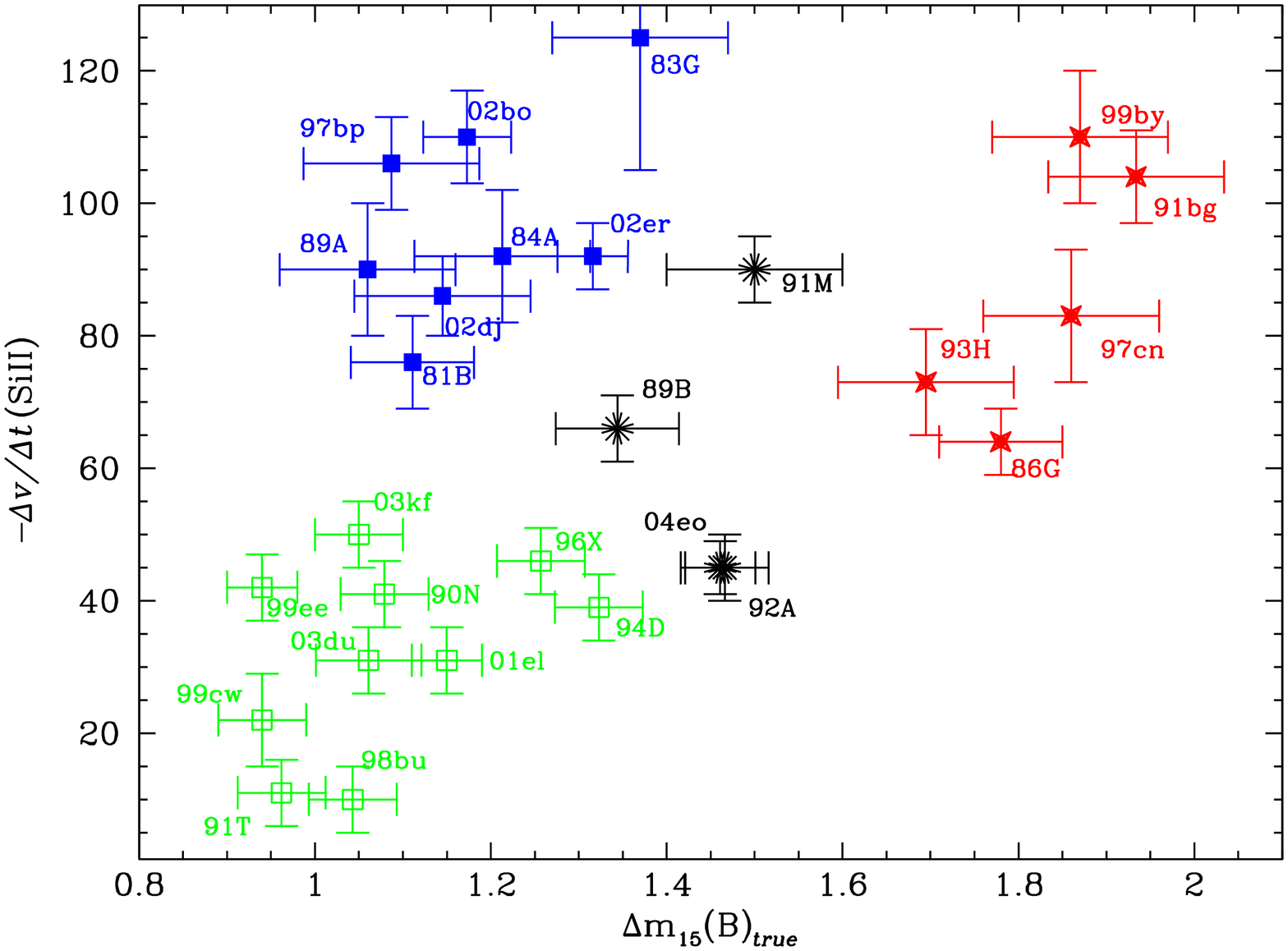}
\caption{Top: $R$(Si~II) vs. $\Delta m_{15}$ diagram.  Bottom:
$dv/dt$(Si~II) vs. $\Delta m_{15}$ diagram.  The nonstandard events
SNe~2004eo, 1989B, 1991M, and 1992A are shown as black asterisks. The other
SNe~Ia are from Benetti et al. (2004). Red stars, {\it faint} SNe~Ia;
filled squares, {\it HVG} SNe~Ia; open squares, {\it LVG} SNe~Ia. 
\label{RSi_vs_DM15} }
\end{figure*}

Branch et al. \shortcite{bran06} have presented an alternative
classification scheme in which SNe~Ia are divided into four 
subgroups on the basis of spectroscopic properties only:
\begin{itemize}
\item {\it Cool} SNe~Ia, whose spectra show red continua and are rich
in metal lines (Ti~II, Sc~II, Cr~II, Mg~I, Ca~I) that are prominent
only when the temperatures are low. They correspond to the {\it faint}
SNe~Ia of Benetti et al. (2005).
\item {\it Broad--line} SNe~Ia, spectroscopically normal, with
relatively strong and broad Si~II $\lambda$6355. They coincide 
roughly with the {\it HVG} SNe~Ia of Benetti et al.
\item {\it Core--normal} events, spectroscopically similar to the
previous group, but with narrower and less prominent Si~II.  They form
a spectroscopically highly homogeneous group and are included mainly
in the {\it LVG} group of the Benetti et al. classification.
\item {\it Shallow--silicon} SNe~Ia are spectroscopically peculiar,
having a high degree of heterogeneity in observed properties.  The
only common feature is the weakness of the Si~II absorption lines.  Some
examples (e.g., the high--temperature SN~1991T) show prominent Fe~III
features. As for the previous subdivision, they are included in the
{\it LVG} group (Benetti et al. 2005).
\end{itemize}
However, some SNe~Ia (e.g.,  SNe 1981B, 1989B, 1991M, and 1992A; see
Branch et al. 2006) are of uncertain classification. We shall now
compare three of these nonstandard events with SN~2004eo.  One of
the interesting issues is whether these transition events demonstrate that
there is actually a continuum of physical properties within the SN~Ia
class, as already established for other SN types (e.g., SNe~II-P; see
Hamuy \& Pinto \shortcite{ham02}, Pastorello \shortcite{pasto03}).

Benetti et al. \shortcite{ben05} found that, while the {\it faint} and
{\it HVG} SNe~Ia obey a plausible relation between $\Delta m_{15}(B)$ and
the depth ratio of the Si~II $\lambda$5972 and Si II $\lambda$6355
features [hereafter $R$(Si~II)], such a relation is weaker or even absent
if the {\it LVG} SNe~Ia are included.  Also, the pre-maximum evolution 
of the $R$(Si~II) parameter is strikingly different in {\it LVG} SNe~Ia
compared with {\it HVG} events.

In Fig. \ref{spec_ev_vel} the evolution of the velocity corresponding
to the Si~II $\lambda$6355 absorption minimum is shown for SN~2004eo
plus a wide range of SN~Ia subtypes.  The figure is an updated version
of that shown by Benetti et al. \shortcite{ben05}.  The symbols are
the same as those adopted by Benetti et al.  The black symbols show 
the data for SNe~2004eo together with the nonstandard events SN~1991M
(Padova--Asiago Supernova Group\footnotemark[7] archive),
\footnotetext[7]{{\it http://web.oapd.inaf.it/supern/}} SN~1989B
\cite{barb90,well94}, and SN~1992A (Padova--Asiago Supernova Group
Archive; Kirshner et al. 1993).  The latter two events were
considered as {\it LVG} SNe~Ia by Benetti et al. \shortcite{ben05}.
It is evident from Fig. \ref{spec_ev_vel} that SN~2004eo is a
low--velocity event, lying roughly midway between the {\it faint} and
{\it LVG} SNe~Ia.

In Fig. \ref{spec_ev_RSi} the time evolution of the $R$(Si~II) 
parameter is shown for SN~2004eo and a subsample of the SNe~Ia in
Fig. \ref{spec_ev_vel}.  The {\it faint} SNe~Ia, which would occupy
the upper part of this diagram, are not shown.  We note that for
SN~2004eo, the overall evolution is similar to that of the {\it HVG}
SNe~Ia, but that it has higher $R$(Si~II) values. A similar behaviour 
is observed for the borderline SN~1992A and (although to a lesser
extent) SN~1989B (Fig. \ref{spec_ev_RSi}).  These diagrams highlight
the considerable heterogeneity in the observed properties of SNe~Ia.

SN~2004eo plus the three nonstandard events seem to follow roughly
the relation between $R$(Si~II) and $\Delta m_{15}(B)$ for {\it faint}
and brighter (either {\it HVG} and {\it LVG}) events (Fig. 
\ref{RSi_vs_DM15}, top), to some extent filling in the gap 
between the three subtypes.  Benetti et al. \shortcite{ben05} 
found that no clear correlation exists between
$dv/dt$(Si~II) and $\Delta m_{15}(B)$, although the three groups of 
SNe~Ia seem to cluster at three different positions in the diagram
(Fig. \ref{RSi_vs_DM15}, bottom).
SN~2004eo plus the three nonstandard events lie roughly in the
middle of the diagram, away from any of the three clusters. In
particular, the position of SN~2004eo is identical to that of
SN~1992A (as measured by Hachinger et al. 2006). 
As already seen in Fig. \ref{spec_ev_vel}, these two SNe~Ia seem
to establish a ``link'' between the {\it faint} and the {\it LVG} 
SNe~Ia.  SN~1989B also contributes to this bridge.  In addition,
SN~1991M links the {\it faint} and the {\it HVG} SNe~Ia, although in
this case, the lack of spectra at phases later than about +30~d
makes the measurement of the velocity gradient of the Si~II
$\lambda$6355 line rather uncertain and possibly
overestimated\footnotemark[8].\\  
\footnotetext[8]{The value of
$dv/dt$(Si~II) for SN~1991M reported by Hachinger et
al. \protect\shortcite{hach06} has been updated with new estimates
using the spectra of Gomez et al. \protect\shortcite{gomez96}
available in the SUSPECT archive} 

We conclude that SN~2004eo should be considered as a member of the
group of nonstandard or ``transitional'' SNe~Ia, along with
SNe~1989B, 1991M, and 1992A.  These relatively rare events are
characterised by a moderate velocity gradient, relatively low ejecta
velocities, peculiar $R$(Si~II) evolution, and moderately high
$\Delta m_{15}(B)$.  Branch et al.  \shortcite{bran06} have also
identified SN~1992A, SN~1989B, and (possibly) the {\it HVG} SN~1991M
(see Hachinger et al. 2006) as transitional SNe. Several other SNe~Ia
with  $\Delta m _{15}(B)$ in the range 1.4--1.7 presented
by Jha et al. \shortcite{jha06} and Reindl et al. \shortcite{rein05}
are potential transitional objects (though those of Jha et al.
\shortcite{jha06} have other peculiarities as well). 
This makes it possible that, as 
the database of well-observed SNe~Ia grows, the clustered distribution 
of SNe~Ia in the subgroups of Benetti et al. (2005) will be less 
evident. Nevertheless, the majority of SNe~Ia may still fall
within the main subgroups.

\section{Conclusions} \label{conc}

We have presented optical and infrared photometric and
spectroscopic observations of the Type Ia SN~2004eo.  The data range
from $-$11~d to about +314~d from the $B$-band maximum light, with 
a gap of about three months when the SN went behind the Sun.

SN~2004eo has an absolute $B$-band magnitude at maximum ($M_B =
-19.08$) which is close to (or only marginally fainter than) the
average for SNe~Ia.  Consistent with its moderate luminosity, modeling
of the bolometric light curve indicates that about 0.45~M$_\odot$ of
$^{56}$Ni was ejected, which is close to the lower limit of the
$^{56}$Ni mass range observed in normal SNe~Ia.  However, its late--time red
colour, low line velocity, high value of $\Delta m _{15}(B)$ (1.46),
and evolution of the $R$(Si~II) parameter are peculiar for a 
``normal'' SN~Ia.  Its behaviour is very similar to that of SN~1992A 
and (to a smaller degree) SN~1989B. These three unusual SNe~Ia 
(together with SN~1991M) exhibit observational properties 
intermediate between the {\it faint} and more typical SNe~Ia. 

We conclude that there may exist a continuous range of physical
properties in SNe~Ia, rather than the discrete clusters proposed by
Benetti et al (2005). However, the resolution of this issue will
require larger numbers of well--observed SNe~Ia.

\section*{Acknowledgments}

The paper is based on observations collected at the 2.2-m and 3.5-m
telescopes of the
Centro Astron\'omico Hispano Alem\'an (Calar Alto, Spain),
the Asiago 1.82-m telescope, and  the AZT--24 Campo Imperatore Telescope 
(Pulkovo Observatory, Russia, and INAF Observatories, Italy), 
the Telescopio Nazionale Galileo, 
the Liverpool telescope, the William Herschel 
Telescope, and the Nordic Optical Telascope (La Palma, Spain), 
the ESO/MPI 2.2-m telescope (La Silla, Chile),
ESO--VLT (Cerro Paranal, Chile; ESO Program ID: 075.D-0346),
the 0.76-m Katzman Automatic Imaging Telescope and the 3-m Shane telescope 
(Lick Observatory, California,
USA), the Anglo--Australian Telescope, the 2.3-m and 
1-m telescopes (Siding Spring Observatory, Australia),
and the Tenagra 0.81-m telescope (Arizona, USA).
We thank the support astronomers at the Telescopio Nazionale Galileo, 
the Liverpool Telescope, the William Herschel Telescope, the Nordic 
Optical Telescope, the 2.2-m and
3.5m telescopes of Calar Alto, and the ESO--VLT and ESO/MPI 2.2-m telescope
for performing the follow--up observations of SN~2004eo.
We also thank R. Viotti for his collaboration during
ToO observations of SN~2004eo with the Asiago 1.82-m telescope.
A.P. thanks Marco Fiaschi for providing a $V$-band observation 
using the 0.41-m Newton telescope of the Gruppo Astrofili di Padova.\\
This work was supported by the European Union's Human
Potential Programme ``The Physics of Type Ia Supernovae,'' 
under contract HPRN-CT-2002-00303.
A.V.F.'s group at the University of California, Berkeley, is supported
by National Science Foundation (NSF) grant AST-0607485 and by the 
TABASGO Foundation. KAIT was made possible by generous donations
from Sun Microsystems, Inc., the Hewlett-Packard Company, AutoScope
Corporation, Lick Observatory, the NSF, the
University of California, and the Sylvia \& Jim Katzman
Foundation. \\
This research used the NASA/IPAC Extragalactic
Database (NED), which is operated by the Jet Propulsion Laboratory,
California Institute of Technology, under contract with the National
Aeronautics and Space Administration. We also made use of the Lyon-Meudon
Extragalactic Database (LEDA), supplied by the LEDA team at
the Centre de Recherche Astronomique de Lyon, Observatoire de Lyon.

\end{document}